\documentclass{libpaper/arxiv}
\pdfoutput=1  %
\usepackage{fullpage}
\usepackage{placeins}

\usepackage{hyperref}
\hypersetup{
    colorlinks,
    citecolor=black,
    filecolor=black,
    linkcolor=black,
    urlcolor=black
}

\makeatletter                   %
\def\mdseries@tt{m}             %
\makeatother                    %

\usepackage[plain]{fancyref}
\usepackage[draft=true]{minted} %
\usepackage{color}
\usepackage{breakurl}           %

 \usepackage{lipsum}
\usepackage[final]{pdfpages}
\usepackage{setspace}
\usepackage{epsf}
\usepackage{epsfig}
\usepackage{graphicx}
\usepackage{algorithmic}
\usepackage{amsmath}
\usepackage{letltxmacro}

\usepackage{enumitem}
\usepackage{tikz}
\usepackage{libpaper/tightenum}

\usepackage{url}
\usepackage{multirow}

  \renewenvironment{thebibliography}[1]{%
    \begin{oldthebibliography}{#1}%
      \setlength{\parskip}{0ex}%
      \setlength{\itemsep}{0ex}%
  }%
  {%
    \end{oldthebibliography}%
  }

\newwrite\arxivdeps
\immediate\openout\arxivdeps=\jobname-arxivdeps.log

\newcommand\verifymarkedforarxivfile[1]{%
\ifdefined\arxivbuild
\else
\IfFileExists{#1}%
{}%
{\GenericWarning{Marked file (#1) for inclusion in arxiv build doesn't exist}}%
\fi%
}

\newcommand\markforarxiv[1]{%
\verifymarkedforarxivfile{#1}%
\write\arxivdeps{IncludeInArxiv: #1}%
}

\newcommand\markforarxivdata[1]{%
\verifymarkedforarxivfile{#1}%
\write\arxivdeps{ArxivData: #1 As: #1}%
}

\newcommand\markforarxivdataas[2]{%
\verifymarkedforarxivfile{#1}%
\write\arxivdeps{ArxivData: #1 As: #2}%
}

\DeclareUrlCommand\UScore{\urlstyle{rm}}

\newcommand\dataref[1]{%
\markforarxivdata{#1}%
\UScore{#1}}

\newcommand\datarefas[2]{%
\markforarxivdataas{#1}{#2}%
\UScore{#2}%
}

\LetLtxMacro\oldincludegraphics\includegraphics
\renewcommand{\includegraphics}[2][]{%
\markforarxiv{#2}%
\oldincludegraphics[#1]{#2}}

\LetLtxMacro\oldincludepdf\includepdf
\renewcommand{\includepdf}[2][]{%
\markforarxiv{#2}%
\oldincludepdf[#1]{#2}}

\def\nfigure[#1,#2,#3]{
\begin{figure}
\vspace*{0mm}
\begin{center}

\includegraphics[width=\columnwidth]{#1} 
\vspace*{-6mm}\caption[]{#2
} \label{#3}

\vspace*{-3mm}
\end{center}
\end{figure}}

\def\cfigure[#1,#2,#3]{
\begin{figure}
\vspace*{0mm}
\begin{center}

\includegraphics[width=3in]{#1} 
\vspace*{-3mm}\caption[]{#2
} \label{#3}
 
\vspace*{-5mm}
\end{center}
\end{figure}}

\def\cfigurefour[#1,#2,#3]{
\begin{figure}
\vspace*{0mm}
\begin{center}

\includegraphics[width=4in]{#1} 
\vspace*{-3mm}\caption[]{#2
} \label{#3}
 
\vspace*{-5mm}
\end{center}
\end{figure}}

\def\cfiguretemp[#1,#2,#3]{
\begin{figure}
\vspace*{0mm}
\begin{center}

\includegraphics[width=3.5in]{#1} 
\vspace*{-3mm}\caption[]{#2
} \label{#3}
 
\vspace*{-5mm}
\end{center}
\vspace*{-2mm}
\end{figure}}

\def\wfigure[#1,#2,#3]{
\begin{figure*}
\vspace*{0mm}
\begin{center}
 \includegraphics[width=\textwidth]{#1} 
 \vspace*{-3mm}\caption[]{#2
} \label{#3}
 
\end{center}
\end{figure*}}

\def\threefigure[#1,#2,#3,#4,#5]{
\begin{figure*}
\vspace*{0mm}
\begin{center}

\begin{tabular}{ccc}
\includegraphics[width=2in]{#1} & \includegraphics[width=2in]{#2} &  \includegraphics[width=2in]{#3} \\
(a) & (b) & (c) \\
\end{tabular}
\vspace*{-3mm}\caption[]{#4
} \label{#5}

\vspace*{-5mm}
\end{center}
\vspace*{-2mm}
\end{figure*}}

\def\dcfigure[#1,#2,#3,#4,#5,#6]{
{
\begin{figure*}
\begin{center}
\begin{minipage}[c]{\columnwidth}{
\includegraphics[width=\columnwidth]{#1} 
\vspace*{0mm}\caption[]{#2} \label{#3} \
}\end{minipage}\hspace*{\columnsep}\
\begin{minipage}[c]{\columnwidth}{
\includegraphics[width=\columnwidth]{#4} 
\vspace*{0mm}\caption[]{#5}\label{#6} \
}\end{minipage}
\end{center}
\end{figure*}
}
}

\def\scfigure[#1,#2,#3]{
{
\begin{figure*}
\begin{center}
\begin{minipage}[c]{3.5in}{
\includegraphics[width=3.5in]{#1} 
}\end{minipage}
\caption[]{#2} \label{#3} \
\end{center}
\end{figure*}
}
}

\def\tableByTable[#1,#2,#3,#4,#5,#6]{
{
\begin{table*}
\begin{center}
\begin{minipage}[c]{3in}{
\centering
{#1}
\vspace*{0mm}\tabcaption[]{#2}\label{#3} \
}\end{minipage}\hspace*{\columnsep}\
\begin{minipage}[c]{3in}{
\centering
{#4}
\vspace*{0mm}\tabcaption[]{#5}\label{#6} \
}\end{minipage}
\end{center}
\end{table*}
}
}

\def\figureByTable[#1,#2,#3,#4,#5,#6]{
{
\begin{figure*}
\begin{center}
\begin{minipage}[c]{3in}{
\centering
\includegraphics[width=\textwidth]{#1}
\vspace*{0mm}\figcaption[]{#2} \label{#3} \
}\end{minipage}\hspace*{\columnsep}\
\begin{minipage}[c]{3.3in}{
\centering
{#4}
\vspace*{0mm}\tabcaption[]{#5}\label{#6} \
}\end{minipage}
\end{center}
\end{figure*}
}
}

\def\tableByFigure[#1,#2,#3,#4,#5,#6]{
{
\begin{figure*}
\begin{center}
\begin{minipage}[c]{4.3in}{
\centering
{#1}
\vspace*{0mm}\tabcaption[]{#2} \label{#3} \
}\end{minipage}\hspace*{\columnsep}\
\begin{minipage}[c]{2.2in}{
\centering
\includegraphics[width=\textwidth]{#4}
\vspace*{-0.35in}\caption[]{#5}\label{#6} \
}\end{minipage}
\end{center}
\end{figure*}
}
}

\def\doublecfigure[#1,#2,#3,#4]{
{
\begin{figure}
\begin{center}
\begin{minipage}[c]{1.5in}{
\begin{center}
\includegraphics[width=1.5in]{#1}%
\end{center}
}\end{minipage}\hspace*{1em}\
\begin{minipage}[c]{1.5in}{
\begin{center}
\includegraphics[width=1.5in]{#2}%
\end{center}
}\end{minipage}
\vspace*{0mm}\caption[]{#3} \label{#4} \
\end{center}
\end{figure}
}
}

\def\qcfigure[#1,#2,#3,#4,#5,#6]{
{
\begin{figure*}
\vspace*{0.2in}\
\begin{center}
\begin{minipage}[c]{3in}{
\includegraphics[width=3in]{#1} 
\vspace*{-3mm}
}
\end{minipage}\hspace*{0.5in}\
\begin{minipage}[c]{3in}{
\includegraphics[width=3in]{#2} 
\vspace*{-3mm}
}\end{minipage}

\begin{minipage}[c]{3in}{
\includegraphics[width=3in]{#3} 
\vspace*{-3mm}
}
\end{minipage}\hspace*{0.5in}\
\begin{minipage}[c]{3in}{
\includegraphics[width=3in]{#4} 
\vspace*{-3mm}
}\end{minipage}
\end{center}
\caption[]{#5}\label{#6}
\end{figure*}
}
}

\def\twfigure[#1,#2,#3,#4,#5]{
{
\begin{figure*}
\vspace*{0.2in}\
\begin{center}
\begin{minipage}[c]{6.5in}{
\includegraphics[width=6.5in]{#1} 
\vspace*{-3mm}
}
\end{minipage}

\begin{minipage}[c]{6.5in}{
\includegraphics[width=6.5in]{#2} 
\vspace*{-3mm}
}\end{minipage}

\begin{minipage}[c]{6.5in}{
\includegraphics[width=6.5in]{#3} 
\vspace*{-3mm}
}
\end{minipage}
\end{center}
\caption[]{#4}\label{#5}
\end{figure*}
}
}

\def\dwfigure[#1,#2,#3,#4]{
{
\begin{figure*}
\vspace*{0.2in}\
\begin{center}
\begin{minipage}[c]{6.5in}{
\includegraphics[width=6.5in]{#1} 
\vspace*{-3mm}
}
\end{minipage}

\begin{minipage}[c]{6.5in}{
\includegraphics[width=6.5in]{#2} 
\vspace*{-3mm}
}\end{minipage}

\end{center}
\caption[]{#3}\label{#4}
\end{figure*}
}
}

\def\dssfigure[#1,#2,#3,#4,#5,#6]{
{
\begin{figure*}
\vspace*{0.2in}\
\begin{center}
\begin{minipage}[c]{4in}{
\includegraphics[width=4in]{#1}
\vspace*{-3mm}\caption[]{#2} \label{#3} \
}\end{minipage}\hspace*{0.5in}\
\begin{minipage}[c]{2in}{
\includegraphics[width=2in]{#4}
\vspace*{-3mm}\caption[]{#5}\label{#6} \
}\end{minipage}
\end{center}
\vspace*{-0.4in}\
\end{figure*}
}
}

\def\dsfigure[#1,#2,#3,#4,#5,#6]{
{
\begin{figure*}
\vspace*{0.2in}\
\begin{center}
\begin{minipage}[c]{3in}{
\includegraphics[width=3in]{#1}
\vspace*{-3mm}\caption[]{#2} \label{#3} \
}\end{minipage}\hspace*{0.5in}\
\begin{minipage}[c]{3in}{
\hspace*{0.5in}\
\includegraphics[height=3in]{#4}
\vspace*{-3mm}\caption[]{#5}\label{#6} \
}\end{minipage}
\end{center}
\vspace*{-0.4in}\
\end{figure*}
}
}

\def\dsyfigure[#1,#2,#3,#4,#5,#6]{
{
\begin{figure*}
\vspace*{0.2in}\
\begin{center}
\begin{minipage}[c]{2.5in}{
\includegraphics[height=2.5in]{#1}
\vspace*{-3mm}\caption[]{#2} \label{#3} \
}\end{minipage}\hspace*{0.5in}\
\begin{minipage}[c]{2.5in}{
\includegraphics[height=2.5in]{#4}
\vspace*{-3mm}\caption[]{#5}\label{#6} \
}\end{minipage}
\end{center}
\vspace*{-0.4in}\
\end{figure*}
}
}

\def\dyfigure[#1,#2,#3,#4,#5,#6]{
{
\begin{figure*}
\vspace*{0.2in}\
\begin{center}
\begin{minipage}[c]{3in}{
\includegraphics[height=3in]{#1} 
\vspace*{-3mm}\caption[]{#2} \label{#3} \
}\end{minipage}\hspace*{0.5in}\
\begin{minipage}[c]{3in}{
\includegraphics[height=3in]{#4} 
\vspace*{-3mm}\caption[]{#5}\label{#6} \
}\end{minipage}
\end{center}
\vspace*{-0.4in}\
\end{figure*}
}
}

\def\dyoldfigure[#1,#2,#3,#4,#5,#6]{
{
\begin{figure*}
\vspace*{0.2in}\
\begin{center}
\begin{minipage}[c]{3in}{
\epsfysize=2.0in\
\hspace{0.5in}\
\epsfbox{#1}
\vspace*{-3mm}\caption[]{#2} \label{#3} \
}\end{minipage}\hspace*{0.25in}\
\begin{minipage}[c]{3in}{
\epsfysize=2.0in\
\hspace{0.5in}\
\epsfbox{#4}
\vspace*{-3mm}\caption[]{#5}\label{#6} \
}\end{minipage}
\end{center}
\vspace*{-0.4in}\
\end{figure*}
}
}

\def\cfiguredouble[#1,#2,#3,#4]{
\begin{figure}
\vspace*{0.2in}\
\begin{center}
\begin{minipage}[c]{1.5in}{
\epsfxsize=1.5in\
\epsfbox{#1}
}\end{minipage}\hspace*{0.1in}\
\begin{minipage}[c]{1.5in}{
\epsfxsize=1.5in\
\vspace{0.1in}\epsfbox{#2}
}\end{minipage}\vspace*{-0.10in} \caption[]{#3}\label{#4}
\end{center}
\vspace*{-0.4in}\
\end{figure}
}

\def\wpfigure[#1,#2,#3,#4]{
\begin{figure*}
\vspace*{4mm}
\begin{center}

\includegraphics[width=#4]{#1} 

\vspace*{-3mm}\caption[]{#2
} \label{#3}

\vspace*{-5mm}
\end{center}
\end{figure*}}

\def\wprfigure[#1,#2,#3,#4,#5]{
\begin{figure*}
\vspace*{4mm}
\begin{center}

\includegraphics[width=#4, angle=#5]{#1} 

\vspace*{-3mm}\caption[]{#2
} \label{#3}

\vspace*{-5mm}
\end{center}
\end{figure*}}

\def\DoubleFigureWSlide[#1,#2,#3,#4,#5,#6,#7,#8,#9]{
\begin{figure*}
\vspace*{#9}
\begin{center}
\begin{minipage}{#4}
\includegraphics[width=#4]{#1}
\vspace*{-3mm}\caption{#2
}\label{#3}
\end{minipage}
\hspace{2em}
\begin{minipage}{#8}
\includegraphics[width=#8]{#5}
\vspace*{-3mm}\caption{#6
}\label{#7}
\end{minipage}
\vspace*{-5mm}
\end{center}
\end{figure*}
}

\def\DoubleFigureW[#1,#2,#3,#4,#5,#6,#7,#8]{
\begin{figure*}
\vspace*{0in}
\begin{center}
\begin{minipage}{#4}
\includegraphics[width=#4]{#1}
\vspace*{-3mm}\caption{#2
}\label{#3}
\end{minipage}
\hspace{2em}
\begin{minipage}{#8}
\includegraphics[width=#8]{#5}
\vspace*{-3mm}\caption{#6
}\label{#7}
\end{minipage}
\vspace*{-5mm}
\end{center}
\end{figure*}
}

\def\DoubleFigureWHack[#1,#2,#3,#4,#5,#6,#7,#8]{
\begin{figure*}
\vspace*{0in}
\begin{center}
\begin{minipage}{3in}
\includegraphics[width=#4]{#1}
\vspace*{-3mm}\caption{#2
}\label{#3}
\end{minipage}
\hspace{2em}
\begin{minipage}{3in}
\includegraphics[width=#8]{#5}
\vspace*{-3mm}\caption{#6
}\label{#7}
\end{minipage}
\vspace*{-5mm}
\end{center}
\end{figure*}
}

\def\ddcfigure[#1,#2,#3,#4]{
\begin{figure*}
\vspace*{0.2in}\
\begin{center}
\begin{minipage}[c]{\columnwidth}{
\includegraphics[width=\columnwidth]{#1} 
}\end{minipage}\hspace{0.5in}\
\begin{minipage}[c]{\columnwidth}{
\includegraphics[width=\columnwidth]{#2} 
}\end{minipage} \caption[]{#3}\label{#4}
\end{center}
\end{figure*}
}

\def\ddcfigureSlide[#1,#2,#3,#4,#5]{
\begin{figure*}
\vspace*{#5}\
\begin{center}
\begin{minipage}[c]{3in}{
\includegraphics[height=3in]{#1} 
}\end{minipage}\hspace{0.5in}\
\begin{minipage}[c]{3in}{
\includegraphics[height=3in]{#2} 
}\end{minipage}\vspace*{-0.10in} \caption[]{#3}\label{#4}
\end{center}
\vspace*{-0.4in}\
\end{figure*}
}

\def\cxfigure[#1,#2,#3]{
\begin{figure}
\vspace*{4mm}
\begin{center}
 
\epsfxsize=2.5in\
\epsfbox{#1}\
 
\vspace*{-0.10in}\caption[]{#2
} \label{#3}
 
\vspace*{-5mm}
\end{center}
\vspace*{-2mm}
\end{figure}}

\newif\ifremark
\long\def\remark#1{
\ifremark%
        \begingroup%
        \dimen0=\columnwidth
        \advance\dimen0 by -1in%
        \setbox0=\hbox{\parbox[b]{\dimen0}{\protect\em #1}}
        \dimen1=\ht0\advance\dimen1 by 2pt%
        \dimen2=\dp0\advance\dimen2 by 2pt%
        \vskip 0.25pt%
        \hbox to \columnwidth{%
                \vrule height\dimen1 width 3pt depth\dimen2%
                \hss\copy0\hss%
                \vrule height\dimen1 width 3pt depth\dimen2%
        }%
        \endgroup%
\fi}

\usepackage{xcolor}

\definecolor{cyanish}{rgb}{0,0.8,1.0}
\definecolor{orange}{rgb}{1.0,0.5,0.0}
\definecolor{pink}{rgb}{1.0,0.47,0.6}
\definecolor{light-gray}{gray}{0.95}
\definecolor{jiancolor}{RGB}{0,153,153}
\definecolor{mygreen}{RGB}{50,200,50}
\definecolor{pink}{rgb}{1.0,0.47,0.6}

\newcommand{\boldparagraph}[1]{\vspace*{1ex}\noindent\textbf{#1}\hspace{1em}}

\newcommand{\ignore}[1]{}

\newcommand{\reffig}[1]{Figure~\ref{#1}}

\newcommand{\refsec}[1]{Section~\ref{#1}}
\newcommand{\reftab}[1]{Table~\ref{#1}}

\newcommand{\us}{$\mu$s}
\newcommand{\x}[1]{$\times$}
\newcommand{\figtitle}[1]{\textbf{#1}}

\newcommand{\nova}[1]{\texttt{NOVA#1}}
\newcommand{\extfs}[1]{\texttt{Ext4#1}}
\newcommand{\extfsDAX}[1]{\extfs{-DAX#1}}
\newcommand{\xfs}[1]{\texttt{XFS#1}}
\newcommand{\xfsDAX}[1]{\xfs{-DAX#1}}

\usepackage{booktabs}
\usepackage{subcaption}
\usepackage{tikz} 
\usepackage{pgfplots,pgfplotstable}
\usepackage{amsthm}
\usepgfplotslibrary{groupplots}

\usepackage{thmtools}
\usepackage{thm-restate}
\usepackage{etoolbox}

\makeatletter
\def\recommend@tions{}
\newcounter{recommend@cnt}

\newenvironment{observation}{%
  \stepcounter{recommend@cnt}%
  \begingroup\edef\x{\endgroup
    \noexpand\restatable{thmrecommend}{recommend\Alph{recommend@cnt}}%
  }\x
}{\endrestatable
  \xappto\recommend@tions{%
    \expandafter\noexpand\csname recommend\Alph{recommend@cnt}\endcsname*%
  }%
}
\newcommand{\allrecommendations}{\recommend@tions}
\makeatother

\newcommand{\takeaway}[2]{%
	\vspace*{4pt}
	\noindent\fbox{%
	\parbox{\textwidth}{%
	\begin{observation}{#1} %
	\normalfont{#2}%
	\end{observation}}}%
	\vspace*{4pt}
}

\definecolor{color-0}{rgb}{0.82,0.10,0.26}
\definecolor{color-1}{rgb}{0.00,0.00,1.00}
\definecolor{color-2}{rgb}{0.00,0.50,0.00}
\definecolor{color-3}{rgb}{1.00,0.65,0.00}
\definecolor{color-4}{rgb}{0.60,0.20,0.80}
\definecolor{color-5}{rgb}{0.57,0.51,0.32}
\definecolor{color-6}{rgb}{0.85,0.20,0.53}
\definecolor{color-7}{rgb}{0.33,0.33,0.33}
\definecolor{color-8}{rgb}{0.29,0.00,0.51}
\definecolor{color-9}{rgb}{0.00,0.29,0.29}

\newcommand{\embargo}[1]{Redacted}
\newcommand{\XP}{Optane DC}
\newcommand{\XPDIMM}{\XP{} PMM}
\newcommand{\XPCommercial}{Intel\textregistered{} Optane\texttrademark{} DC Persistent Memory Module}
\newcommand{\APC}{controller}
\newcommand{\lattest}{LATTester}

\newcommand{\MM}{MM}
\newcommand{\PM}{PM}

\newcommand{\MMLDRAM}{\MM{}-LDRAM}
\newcommand{\MMPMEMC}{\MM{}-Optane-Cached}
\newcommand{\MMPMEMUC}{\MM{}-Optane-Uncached}

\newcommand{\PMLDRAM}{\PM{}-LDRAM}
\newcommand{\PMRDRAM}{\PM{}-RDRAM}
\newcommand{\PMLPMEM}{\PM{}-Optane}

\newcommand{\SSDOptane}{SSD-Optane}
\newcommand{\SSDSATA}{SSD-SATA}

\newcommand{\emusize}{80}

\newcommand{\KernelVersion}{4.13.0}

\newcommand{\aeppersocketsizetb}{1.5}
\newcommand{\aepdimmsizegb}{256}
\newcommand{\aeptotalsizetb}{3}
\newcommand{\drampersocketsizegb}{192}
\newcommand{\aepfirmware}{01.01.00.5253}

\newcommand\seedatain[1]{(see data in~\dataref{#1})}
\newcommand\seedatainas[2]{(see data in~\datarefas{#1}{#2})}
\def\ntwfigure[#1,#2,#3]{
\begin{figure*}
\vspace*{0mm}
\begin{center}
 \includegraphics[width=5in]{#1} 
 \markforarxiv{#1}
 \vspace*{-3mm}\caption[]{#2
} \label{#3}
 
\end{center}
\end{figure*}}

\def\doublerfigure[#1,#2,#3,#4]{
\begin{figure*}
\begin{center}
\begin{subfigure}[c]{0.92\columnwidth}{
\includegraphics[width=\columnwidth]{#1} 
\caption{}
}\end{subfigure}
\begin{subfigure}[c]{0.88\columnwidth}{
\includegraphics[width=\columnwidth]{#2}
\caption{}
}\end{subfigure}
\caption[]{#3}
\label{#4}
\end{center}
\end{figure*}
}

\newcommand{\gitcommit}{2019-08-09 7041bc9}
\def\arxivbuild{}

\usepackage{fancyhdr}
\newcommand{\redact}[1]{}

\let\Oldsection\section %
\renewcommand{\section}{\clearpage \FloatBarrier \Oldsection}

\let\Oldsubsection\subsection
\renewcommand{\subsection}{\FloatBarrier \Oldsubsection}

\let\Oldsubsubsection\subsubsection
\renewcommand{\subsubsection}{\FloatBarrier \Oldsubsubsection}

\makeatletter
\renewcommand*{\fps@figure}{htbp!}
\makeatother

\makeatletter
\renewcommand*{\fps@table}{htbp!}
\makeatother

\remarktrue

\begin{document}

\title{Basic Performance Measurements of the\\ Intel Optane DC Persistent Memory Module\\{\small Or: It's Finally Here!  How Fast is it?}}

\author{
  Joseph Izraelevitz \and 
  Jian Yang \and 
  Lu Zhang \and
  Juno Kim \and 
  Xiao Liu \and 
  Amirsaman Memaripour \and 
  Yun Joon Soh \and 
  Zixuan Wang \and 
  Yi Xu \and 
	Subramanya R.\ Dulloor \and
  Jishen Zhao \and 
  Steven Swanson\footnote{Correspondence should be directed to swanson@cs.ucsd.edu.}\\
\and 
\includegraphics[width=2in]{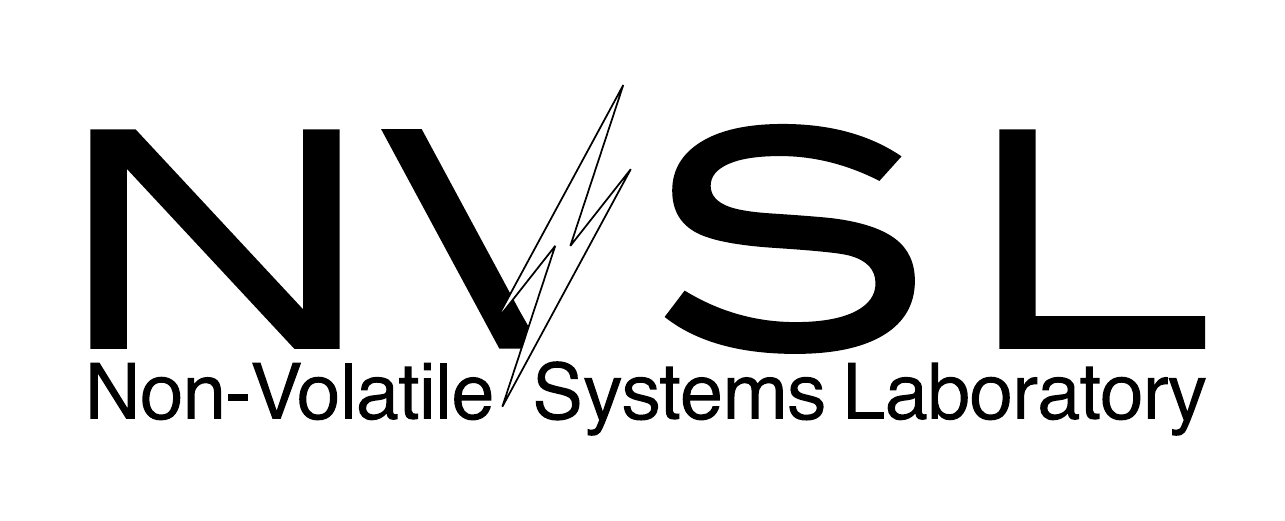}\\ 
Computer Science \& Engineering\\
University of California, San Diego}

\date{}

\maketitle
\thispagestyle{fancy}

\begin{abstract}
After nearly a decade of anticipation, scalable 
nonvolatile memory DIMMs are finally commercially available
with the release of the \XPCommercial{} (or just ``\XPDIMM{}'').
This new nonvolatile DIMM supports byte-granularity
accesses with access times on the order of DRAM, while
also providing data storage that survives power outages.

This work comprises the first in-depth, scholarly, performance
review of Intel's \XPDIMM{}, exploring its
capabilities as a main memory device, 
and as persistent, byte-addressable
memory exposed to user-space applications.  For the past
several months, our group has had access to machines with
\XP{} memory and has investigated the \XPDIMM{}'s performance
characteristics.  This report
details the chip's performance under a number of modes
and scenarios, and across a wide variety of both
micro- and macro-scale benchmarks.  In total,
this report represents approximately
330 hours of machine time.

\XP{} memory occupies
a tier in-between SSDs and DRAM.  It has higher
latency (346 ns) than DRAM but lower latency than 
an SSD.  Unlike DRAM, its bandwidth is
asymmetric with respect to access type: 
for a single \XPDIMM{}, 
its max read bandwidth is 6.6~GB/s, whereas its max write bandwidth is
2.3~GB/s.  However, the expected price
point of \XP{} memory means that machines with large quantities
of \XP{} memory are feasible --- our test machine has
\aeptotalsizetb~TB of \XP{} memory across two sockets.

\XPDIMM{}s can be used as large memory devices
with a DRAM cache to hide their lower bandwidth and 
higher latency.  When used in this Memory (or cached) mode,
\XP{} memory has little impact on applications with small
memory footprints.  Applications with larger memory footprints
may experience some slow-down relative to DRAM, but are now able
to keep much more data in memory.

In contrast, in App Direct (or uncached) mode, 
\XPDIMM{}s can be used as a persistent
storage device.  When used under a file system, this configuration
can result in significant performance gains, especially when the file system
is optimized to use the load/store interface of the \XPDIMM{} and the application
uses many small, persistent writes.  For instance, 
using the NOVA-relaxed NVMM file system, we can improve the performance of
Kyoto Cabinet by almost 2\x{}.

In App Direct mode, \XPDIMM{}s can also be used to enable user-space persistence
where the application explicitly controls its writes into persistent \XP{}
media.  By modifying the actual application, application programmers
can gain additional performance benefits 
since persistent updates bypass both the kernel and file system.  
In our experiments, modified applications that used user-space \XP{}
persistence generally outperformed their file system counterparts;
for instance, the user-space persistent version of RocksDB 
performed almost 2\x{} faster than the equivalent program utilizing
an NVMM-aware file system.

This early report is only the beginning in an effort to understand
these new memory devices.
We hope that these results
will be enlightening to the research community in general and 
will be useful in guiding future work into nonvolatile memory systems.

\end{abstract}

\section*{How to Use this Document}
\label{sec:howtouse}

Specialists in different
areas will be interested in different 
sections. Researchers who are interested in the basic characteristics
of \XP{} memory should pay close attention to 
Section~\ref{sec:basic}. 
Application developers that use large amounts of memory
should read Section~\ref{sec:big} to see how
\XP{} memory performs when used as a very large main memory device.
File systems and storage researchers
should head to Section~\ref{sec:storage} to see how \XP{} memory
affects file systems.  Persistent memory
researchers should see Section~\ref{sec:pmem} to see how
prototype persistent memory libraries perform when 
run on real \XPDIMM{}s and how prior methods
of emulation compare.

All data presented in this document is included in the
arxiv directory under the \texttt{anc} folder.  All
figures are tagged with a reference to their data files.

We have called out ``observations'' in boxes throughout this document. 
These observations represent key facts or findings about the Intel's
\XPDIMM{}.  In general, we highlight findings that are useful to a
wide group of readers, or that represent key statistics about the
device.

We welcome and will try to answer any questions about
the data or our methodology.  However, many aspects of Intel's design
are still not publicly available, so we may be limited on
the information that we can provide.

\section*{Versions}

This is Version 1.0.1 of this document. 
\\

\begin{description}
\item[Version 0.1.0 (3/13/2019)] The initial release of this document, with a number of results still under embargo.
\item[Version 1.0.0 (4/3/2019)] The first full release of this document, with all results released from embargo and including non-normalized results.
\item[Version 1.0.1 (8/9/2019)] Updated release of this document. Includes minor textual clarifications, new experiments on device bandwidth, and errata.  In particular, earlier versions stated that the cache-line size for data movement in Memory Mode between \XP{} and DRAM memory was 4~KB --- it is in fact 64~bytes.
\end{description}

\section*{Executive Summary}

For the last ten years, researchers have been anticipating the arrival of
commercially available, scalable non-volatile main memory (NVMM) technologies
that provide byte-granularity storage and survive power outages.  Recently,
Intel released a product based on one of these
technologies: the \XPCommercial{} (or just ``\XPDIMM{}'').

Researchers have not waited idly for real nonvolatile DIMMs (NVDIMMs) to arrive\footnote{\XPDIMM{} are not technically NVDIMMs since they do not comply with any of the NVMM-F, -N, or -P JEDEC standards.}.
Over the past decade, they have written a slew of papers proposing new
programming models~\cite{nvheaps,mnemosyne,pmdk}, file
systems~\cite{novapaper,BPFS,pmfs}, and other tools built to exploit the
performance and flexibility that NVDIMMs promised to deliver.

Now that \XPDIMM{}s are finally here, researchers can begin to grapple
with their complexities and idiosyncrasies.  We have started that
process over the last several months by putting \XP{} memory through its
paces on test systems graciously provided by Intel.

This report describes how \XPDIMM{}s attach to the processor and
summarizes our findings about basic \XP{} performance as an extension of
volatile DRAM, as a fast storage medium in a conventional storage stack, and as
non-volatile main memory.
The goal of this report is to help the computer architecture and systems
research community develop intuition about this new memory technology behaves.

This executive summary presents our key findings and provides a snapshot of the
data on \XP{} that we think are most useful.  The full report provides more
detail, a comparison to multiple memory technologies (e.g., DRAM used to
emulated \XP{}), data for additional software components, and much more data.
It also provides pointers to the raw data underlying each of the graphs.

\subsection*{Background~(\refsec{sec:method})}

Like traditional DRAM DIMMs, the \XPDIMM{} sits on the memory bus and connects
to the processor's on-board memory controller.  Our test systems use Intel's
new second generation Xeon Scalable processors 
(codenamed Cascade Lake).  A single CPU can host six \XPDIMM{}s
for a total of 3~TB of \XP{} memory.  The memory controller communicates with
the \XPDIMM{} via a custom protocol that is mechanically and electrically
compatible with DDR4 but allows for variable-latency memory transactions.
\XPDIMM{}s currently come in three capacities: 128~GB, 256~GB, and 512~GB.  In
this article, we report numbers for 256~GB \XPDIMM{}s.

Cascade Lake includes a suite of instruction to enforce ordering constraints
between stores to \XP{}.  Some of these have existed for a long time (e.g.,
\texttt{sfence} and non-temporal stores that bypass the caches), but others are
new.  For example, \texttt{clwb} writes back a cache line without necessarily
invalidating it.

\XPDIMM{}s can operate in two modes:
\emph{Memory} and \emph{App Direct} modes.

Memory mode uses \XP{} to expand main memory capacity without
persistence.  It combines a \XPDIMM{} with a conventional DRAM DIMM that
serves as a direct-mapped cache for the \XPDIMM{}.  The cache block size is 64 bytes, 
and the CPU's memory controller manages the cache transparently.  The
CPU and operating system simply see a larger pool of main memory.  In graphs,
for brevity and clarity, we refer to this mode as \emph{cached}.

App Direct mode is useful for building storage systems out of \XP{}.  The \XPDIMM{} appears
as a separate, persistent memory device.  There is no DRAM cache.  Instead, the
system installs a file system to manage the device.  \XP{}-aware
applications and the file system can access the \XPDIMM{}s with load and store
instructions and use the ordering facilities mentioned above to enforce ordering
constraints and ensure crash consistency.  In graphs, for brevity and clarity,
we refer to App Direct mode as \emph{uncached}.

Our experiments explore a number of \XP{} memory configurations and modes.
In addition to the cached/uncached option, \XP{} memory can be integrated
in two ways.  It can be used as the main memory of the system as a direct
replacement for DRAM, an option we refer to as \emph{main memory} or \emph{MM};
or it can be used as a storage tier underlying the file system, an option we refer
to as \emph{persistent memory} or \emph{PM}.

\subsection*{Basic \XP{} Performance~(\refsec{sec:basic})}

The most critical difference between \XP{} and DRAM is that \XP{} has longer
latency and lower bandwidth.  Load and store
performance is also asymmetric.

\boldparagraph{Latency} To measure \XP{} load latency, we disable the DRAM
cache and issue single load instructions with a cold cache.  On average,
random loads take 305~ns compared to 81~ns for DRAM accesses on the same
platform.  For sequential loads, latencies are 169~ns, suggesting some
buffering or caching inside the \XPDIMM{}.

Measuring write latency is difficult because we cannot detect when a store
physically reaches the \XPDIMM{}.  We can, however, detect when the store reaches
the processor's \emph{asynchronous DRAM refresh (ADR)} domain, which guarantees
that the store's effects are persistent.  To measure that latency, we issue a
store followed by a cache flush instruction and a fence.  That latency is
94~ns for \XP{} compared to 86~ns for DRAM.

\boldparagraph{Bandwidth} Our measurements show that \XP{} bandwidth is lower
than DRAM bandwidth, especially for stores.  \reffig{fig:bw} plots sequential
access bandwidth to six \XPDIMM{}s for between 1 and 23 threads and compares its
bandwidth to six DRAM DIMMs (on a local and remote NUMA node).  
For reads (at left), bandwidth peaks at
39.4~GB/s.  For writes (at right), it takes just four threads to reach
saturation at 13.9~GB/s.  For a single \XPDIMM{}, 
its max read bandwidth is 6.6~GB/s, whereas its max write bandwidth is
2.3~GB/s.

\ntwfigure[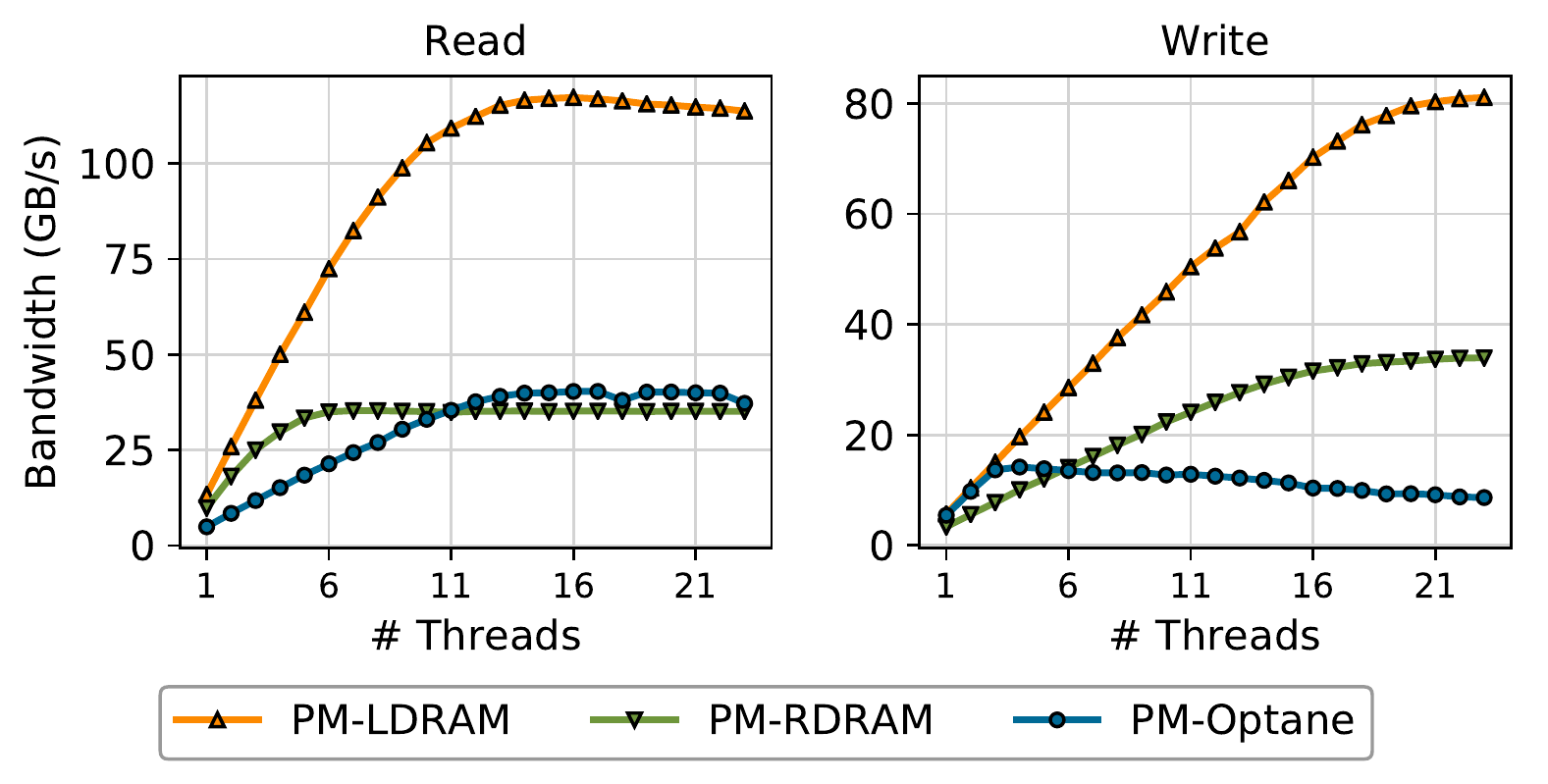,{\figtitle{\XP{} Sequential Bandwidth} The
    data show read (left) and write (right) bandwidth for an array of six
    \XPDIMM{}s compared to a similar array of six DRAM DIMMs.  \XP{} bandwidth
    is lower and, for writes, reaches saturation with fewer threads.},fig:bw]

\reffig{fig:randombw} plots bandwidth for random accesses of different sizes to
a single DIMM issued by one thread.  The left edge of the graph corresponds to
small (64~B) random updates, while the right edge measure accesses large enough
(128~kB) to be effectively sequential.  

\cfigure[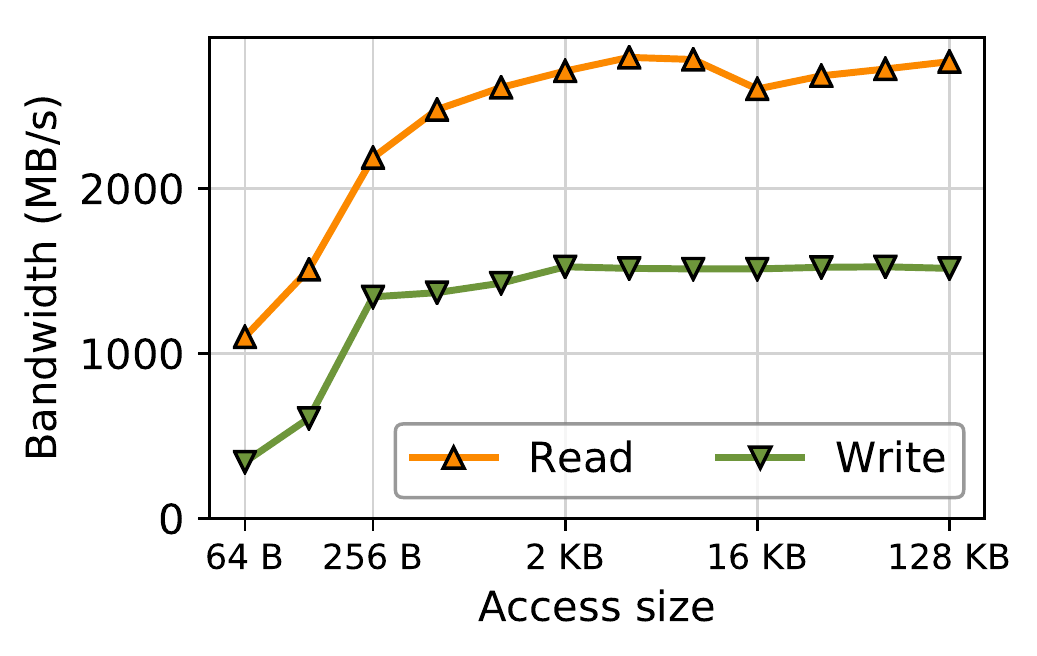,{\figtitle{\XP{} Random Access Bandwidth}
    Bandwidth for small accesses rises quickly but begins to taper off at
    256~B.  This data is for one thread accessing one DIMM.},fig:randombw]

Performance for read and write rises quickly until access size reaches 256~B
and slowly climbs to a peak of 1.5~GB/s for stores and 2.8~GB/s for loads.
256~B is \XP{}'s internal block size.  It
represents the smallest efficient access granularity for \XP{}.  Loads and
stores that are smaller than this granularity waste bandwidth as they have the same
latency as a 256~B access.  Stores that are smaller also result in write
amplification since \XP{} writes at least 256~B for every update, incurring
wear and consuming energy.

\subsection*{\XP{} as Main Memory~(\refsec{sec:big})}

When used as main memory, we expect that the \XPDIMM{} will be used
in Memory Mode (that is, with a DRAM cache) 
in order to provide a large main memory device.

\wfigure[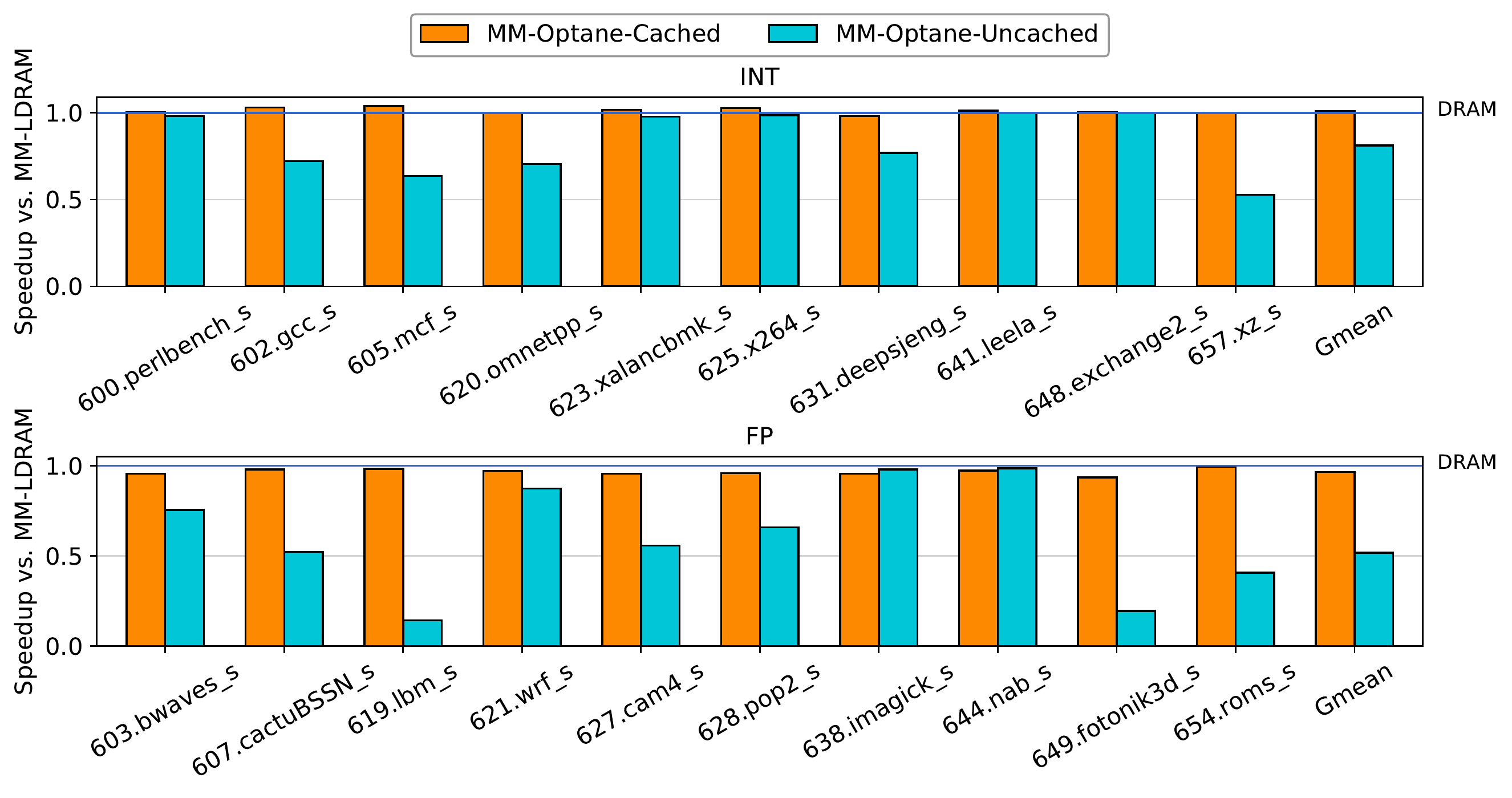,{\figtitle{SPEC 2017 Performance on \XP{}}
    Using cached \XP{} instead of normal DRAM does not affect performance for
    the integer workloads in SPEC 2017.  Without caching, performance drops
    38\%.  For floating point, cached \XP{} drops performance for the
    floating point workloads by 15\%, probably due to bandwidth
    limitations.  Uncached drops it by 61\%. 
  },fig:spec17]

\reffig{fig:spec17} compares the performance of SPEC 
2017~\cite{Bucek:2018:SCN:3185768.3185771}
running on cached \XP{} and uncached \XP{} normalized to performance using
DRAM.  The working sets of the SPEC 2017 applications are all small enough to
fit in the DRAM cache.  For the integer workloads, cached \XP{} is just as fast
as DRAM, and uncached \XP{} is 38\% slower.  Floating point performance,
however, drops 15\% with cached \XP{} and 61\% with uncached.  The poor
performance in cached mode is surprising, and we do not yet sure of the root
cause.  One likely candidate is the limited write bandwidth of \XP{}.

The caching mechanism works for larger memory footprints as
well.  \reffig{fig:memcachedredis} measures performance for Memcached and Redis
(each configured as a non-persistent key-value store) each managing a 96~GB
data set. Memcached serves a workload of 50\% 
SET operations, and Redis
serves a workload with pure SETs.  It shows that for these two applications, replacing 
DRAM with uncached \XP{} reduces
performance by 20.1\% 
and 23.0\%
for memcached and Redis, respectively.  Enabling
the DRAM cache, as would normally be done in system deployment,
means performance drops only between 8.6\%
and 19.2\%.
Regardless of performance losses, \XP{} memory is far denser; our machine can fit
\drampersocketsizegb~GB of DRAM but \aeppersocketsizetb~TB of \XP{} memory on a socket,
giving us the ability to run larger workloads than fit solely in DRAM.

\cfigure[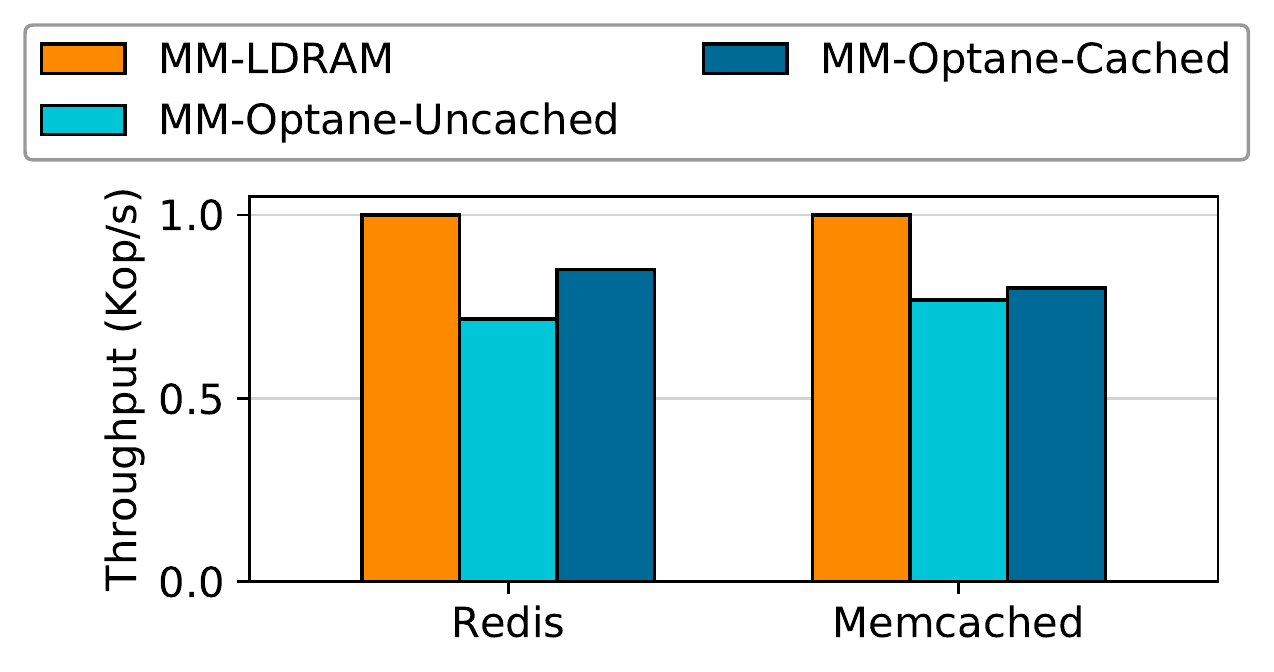,{\figtitle{Large Key-Value Store Performance}
    \XP{} can extend the capacity of in-memory key-value stores like Memcached
    and Redis, and Cascade Lake can use normal DRAM to hide some of \XP{}'s
    latency. The performance with uncached \XP{} is 4.8-12.6\%
    lower than cached \XP{}.  Despite performance losses, \XP{} memory
		allows for far larger sized databases than DRAM due to its density --- we cannot fit
		the larger workloads in DRAM.},fig:memcachedredis]

\subsection*{\XP{} as Persistent Storage~(\refsec{sec:storage})}

\XP{} will profoundly affect the performance of storage systems.  Using
\XPDIMM{}s as storage media disables the DRAM cache and exposes the \XP{} as
a persistent memory block device in Linux.  Several persistent-memory file
systems are available to run on such a device: Ext4 and XFS were built for disks but have direct access
(or ``DAX'') modes, while NOVA~\cite{novapaper} is purpose-built for 
persistent memory.

\reffig{fig:fio} summarizes performance for several file systems performing
random reads and random writes with between one and sixteen threads.  It also
compares performance to a SATA flash-based SSD and an Optane SSD that exposes
\XP{} as block device via the PCIe bus.  The data show that \XP{} improves
basic storage performance over both of those storage devices by a wide margin.

\wfigure[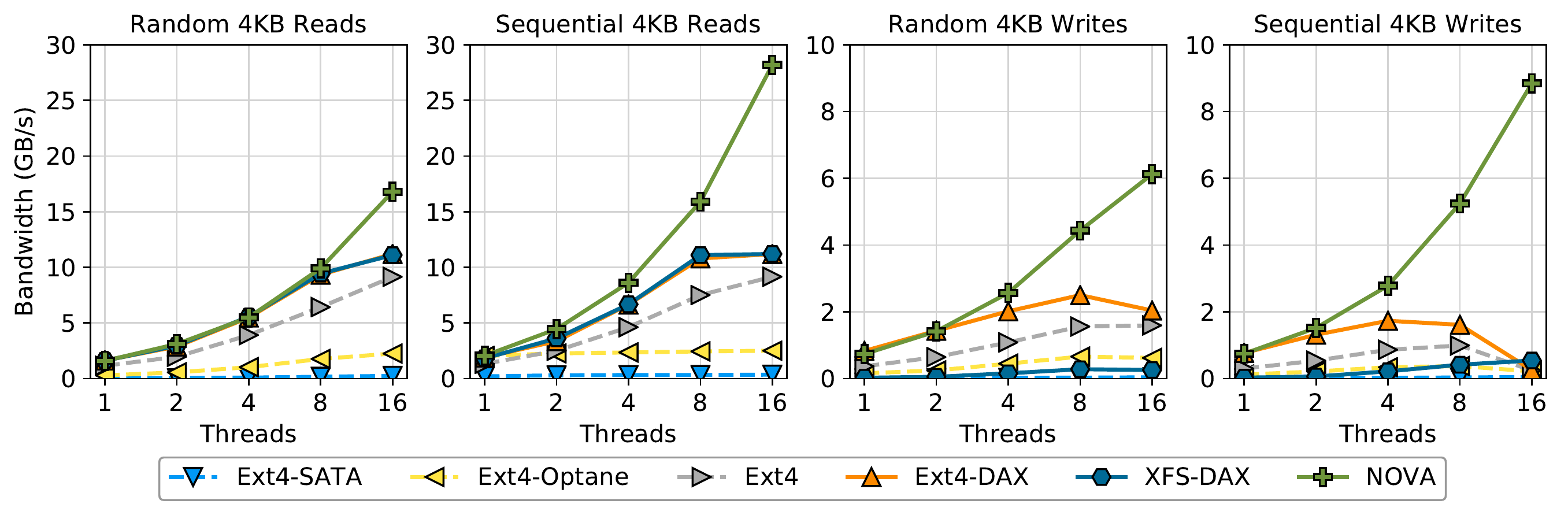,{\figtitle{Raw Performance in Persistent Memory
      File Systems} \XPDIMM{}s provides a big boost for basic file access
    performance compared to SATA SSDs (``Ext4-SATA'') and Optane-based SSDs
    (``Ext4-Optane'').  The data also show the improved scalability that NOVA
    offers relative to legacy file systems in DAX mode.},fig:fio]

The data also demonstrate the benefits of designing software specifically for
persistent memory.  NOVA outperforms the legacy file systems and provides much
better performance scaling.

\reffig{fig:apps} shows how \XP{} affects application-level performance for
RocksDB~\cite{rocksdb}, Redis~\cite{redis}, MySQL~\cite{mysql},
SQLite~\cite{sqlite}, MongoDB~\cite{mongodb}, Kyoto Cabinet~\cite{kyotocabinet}, 
and LMDB~\cite{lmdb}.  MySQL is running TPC-C; the
others are running workloads that insert key-value pairs.

The impact at the application level varies widely.  Interestingly, for MongoDB, the legacy version of Ext4 outperforms the DAX version. We suspect
this result occurs because DAX disables the DRAM page cache,
but the cache is still useful since DRAM is faster than \XP{}.

\wfigure[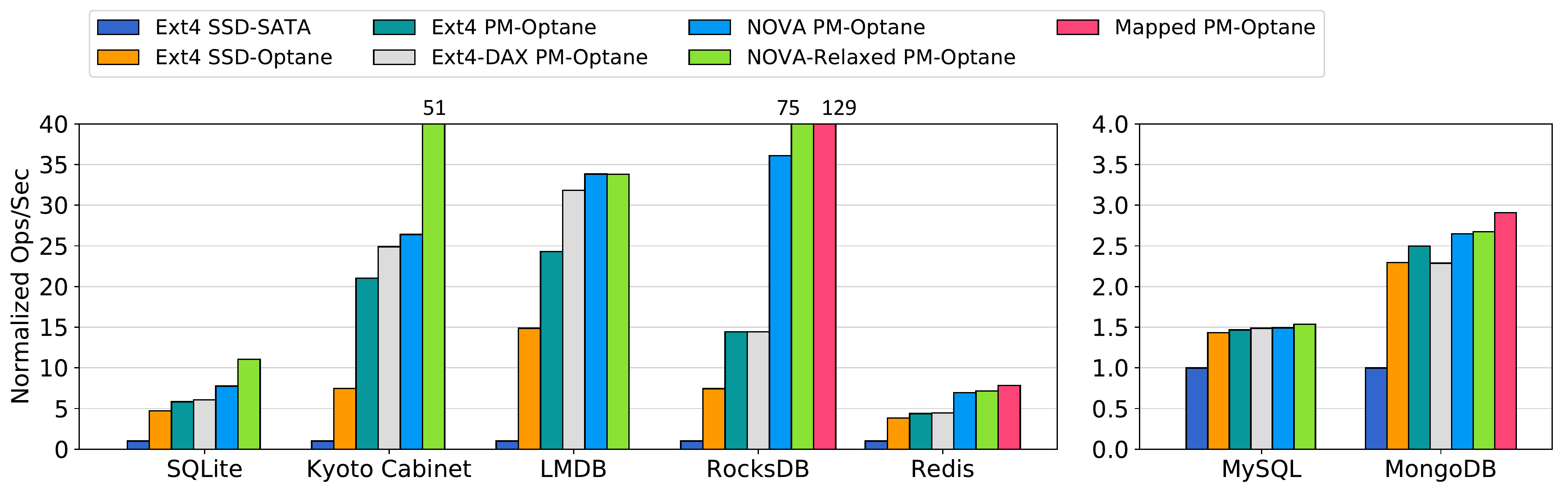,{\figtitle{Application Performance on \XP{} and
      SSDs} These data show the impact of more aggressively integrating \XP{}
    into the storage system.  Replacing flash memory with \XP{} in the SSD
    gives a significant boost, but for most applications deeper integration
    with hardware (i.e., putting the \XP{} on a DIMM rather than an SSD) and
    software (i.e., using an PMEM-optimized file system or rewriting the
    application to use memory-mapped \XP{}) yields the highest
    performance.},fig:apps]

\subsection*{\XP{} as Persistent Memory~(\refsec{sec:pmem})}

\XP{}'s most intriguing application is as a byte-addressable persistent memory
that user space applications map into their address space (with the mmap()
system call) and then access directly with loads and stores.

Using \XP{} in this way is more complex than accessing it through a conventional
file-based interface because the application has to ensure crash consistency
rather than relying on the file system.  However, the potential performance
gains are much larger.

\reffig{fig:apps} includes results for versions of  Redis and
RocksDB modified to use \XP{} in this manner.  The impact varies widely:
performance for RocksDB increases by 3.5\x{}, while Redis~3.2 gains just 20\%.
Understanding the root cause of the difference in performance and how to
achieve RocksDB-like results will be fertile ground for developers and
researchers.

\subsection*{Conclusion}

Intel's \XP{} is the first new memory technology to arrive in the processor's
memory hierarchy since DRAM.  It will take many years to fully understand how
this new memory behaves, how to make the best use of it, and how applications
should exploit it.

The data we present are a drop in the bucket compared to our understanding
of other memory technologies.  The data are exciting, though, because they
show \XP{}'s strengths and its weaknesses, both where it can have an
immediate positive impact on systems and where more work is required.

We are most excited to see what emerges as persistent main memory moves from a
subject of research and development by a small number engineers and academics
to a mainstream technology used by, eventually, millions of developers.  Their
experiences and the challenges they encounter will give rise to the most
innovative tools, the most exciting applications, and the most challenging
research questions for \XP{} and other emerging NVMM technologies.

\newpage
\tableofcontents
\newpage
\section{Introduction}
\label{sec:introduction}

Over the past ten years,
researchers have been anticipating the arrival of commercially available, scalable non-volatile main memory (NVMM) technologies that provide byte-granularity storage that survives power outages.  In the near future, Intel is expected to release a enterprise product based on one of these technologies: the \XPCommercial{} (or just ``\XPDIMM{}'').

Researchers have not idly waited for real nonvolatile DIMMs (NVDIMMs) to arrive\footnote{\XPDIMM{} are not technically NVDIMMs since they do not comply with any of the NVMM-F, -N, or -P JEDEC standards.}.  Over the past decade, they have written a slew of papers proposing new programming models \cite{nvheaps,mnemosyne}, file systems \cite{novapaper,nova-fortis,scmfs,aerie,BPFS,pmfs}, libraries \cite{oracle-nvm-direct,atlas,pmdk}, and applications built to exploit the performance and flexibility that NVDIMMs promised to deliver.

Those papers drew conclusions and made design decisions without detailed knowledge of how real NVDIMMs would behave, what level of performance they would offer, or how industry would integrate them into computer architectures.  In its place, researchers have used a variety to techniques to model the performance of NVDIMMs, including custom hardware \cite{pmfs}, software simulation \cite{nvheaps}, slowing the DRAM frequency~\cite{justdo}, exploiting NUMA effects \cite{duan-date-2018}, or simply pretending that DRAM is persistent.

Now that \XPDIMM{}s are actually here, we can begin to grapple with their complexities and idiosyncrasies.
The first step in understanding \XPDIMM{} performance is to conduct measurements that explore fundamental questions about the \XP{} memory technology and how Intel has integrated it into a system.
This report provides some of those measurements.   

We have attempted to answer several questions, namely:
\begin{enumerate}[noitemsep]
\item What are the basic performance characteristics of \XP{} memory?
\item What are the basic performance characteristics of persistent memory-specific instructions?
\item How does \XP{} memory affect the performance of applications when used as an extension of (non-persistent) DRAM?
\item How does \XP{} memory affect the performance of applications when used as storage?
\item How does \XP{} memory affect the performance of system software (e.g.\ file systems)?
\item How does custom software written for NVMMs perform on \XP{} memory?
\item How does the performance of \XP{} compare to prior methods used to emulate \XP{}?
\end{enumerate}

This report presents measurements over of a wide range of applications, 
benchmark suites, and microbenchmarks, representing over 330 hours
of machine time.  All of the underlying data 
is freely available in the attached \texttt{anc/} directory.  
We hope that the community finds this data useful.

\section{Background and Methodology}
\label{sec:method}

In this section, we provide background on the \XPCommercial{}, describe the test system, and then
describe the configurations we use throughout the rest of the paper.

\subsection{\XP{} Memory}
The \XPCommercial{}, which we term the \XPDIMM{} for shorthand, 
is the first commercially available NVDIMM that creates a new tier between volatile DRAM
and block-based storage. Compared to existing storage devices (including the related
Optane SSDs) that connect to
an external interface such as PCIe, the \XPDIMM{} has better performance and uses a
byte-addressable memory interface.
Compared to DRAM, it has higher density and 
persistence.
At its debut, the \XPDIMM{} is available in 3 different capacities: 128~GB, 256~GB, and 512~GB.

\subsubsection{Intel's \XPDIMM{}}

Like traditional DRAM DIMMs, the \XPDIMM{} sits on the memory bus, and
connects to the integrated memory controller~(iMC) on the CPU.
The \XPDIMM{} debuts alongside the new 
Intel second generation Xeon Scalable processors 
(codenamed Cascade Lake).
On this platform, each CPU has two iMCs, and each iMC supports three channels.
Therefore, in total, a CPU socket can support a total of six \XPDIMM{}s, for a maximum
of 6~TB of \XP{} memory.

For ensuring data persistency, the iMC sits within the \emph{asynchronous DRAM
refresh (ADR)} domain --- Intel's ADR feature ensures that CPU stores that
reach the ADR domain will survive 
a power failure (i.e.\ will be flushed to 
the NVDIMM within the hold-up time, $< 100$~\us{}).  
The ADR domain does not include the processor caches, so stores will only
be persistent after they reach the iMC.

The iMC communicates with the \XPDIMM{} using the
DDR-T interface.  This interface shares a mechanical and electrical interface with DDR4 
but uses a different protocol that allows for variable latencies,
since \XP{} memory access latencies are not deterministic.
Like DDR4 (with ECC), it uses 72-bit data bus and transfers data
in cache-line (64-byte) granularity for CPU loads and stores.

When a memory access request arrives on the NVDIMM, it
is received by the on-DIMM Controller.
This central controller handles most of the processing required
on the NVDIMM and coordinates access to the banks of \XP{} media.

After an access request reaches the controller, the address is internally translated.
Like SSDs, the \XPDIMM{} performs an internal address translation for wear-leveling and bad-block
management. The \emph{address indirection table} (AIT) translates from the DIMM physical address
to an internal \XP{} media device address.
The AIT resides in \XP{} media, though on-DIMM DRAM 
keeps a copy of the AIT entries.

After the request is translated, the actual access to storage media
occurs.  As \XP{} media access granularity is 256 bytes,
the \APC{} will translate 64-byte load/stores into larger 256 byte accesses.
As a consequence, write amplification occurs
as smaller stores issued by the CPU are handled as 
read-modify-write operations on \XP{} memory by the \APC{}.

Unlike DRAM, \XP{} memory does not need constant refresh for data retention; consequently
it consumes less power when idle. 
The \XPDIMM{} has two configurable power budgets. The
\emph{average power budget}
controls the power budget allowed for contiguous workloads,
and the \emph{peak power budget} controls the maximum
power usage under burst traffic.
Both budgets are configurable by the user. 

\subsubsection{Operation Modes}

Each \XPDIMM{} can be configured into one of the following two modes, 
or can be partitioned and used in both modes respectively:

\begin{itemize}
\item \emph{Memory mode}: 
In this two-level mode, the
DDR4 DIMMs connected to the same iMC operate as caches for slower \XP{} memory.
In this scheme, the DDR4 DIMM acts as a direct mapped write-back cache for the \XPDIMM{}, where
each cache line is 64 bytes. The \XPDIMM{}s are exposed to the
operating system as large volatile regions of memory and do not have persistence
(since updates may not be written back all the way into \XP{} memory).

\item \emph{App Direct mode}:
In this one-level mode, the \XPDIMM{}s are directly exposed to the CPU and operating
system and can consequently be used as persistent storage.
Both \XPDIMM{}s and their adjacent DDR4 DIMMs are visible to the operating system
as memory devices.
In App Direct mode, the \XPDIMM{} is exposed to operating system via configurable \emph{regions} 
on contiguously-addressed ranges. 
\end{itemize}

In this paper, for simplicity, we only evaluate a single operation mode at
a time and use the same NVDIMM configuration
across across all \XP{} memory.  That is, for a given configuration,
we allocate all \XP{} memory in the same mode 
(i.e.\ Memory / App Direct),
and, when using Memory mode, share a single fsdax namespace
across all NVDIMMs on a socket. 

\subsection{System Description}

We perform our experiments on a dual-socket evaluation 
platform provided by Intel Corporation.
The hardware and software configuration is shown in \reftab{table:system}.

Two CPUs are installed on the evaluation platform. They are
Intel's new second generation Xeon Scalable processors (codenamed Cascade Lake), 
and are engineering samples with an obfuscated model number.
The overall specifications of these CPUs are close to Xeon Platinum 8160 (former-gen Skylake) 
with higher base clock at 2.2~GHz, and
same single-core turbo boost frequency at 3.7~GHz.
Each CPU has 24 cores, each with exclusive 32~KB L1 instruction 
and data caches, and 1~MB L2 caches.  All cores share
a 33~MB L3 cache. Each CPU has two iMCs and 
six memory channels (three channels per iMC). A 32~GB Micron DDR4 DIMM and 
a \aepdimmsizegb{}~GB Intel \XPCommercial{} are attached to each of the memory channels. 
Thus the system has 384~GB (2 socket x 6 channel x 32~GB/DIMM)
of DRAM, and \aeptotalsizetb{}~TB (2 socket x 6 channel x \aepdimmsizegb{}~GB/DIMM) of NVMM. 
To compare \XP{} memory with traditional blocked based storage,
we use an NVMe Optane SSD (NVMe interface) and an NAND SSD (SATA interface) as baselines.

On this system, 
we run Fedora 27 with Linux kernel version \KernelVersion{} built from source. 
For all of the experiments,
we disable hyper-threading and set the CPU power governor to performance mode, 
which forces the CPU to use
the highest possible clock frequency.
All \XPDIMM{}s have the same firmware (version \aepfirmware{}) and 
use the default power budget (average 15~W / peak 20~W).

In all experiments, transparent huge pages (THP) are enabled unless explicitly mentioned.
We do not apply security mitigations (KASLR, KPTI, Spectre and L1TF patches) because
Cascade Lake fixes these vulnerabilities at the hardware level~\cite{intell1tf}.

\begin{table}

		\centering
		\begin{tabular}{lc}
            \\\hline
            \# Sockets & 2 \\
            Microarch & Intel Cascade Lake-SP (engineering sample) \\
            CPU Spec. &  24 Cores at 2.2~GHz (Turbo Boost at~3.7 GHz) \\
            L1 Cache& 32~KB i-Cache \& 32~KB d-Cache (per-core)\\
            L2 Cache& 1~MB (per-core) \\
            L3 Cache& 33~MB (shared) \\
\ignore{           Chipset & Unknown (Intel C624) \\}
            DRAM Spec.& 32~GB Micron DDR4 2666~MHz (36ASF4G72PZ)\\
            Total DRAM&  384~GB [2 (socket) \x{}6 (channel) \x{} 32~GB] \\
            NVMM Spec.& \aepdimmsizegb{}~GB Intel Optane DC 2666~MHz QS (NMA1XXD256GQS) \\
            Total NVMM& \aeptotalsizetb{}~TB  [2 (socket) \x{}6 (channel) \x{} \aepdimmsizegb{}~GB] \\
            Storage (NVMe)& Intel Optane SSD DC P4800X 375 GB \\
            Storage (SATA)& Intel SSD DC S3610 1.6 TB (MLC)
            \\\hline
            GNU/Linux Distro & Fedora 27 \\
            Linux Kernel & \KernelVersion{} \\
            CPUFreq Governor & Performance \\
            Hyper-Threading & Disabled \\
            NVDIMM Firmware & \aepfirmware{} \\
            Avg. Power Budget & 15~W \\
            Peak Power Budget & 20~W
\ignore{            Infiniband HCA & Mellanox Connect X-3}
            \\\hline
            Transparent Huge Page (THP) & Enabled \\
            Kernel ASLR & Disabled \\
            KPTI \& Security Mitigations & Not Applied
            \\\hline
		\end{tabular}
	\vspace*{1mm}
	\caption{\figtitle{Evaluation platform specifications}}
	\label{table:system}
	\vspace{-2mm}
\end{table}

\subsection{Configurations}

As the \XPDIMM{} is both persistent and byte-addressable,
it can fill the role of either a main memory
device (i.e.\ replacing DRAM) or as a persistent device
(i.e.\ replacing disk).  Both use cases
will be common.  To fully examine
the performance of \XP{} memory, we test its performance
in both roles.  We examine six
system configurations --- three that explore the main memory
role and three that explore the persistence role. They are
shown in \reftab{table:modes}.

\begin{table}
		\centering
        \begin{tabular}{lccccc}
             & DRAM & NVDIMM & \multirow{2}{*}{Persistence} & \multirow{2}{*}{Namespace} & Size \\
             & Mode & Mode & & & \textit{(per-socket)} \\\hline
            \MMLDRAM{} & Memory & n/a &  No & unmanaged & 192~GB \\
            \MMPMEMC{} & Cache & Memory &  No & unmanaged & \aeppersocketsizetb{}~TB \\
            \MMPMEMUC{} & Memory & App Direct &  No & unmanaged & \aeppersocketsizetb{}~TB \\
            \PMLPMEM{} & Memory & App Direct & Yes & fsdax & \aeppersocketsizetb{}~TB\\
            \PMLDRAM{} & Fake PMem & n/a & Emulated & fsdax & \emusize{}~GB \\
            \PMRDRAM{} & Fake PMem & n/a & Emulated & fsdax & \emusize{}~GB \\
						\SSDOptane{} & Memory & n/a & Yes & n/a & 375~GB (total)\\
						\SSDSATA{} & Memory & n/a & Yes & n/a & 1.6~TB (total)
            \\\hline
        \end{tabular}
	\vspace*{1mm}
	\caption{\figtitle{Evaluation modes summary} A summary of our experimental configurations.  
	Modes that begin with \textbf{MM-} represent systems where we vary the type of memory
	attached behind the traditional DRAM interface.  
	Modes that begin with \textbf{PM-} or \textbf{SSD-} represent systems where system memory is in DRAM, 
	but we vary the device underneath the file system.}
	\label{table:modes}
	\vspace{-2mm}
\end{table}

\subsubsection{Memory Configurations}
We use our first set of configurations to examine the
performance of \XP{} as memory; they therefore vary the
type of memory attached behind the traditional DRAM interface.
In these configurations, 
the main memory used by the system
is of a single type.  These configurations, 
prefixed by \textbf{\MM{}} (main memory), 
are explored in
detail in Section~\ref{sec:big}.  They are:

\boldparagraph{\MMLDRAM{}} 
Our baseline configuration simply uses the DRAM in the system as DRAM
and ignores the \XPDIMM{}s. This configuration is our control configuration
and represents an existing system without NVDIMMs.
It provides a DRAM memory capacity of 192~GB per socket.

\boldparagraph{\MMPMEMC{}}
This configuration uses cached \XP{} as the system memory.
That is, all memory in the system is comprised of \XPDIMM{}s but with the adjacent
DRAM DIMMs as caches.
This configuration represents the likely system configuration used when \XPDIMM{}s are
utilized as large (but volatile) memory.
In this configuration, we set \XP{} into Memory mode, so each \XPDIMM{} uses
volatile DRAM as a cache. 
This configuration provides \aeppersocketsizetb~TB of \XP{} per socket.
The 192~GB per-socket DRAM functions as a cache and is transparent to the operating system.

\boldparagraph{\MMPMEMUC{}}
In this configuration, we use uncached \XP{} as the system memory,
that is, without DRAM caching \XP{}.
This configuration represents a \XP{} system configuration where raw,
uncached \XP{} is used as the main memory device.  We include this configuration
since the DRAM cache in \MMPMEMC{} obscures the raw performance of the \XP{} media ---
we do not expect this to be a common system configuration.
To build this configuration, we configured the \XPDIMM{} into App Direct mode and let the Linux kernel
consider \XP{} to be DRAM. 
The kernel considers \XP{} to be slower memory and DRAM to be 
faster memory, and puts them in two separate NUMA nodes. 
Although it would be interesting to measure the performance when the whole system running directly
on the NVMM, we cannot boot 
the operating system without any DRAM. Therefore, to run the tests, we configure applications 
to bind their memory to a NUMA node with exclusively \XP{} memory.

\subsubsection{Persistence Configurations}

Our second set of configurations explores the persistence
capabilities of \XPDIMM{}, and we explore the persistence
performance of the \XPDIMM{} in 
Sections~\ref{sec:storage} and~\ref{sec:pmem}.
As such, these configurations assume a machine
model in which system memory resides in DRAM and we vary
the device underlying the file system.  
These configurations use fsdax mode,
which exposes
the memory as a persistent memory device under \texttt{/dev/pmem}.  
This arrangement allows both
DAX (direct access) file systems and user-level libraries to directly 
access the memory using a load/store interface while still supporting
block based access for non-DAX file systems.

These configurations, which we prefix by \textbf{\PM{}} (persistent memory), vary the 
memory media
underneath the file system.  The configurations are:

\boldparagraph{\PMLPMEM}
This configuration uses \XP{} as persistent memory.  
The configuration represents a system with both DRAM and large quantities of NVMM
used for storage.
In it, we set \XP{} to be persistent memory running in App Direct mode.
Each persistent memory device has a capacity of \aeppersocketsizetb~GB.

\boldparagraph{\PMLDRAM{}}
This configuration uses local DRAM as an emulated persistent memory device.
Pretending that DRAM is persistent is a simple 
way of emulating \XP{},
and has served as a common baseline for research in the past decade. 
This configuration helps us understand how existing methods of emulating
NVMM compare to real \XP{} memory.  For this configuration,
we create \emusize{}~GB emulated \texttt{pmem} devices 
on the same platform using DRAM. 
In this setup, \XP{} memory is configured in App Direct mode
but not used.

\boldparagraph{\PMRDRAM{}}
Like the previous configuration, this configuration uses DRAM
(but in this case remote DRAM) to emulate a persistent memory device.
Using DRAM on a remote NUMA node simulates the delay when accessing
slower NVMM, and researchers used this configuration to investigate the costs
of integrating NVMM into real systems before real \XPDIMM{}s were available.
Like the previous configuration, we use this configuration to examine how
prior emulation methods used in research compare to real NVDIMMs.  In this configuration,
we allocate a simulated \texttt{pmem} device on one socket, and ensure
all applications are run on the other.

For experiments that run on file systems, we can also compare
\XPDIMM{}s against traditional block based storage devices
(see Section~\ref{sec:storage}).  For these experiments,
we also use the following block-based devices underneath the
file system:

\boldparagraph{\SSDOptane{}}
This configuration loads an Intel Optane
drive underneath the file system using the NVMe
interface.  This PCIe
device uses Optane media as the underlying storage technology,
but is optimized for block-based storage.  We use this
configuration to compare the load/store interface of
the \XPDIMM{} with a comparable block-based device
using the same storage technology.

\boldparagraph{\SSDSATA{}}
This configuration loads a NAND flash
solid state drive beneath the file system using the SATA
interface.  We use this
configuration to compare novel storage devices and interfaces
with a more established technology.

\section{Basic Performance}
\label{sec:basic}

The impact of \XP{} on the performance of a particular application 
depends on the details of \XP{}'s basic parameters.
Since \XP{} is persistent, the landscape of its performance characteristics is 
more complex than DRAM's.  In this section, we measure the performance of 
reads and writes of various
sizes, compare the performance different types of stores, 
and quantify the costs of enforcing persistence with \XP{}.

In particular, we focus on three questions:

\begin{tightenum}
\item What are the performance characteristics of \XP{} memory and 
how do they differ from local and remote DRAM?
\item What is the cost of performing a persistent write to \XP{} memory?
\item How do access patterns impact the performance of \XP{} media?
\end{tightenum}

In order to answer these questions, we experiment
on system configurations that load persistent memory underneath a file system,
namely \textbf{\PMLDRAM{}}, \textbf{\PMRDRAM{}},
and \textbf{\PMLPMEM{}}, and use two different experimental tools,
the Intel Memory Latency Checker (MLC)~\cite{intel-mlc},
and a self-built tool called \lattest{}.

Intel MLC is a tool designed to accurately 
measure memory latencies and bandwidth. 
In particular,
every thread in MLC accesses its own mmapped file created on
a \texttt{pmem} device mounted with ext4 file system with direct access (DAX) enabled.
MLC also disables hardware prefetching to get accurate memory latency measurements.
Finally, for multithreaded microbenchmarks, 
each thread writes to its own 400~MB buffer.

\lattest{} is our tool for running microbenchmarks that measure the latency and bandwidth 
through a combination of load/store (64-byte, 128-byte and 256-byte granularity) and
flush (\texttt{clflush}, \texttt{clflushopt}, \texttt{clwb} and non-temporary store) instructions. 
To ensure timing results are accurate, we take a number of precautions.
To avoid interference with other software, we built \lattest{} as a kernel module that
exposes a mock DAX file system interface and works on a \texttt{pmem} device in fsdax mode.
It runs kernel threads that pin to specific cores with IRQ disabled, 
so it will not be descheduled.
Like MLC, we disable prefetching. 
For latency tests, an \texttt{mfence} instruction is inserted after each load/store to 
prevent pipelining from affecting timing results.

\subsection{Latency}

In this subsection, we investigate the latency of operations
to \XP{} memory under several conditions and explore how
its latency changes in comparison to DRAM.

\subsubsection{Read Latency}
The most basic latency measurement is the latency of a single access to
a local \XPDIMM{}.  Such a measurement assumes that the access misses in the
entire cache hierarchy as would happen when the system is idle and the cache is cold. 
We use MLC to measure the raw \emph{read latency} by timing the latency of reading
a random cacheline on a single core.  Since MLC turns off prefetching, this load
access is guaranteed to touch \XP{} memory. 

Our next experiment investigates the \emph{sequential read latency}
of the \XPDIMM{}.  
We again use MLC to measure the latency of sequential loads with
prefetching turned off.  Like the previous read latency test,
we measure this latency using a single core.
\reffig{fig:raw_latency} shows the evaluation result for both random and sequential accesses, 
for local \XP{} memory and demonstrates that
\XP{} memory improves its performance under sequential accesses,
indicating some amount of on-DIMM prefetching and batching logic.

\cfigure[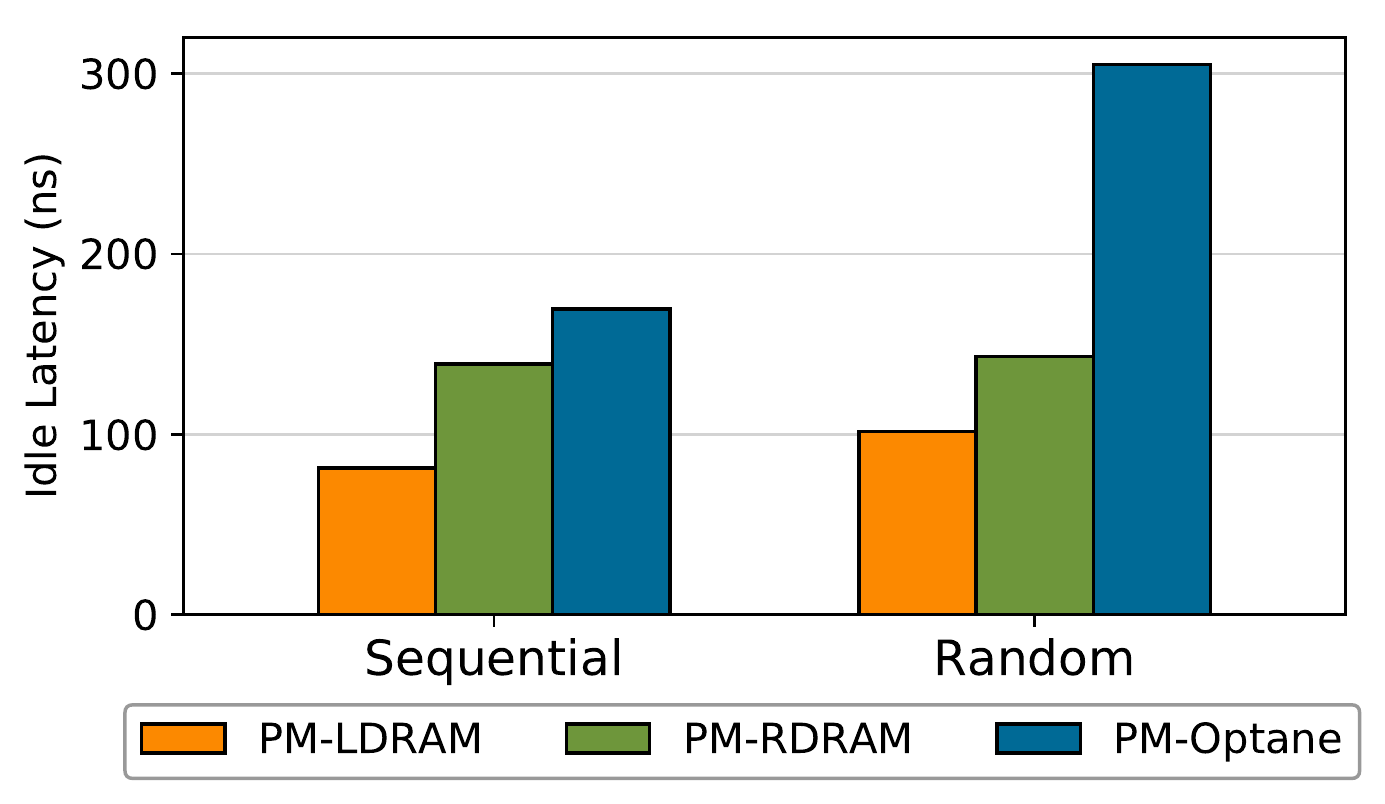,{\figtitle{Read latency} An experiment showing
random and sequential read latency to \XP{} memory on a cache miss. \XP{} memory is about 3\x{}
slower than DRAM for random accesses and about 2\x{} slower for sequential access (see data in~\dataref{csvroot/basic/idle_latency.csv}).},fig:raw_latency]

\takeaway{The read latency of random \XP{} memory loads is 305 ns}{
This latency is about 3\x{} slower than local DRAM. \
}

\takeaway{\XP{} memory latency is significantly better (2\x{}) when accessed in a sequential pattern.}{
This result indicates that \XPDIMM{}s merge adjacent requests into a single 256
byte access.
}

\subsubsection{Memory Instruction Latency}

Our next experiment
shows the raw latency for a variety of ways of accessing \XP{} memory.
Using our \lattest{} tool,
this microbenchmark measures the latency of memory access instructions
for a combination of load/store, flush and fence 
instructions used to access a persistent
memory region.  The test runs on a single kernel thread
over the first 16~GB region of a \texttt{pmem} device. 
We use 256-bit SSE instructions 
(64-bit instructions show similar results),
and record the timing for each set of instructions executed 
one million times over sequential and pre-calculated random addresses.
We drain the CPU pipeline before each access and issue an \texttt{mfence} after each access.

\reffig{fig:klattest} shows the median latency for a variety
of instruction sequences. Along the x-axis,
we use \texttt{Size\_Type} to identify
different types of persistent memory accesses, where the type can be one of the following:
\texttt{L} stands for regular loads, \texttt{LN} stands for non-temporal loads, 
\texttt{SF} stands for stores followed by \texttt{clflush}, \texttt{SN}
stands for non-temporal stores, \texttt{SO} stands for stores followed by \texttt{clflushopt}, 
and \texttt{SW} stands for stores followed by \texttt{clwb}. 

As shown in the data, 
the load latency of \PMLPMEM{} is higher, whereas the store latency 
of \PMLPMEM{} is similar to \PMLDRAM{} since it is hidden within the
ADR domain. Finally, for small accesses, 
\texttt{clflushopt} and \texttt{clwb} give better performance 
than \texttt{clflush} or non-temporal stores.

\takeaway{For small accesses, \texttt{clflushopt} and \texttt{clwb} give better performance 
than \texttt{clflush} or non-temporal stores.}{
This result demonstrates the utility of ISA modifications in support
of persistent memory.
}

\wfigure[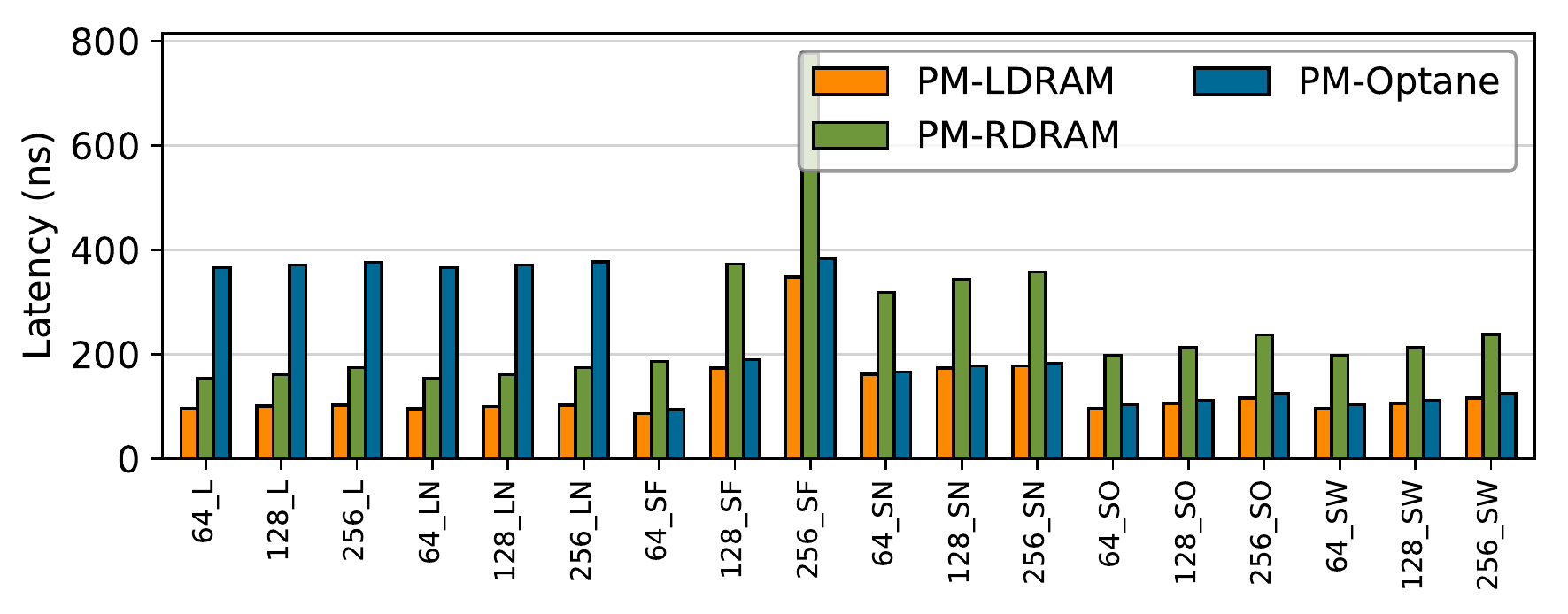,{\figtitle{Memory Instruction Latency} This graph
shows the median latency for a variety of ways of accessing persistent memory.
Note that for store instructions followed by flushes, 
there is little performance difference between PM-LDRAM and PM-Optane, whereas
DRAM outperforms \XP{} memory for load sequences (see data in~\dataref{csvroot/basic/instruction_latency.csv}).},fig:klattest]

\subsection{Bandwidth}

This subsection investigates the bandwidth of the \XPDIMM{}
under varying loads and conditions.

\subsubsection{Maximum Bandwidth}
\label{sec:max_bw}

Our first bandwidth experiment explores the
maximum read and write bandwidth of the memory
device.  We use the MLC tool to spawn threads
that spin issuing sequential reads or writes.  By gradually
increasing the thread count, we find the
point at which the memory device's bandwidth becomes saturated.
We use up to 23~threads (leaving one core of the CPU idle to avoid contention).  
Figure~\ref{fig:mlc_bwcnt}
has the result.

For read accesses, \PMLPMEM{} keeps 
scaling with the thread count but at a lower rate
than \PMLDRAM{}.
For non-cached writes,
\PMLPMEM{} peaks at four threads and then stops scaling, whereas 
both \PMLDRAM{} and \PMLPMEM{} show better scalability.

\ntwfigure[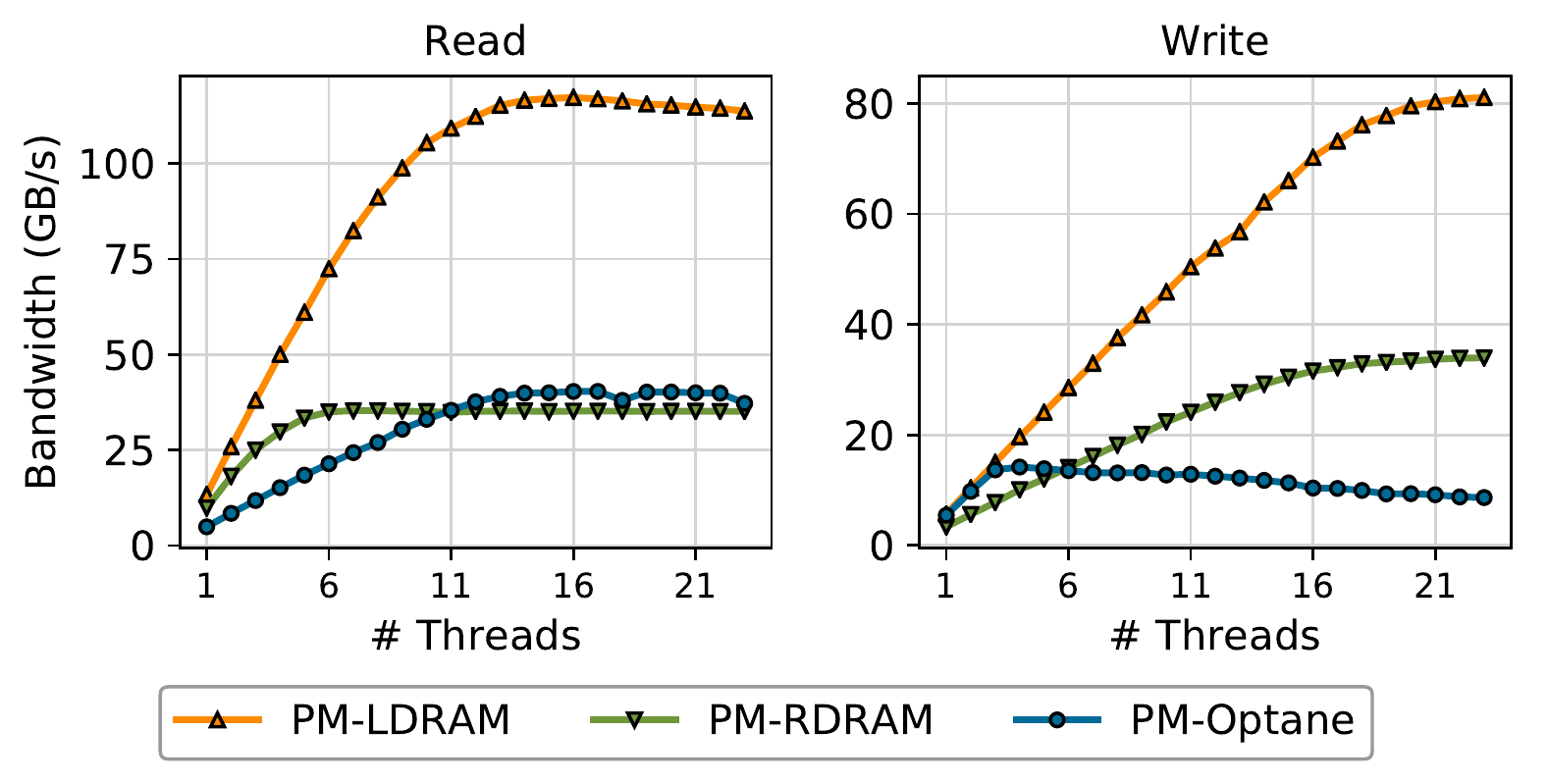,{\figtitle{Sequential memory bandwidth with different \# threads}
This graph shows memory performance under a varying number of threads performing
reads (left) or non-temporal stores (right).  Note that
\XP{} reads scale well with the number of threads, whereas write 
bandwidth is saturated with only four threads.  Remote DRAM has an interesting
access pattern that peaks around 35 GB/sec due to the bus bandwidth (see 
data in~\dataref{csvroot/basic/read_bandwidth.csv}
and~\dataref{csvroot/basic/write_bandwidth.csv}).},fig:mlc_bwcnt]

\takeaway{Our six interleaved 
\XPDIMM{}s' maximum read bandwidth is 39.4~GB/sec, and their
maximum write bandwidth is 13.9~GB/sec.} {This experiment
utilizes our six interleaved \XPDIMM{}s, so accesses are spread across the devices.}

\takeaway{\XP{} reads scale with thread count; whereas writes do not.} { 
\XP{} memory bandwidth scales with thread count, achieving maximum throughput
at 17 threads. However, four threads are enough to saturate \XP{} memory write bandwidth.}

\subsubsection{Concurrency and Bandwidth}

In this section, we measure \XP{} and DRAM bandwidth for random and sequential reads and writes
under different levels of concurrency.  
\reffig{fig:bwthreads} shows the
bandwidth achieved at different thread counts for sequential accesses with 256~B access granularity.
We show loads and stores (\texttt{Write(ntstore)}),
as well as cached writes with flushes (\texttt{Write(clwb)}).  All experiments use AVX-512 instructions and access
the data at 64~B granularity.  The left-most graph plots performance for interleaved DRAM accesses,
while the center and right-most graphs plot performance for non-interleaved and interleaved \XP{}. In the non-interleaved measurements all the accesses go to a single DIMM.

The data shows that DRAM bandwidth is both significantly higher than \XP{}
and scales predictably (and monotonically) with thread count until
it saturates the DRAM's bandwidth and that bandwidth is mostly independent of
access size.  

The results for \XP{} are wildly different. First, for a single DIMM, the maximal read bandwidth
is 2.9\x{} of the maximal write bandwidth (6.6~GB/s and 2.3~GB/s respectively), where DRAM has a smaller gap (1.3\x{}) between read and write bandwidth. 

Second, with the exception of interleaved reads, \XP{} performance is non-monotonic
with increasing thread count.  For the non-interleaved (i.e., single-DIMM)
cases, performance peaks at between one and four threads and then tails off.
Interleaving pushes the peak to twelve threads for \texttt{store+clwb}.

\wfigure[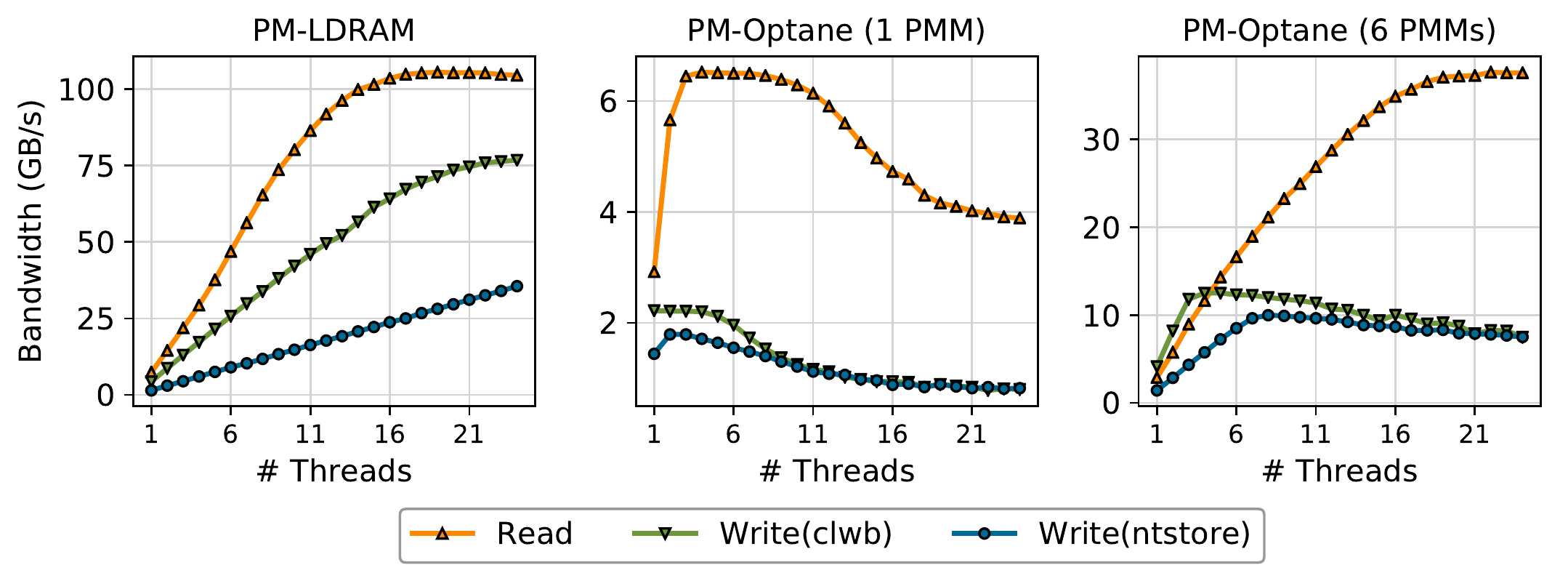,{\figtitle{Bandwidth vs. thread count} An experiment showing
maximal bandwidth as thread count increases (from left to right) on local DRAM, on a single \XPDIMM{},
and interleaved \XP{} memory across six \XPDIMM{}s. All threads use a 256~B access size. (Note the difference in vertical scales).
(see data in~\dataref{csvroot/basic/bandwidth_dram.csv},
~\dataref{csvroot/basic/bandwidth_optane_1pmm.csv} and
~\dataref{csvroot/basic/bandwidth_optane_6pmms.csv})
\vspace{-8mm}} ,fig:bwthreads]

\subsubsection{Access Size and Bandwidth}
\ignore{

Our next experiment measures the correlation between access size and bandwidth.
To exclude the interleaving factor, we use the \PMLPMEM{} scheme but create a 
non-interleaved region on a single DIMM.
We use a single thread to access regions at random locations, and the access size varies from
64~bytes to 128~KB (using 64 byte SSE instructions). We plot the result in~\reffig{fig:onedimm}.
   
As expected, for both read and write access, \XP{}'s bandwidth is smaller for 
accesses smaller than 256~B.
Since the access granularity for \XP{} memory media is 256~bytes, 
small writes become read-modify-write
updates at the DIMM level. For an individual DIMM, accesses larger 
than 2~KB do not increase the overall bandwidth.

\cfigure[Graphs/Jian/micro_bwcnt_1dimm.pdf,{\figtitle{Single threaded \XP{} random access bandwidth
on a single DIMM}
    Bandwidth for small accesses on a single DIMM rises quickly but begins to taper off at
    256~bytes for both reads and writes (see data in~\dataref{csvroot/basic/singledimm_bandwidth.csv}).},fig:onedimm]
		
		\takeaway{The application-level \XP{} bandwidth is affected by access size.} {To fully utilize the \XP{} device bandwidth, 256~byte or larger accesses are preferred.}
}

\reffig{fig:bwsz} shows how performance varies with access size.  The
graphs plot aggregate bandwidth for random accesses of a given size.  We use
the best-performing thread count for each curve (given as ``load thread
count/ntstore thread count/store+clwb thread count'' in the figure).
Note that the best performing thread count for \PMLPMEM{}(Read) varies with different access sizes for random accesses,
where 16 threads show good performance consistently.

Note that \XP{} bandwidth for random accesses under 256~B is poor.  This ``knee'' corresponds to the
\XP{}'s access granularity of 256~bytes.
DRAM bandwidth does not exhibit a similar ``knee''
at 8~kB (the typical DRAM page size), because the cost of opening a page of
DRAM is much lower than accessing a new page of \XP{}.

Interleaving (which spreads accesses across all six DIMMs) adds further complexity: 
\reffig{fig:bwsz}(right) measures bandwidth across six interleaved NVDIMMs as a function
of access size.
Interleaving improves
peak read and write bandwidth by 5.8\x{} and 5.6\x{}, respectively.
These speedups match the number of DIMMs in the system (6) and highlight the
per-DIMM bandwidth limitations of \XP{}. The most striking feature of the graph is
a dip in performance at 4~KB --- we believe 
this dip is an emergent effect caused by contention at the iMC,
and it is maximized when threads perform random accesses close to the interleaving size. 

\wfigure[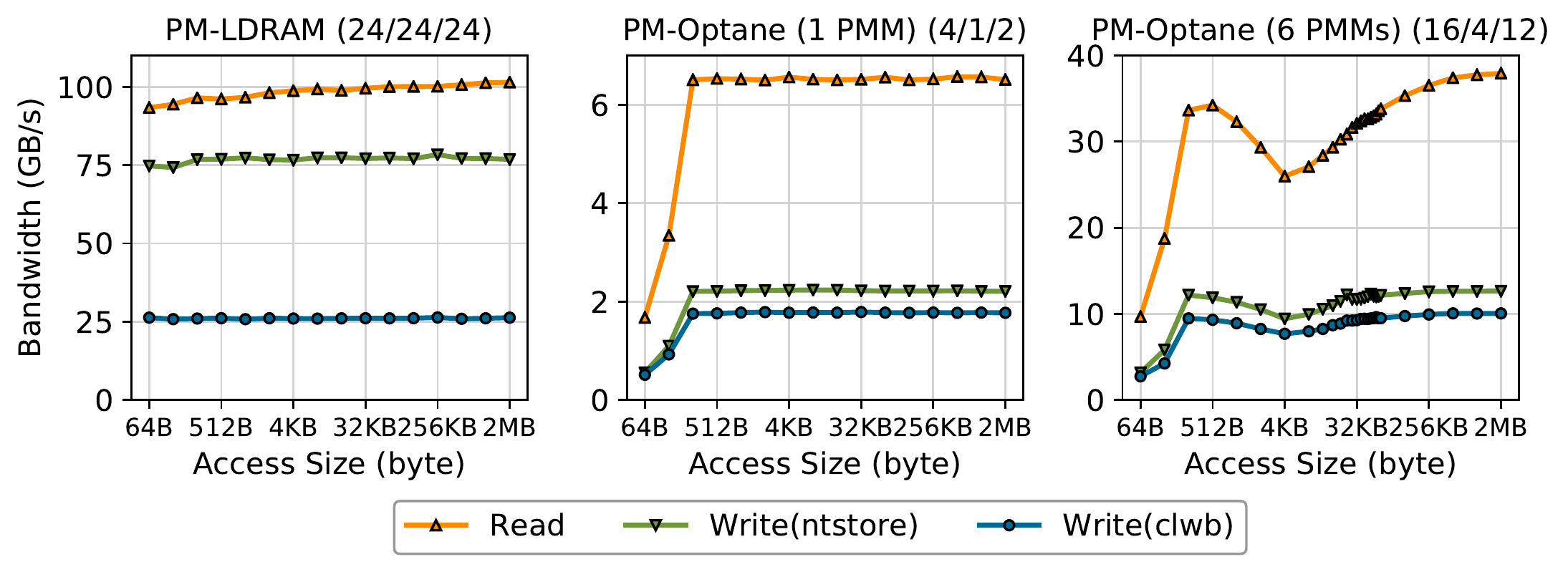,{\figtitle{Bandwidth over access size} An experiment showing
    maximal bandwidth over different access sizes on (from left to right) local DRAM, 
    on a single \XPDIMM{}, and interleaved \XP{} memory across six \XPDIMM{}s. Graph titles include the number of threads used
    in each experiment (\texttt{Read/Write(ntstore)/Write(clwb)})
    (see data in~\dataref{csvroot/basic/bw_access_size_dram.csv},
    ~\dataref{csvroot/basic/bw_access_size_optane_1pmm.csv} and
    ~\dataref{csvroot/basic/bw_access_size_optane_6pmms.csv})
    . \vspace{-8mm}} ,fig:bwsz]

\takeaway{The application-level \XP{} bandwidth is affected by access size.} {To fully utilize the \XP{} device bandwidth, 256~byte or larger accesses are preferred.}

\subsubsection{Bandwidth under Mixed Workloads}

This experiment investigates the device's memory bandwidth
under varying patterns of reads and writes.  
The experiment, using Intel's MLC tool, measures the bandwidth using multiple threads accessing
memory in a sequential pattern, where
each thread is performing one configuration that issues reads, 
writes or non-temporal writes (writes that bypass the CPU caches), 
or a combination of two types of instructions.  Figure~\ref{fig:mlc_bwcnt_mix}
shows the results.

Mixing reads and writes hurts \XP{} performance more than DRAM.
For pure reads and writes (see Section~\ref{sec:max_bw}), 
the \PMLDRAM{} outperforms
\PMLPMEM{} by 2.90\x{} and 5.72\x{}, and \PMRDRAM{} achieves 0.87\x{} performance to \PMLPMEM{} on read
 and outperforms by 2.39\x{} on write.
whereas for mixed workloads, 
both \PMLDRAM{} and \PMRDRAM{} 
outperform \PMLPMEM{} by a large
margin (up to 12.0\x{} and 7.5\x{} respectively).

\wfigure[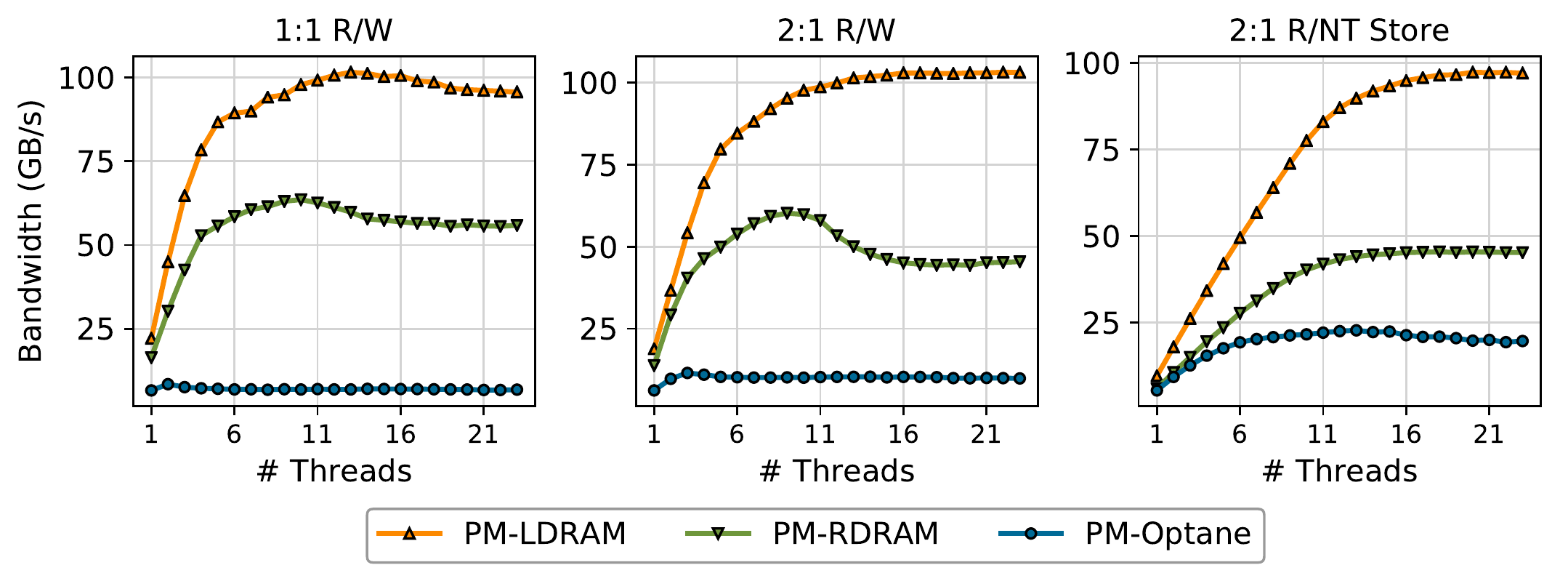,{\figtitle{Sequential memory bandwidth with different \# threads for mixed workloads}
This graph shows memory performance under a varying number of threads performing (from left to right)
reads and writes with 1:1 ratio, reads and writes with 2:1 ratio, or reads and non-temporal stores with 2:1 ratio
(see data in~\dataref{csvroot/basic/mixed_bandwidth_1r1w.csv},
~\dataref{csvroot/basic/mixed_bandwidth_2r1w.csv} and
~\dataref{csvroot/basic/mixed_bandwidth_2r1nt.csv}).},fig:mlc_bwcnt_mix]

\takeaway{\XP{} is more affected than DRAM by access patterns.} {\XP{} memory 
is vulnerable to workloads with mixed reads and writes.}

\wfigure[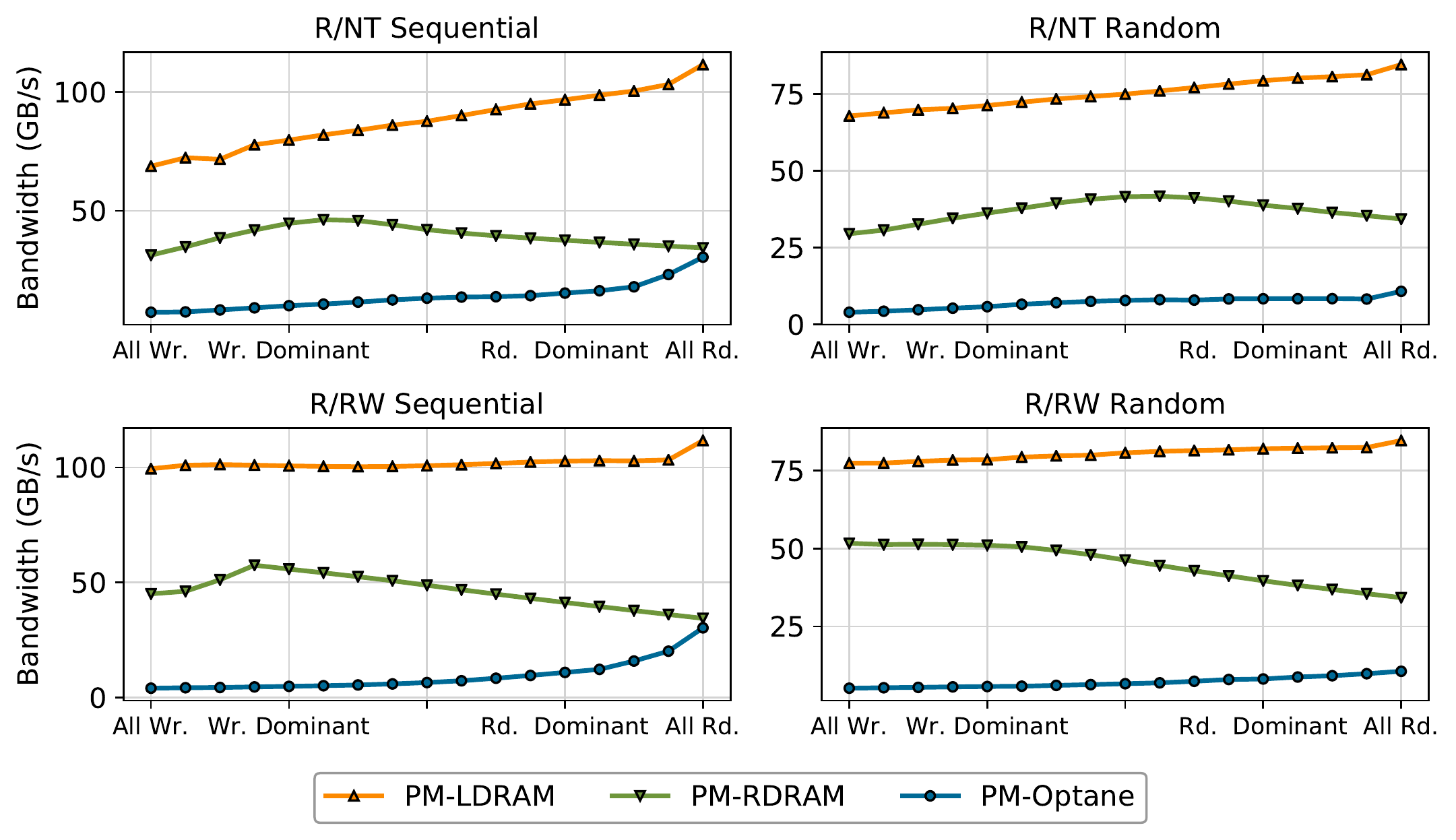,{\figtitle{Memory bandwidth with different mix of read and write threads}
This graph shows how bandwidth varies under a mix of read and write threads.  On the left of each graph,
all threads are performing some write instruction (either a non-temporal store, abbreviated ``NT'', or a
regular write, abbreviated ``W''), whereas on the right of each graph, all threads are performing reads
(see data in~\dataref{csvroot/basic/bandwidth_readwrite_rw_seq.csv}, 
\dataref{csvroot/basic/bandwidth_readwrite_rw_rand.csv},
\dataref{csvroot/basic/bandwidth_readwrite_nt_seq.csv} and
\dataref{csvroot/basic/bandwidth_readwrite_nt_rand.csv}).},fig:mlc_rwmix]

Finally, we run a mix of read and write workloads. For each run, we use sixteen total threads and 
change the number of read and write threads. \reffig{fig:mlc_rwmix}
shows the result. As expected, \PMLPMEM{} shows best result on all-read scenarios, and 
performs much better on sequential workloads.

\subsubsection{Performance under Load}
Our final experiment examines how latency and bandwidth vary under load
by gradually increasing the load on the device.
In this test, we use MLC and use 23 threads for loads,
and, 
for non-temporal stores, we use 12 threads.
Each of the worker threads repeatedly accesses memory.
Each thread performs memory accesses to cache lines
and delays for a set interval between two accesses.  For each delay interval,
varying from 0 to 80~\us{},
we plot the latency and bandwidth in \reffig{fig:loaded_latency}.
When the delay time is zero (corresponding to the right side of the graph), 
the bandwidth is close to the maximum bandwidth and latency skyrockets
as queuing effects dominate. When the delay time
is high enough (80~\us{}, corresponding to the left side of the graph), 
the latency is close to the raw, unloaded, latency.
The ``knee'' in the graph shows the point at which the device is able to 
maintain steady bandwidth without suffering from queuing effects.
The experimental results show that the \XPDIMM{}'s performance deviates 
significantly from DRAM.  In particular, the \XPDIMM{}'s read bandwidth tops
out much lower limits than DRAM: 38.9 GB/sec for \XP{} vs 105.9 GB/sec for DRAM
on sequential accesses, and 10.3 GB/sec for \XP{} vs 70.4 GB/sec for DRAM on random
accesses. We observe a similar effect on sequential writes: \XPDIMM{}'s write bandwidth
tops out around 11.5 GB/sec while the DRAM achieves 52.3GB/sec when fully loaded.

\takeaway{\XP{} bandwidth is significantly higher (4\x{}) when accessed in a sequential pattern.}{
This result indicates that \XPDIMM{}s contain access to merging logic to merge overlapping
memory requests --- merged, sequential, accesses do not pay the write amplification cost associated
with the NVDIMM's 256 byte access size.
}

\wfigure[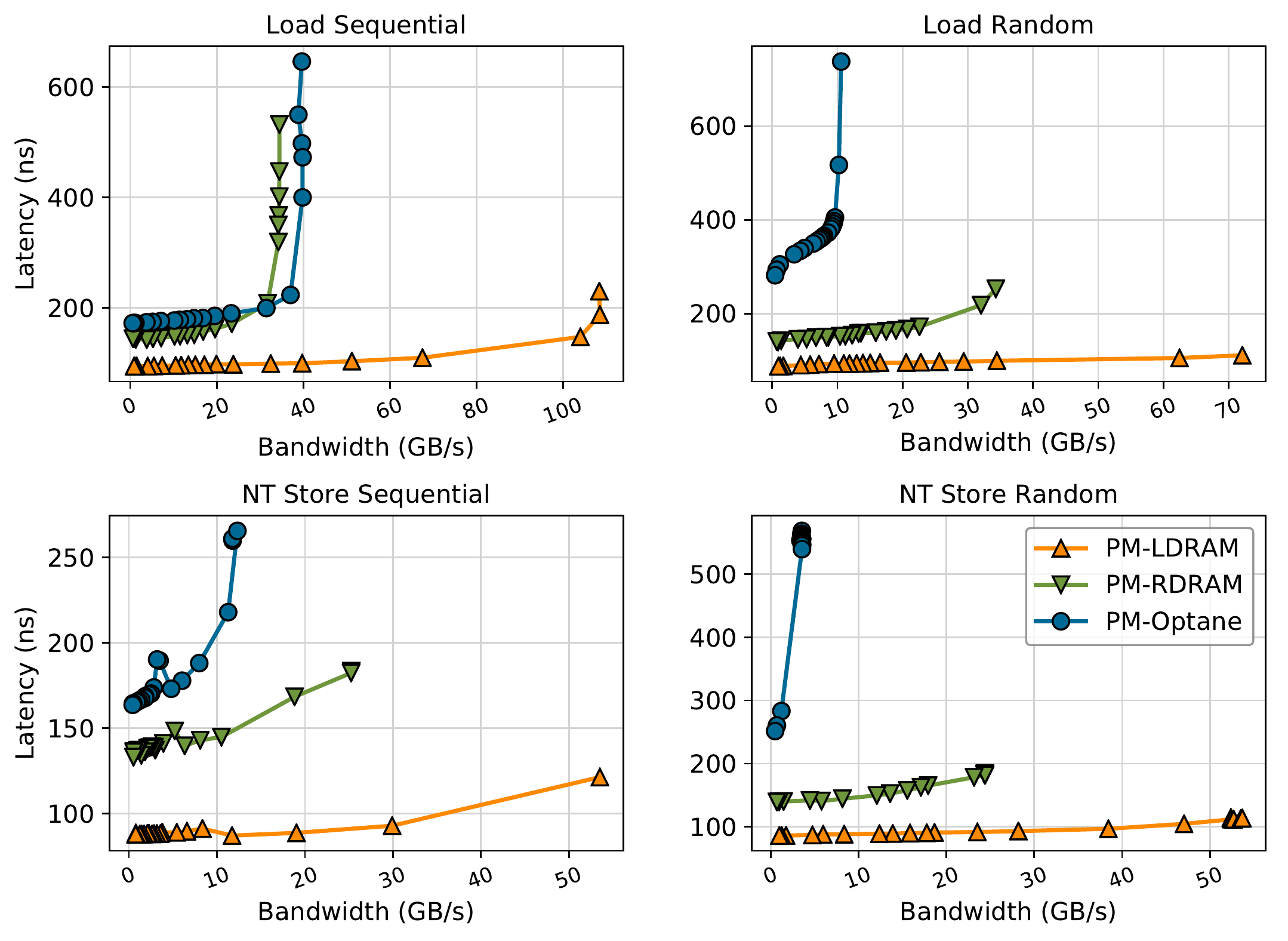,{\figtitle{Performance under load}
 This experiment shows memory latency and bandwidth under varying load.  The ``knee''
 in the graph occurs when the device begins to suffer from queuing effects and maximum
 bandwidth is reached.  Note that
 DRAM memory can support significantly higher bandwidth than \XP{} memory, and that
 \XP{} memory is much more sensitive to the access pattern
 (see data in~\dataref{csvroot/basic/bandwidth_loadedlat_load_seq.csv},
 \dataref{csvroot/basic/bandwidth_loadedlat_load_rand.csv},
 \dataref{csvroot/basic/bandwidth_loadedlat_nt_seq.csv} and
 \dataref{csvroot/basic/bandwidth_loadedlat_nt_rand.csv}).
 },fig:loaded_latency]

\section{\XP{} as Main Memory}
\label{sec:big}

The advent of \XPCommercial{}s means that large memory devices
are now more affordable --- the memory capacity
of a single host has increased, and the unit cost of memory has decreased.
By using \XPDIMM{}s, customers can pack larger datasets into main memory
than before.

In this section, we explore \XP{} memory's performance when placed
in the role of a large main memory
device, and therefore use system configurations that vary the 
device underlying the DRAM interface (\textbf{\MMLDRAM{}}, 
\textbf{\MMPMEMC{}}, and \textbf{\MMPMEMUC{}}).
Naturally, two questions arise:

\begin{tightenum}
\item How does slower \XP{} memory affect application performance?
\item Is the DRAM cache effective at hiding \XP{}'s higher latency
and lower bandwidth?
\end{tightenum}

To understand how slower \XP{} memory impacts performance when used as
application memory,
we run existing applications on \XP{} without modifying them.
These applications generate memory footprints that fit in both DRAM and \XP{},
and can be found in Sections~\ref{sec:spec} and~\ref{sec:parsec}.
In these tests, \XP{} memory can be considered to be a larger capacity
(but slower) alternative to DRAM.
These tests include three standardized benchmark suites: 
SPEC CPU 2006~\cite{Henning:2006:SCB:1186736.1186737},
SPEC CPU 2017~\cite{Bucek:2018:SCN:3185768.3185771},
and PARSEC~\cite{Zhan:2017:PMB:3053277.3053279}.

To investigate the effectiveness of using DRAM as a cache for \XP{} memory, 
we run applications with workloads that exceed the DRAM capacity of the system.
We use two in-memory data stores (Memcached~\cite{memcached} and Redis~\cite{redis}), 
and adjust their workset size to exceed DRAM capacity.  These experiments
can be found in Sections~\ref{sec:memcached} and~\ref{sec:redis-mm}

\subsection{SPEC CPU 2006 and 2017}
\label{sec:spec}

SPEC CPU~\cite{Henning:2006:SCB:1186736.1186737, Bucek:2018:SCN:3185768.3185771}
is a popular benchmark suite for measuring CPU performance. We use SPEC
CPU to investigate how \XP{} memory impacts system performance when used
as the primary main memory device, since SPEC CPU exercises both
the processor and memory hierarchy. 
SPEC CPU contains a wide range of benchmarks, and these benchmarks can be divided into 
integer and floating point groups, indicating what the majority of their
computation handles. The included benchmarks are different between the 2006 and
2017 versions of SPEC --- as such we used both for our tests. 
The memory footprint of SPEC CPU
workloads varies from several megabytes to several gigabytes~\cite{limaye2018workload},
so they easily fit into memory on our test system. 

Both SPEC suites include \emph{speed} and \emph{rate} variants,
called SPECspeed and SPECrate respectively.
SPECspeed tests the system with a single instance
of the benchmark (all benchmarks are single threaded). In contrast, SPECrate
measures the system throughput with multiple task copies running as 
separate processes.  SPEC 2006 uses the
same benchmarks for both rate and speed, whereas SPEC 2017 creates different
versions of a benchmark for rate and speed (therefore SPEC 2017 contains four
benchmark sub-suites: int speed, int rate, float speed and float
rate). For SPECrate, we use 24 copies of each benchmark to fully occupy all the
cores on a single socket. We use the default configuration throughout all
tests.

We report our results in terms of speedup
relative to the execution time of the local DRAM configuration (\MMLDRAM{}).
These results can be found in Figures~\ref{fig:spec06fp}
through~\ref{fig:spec17fp}. 

\doublerfigure[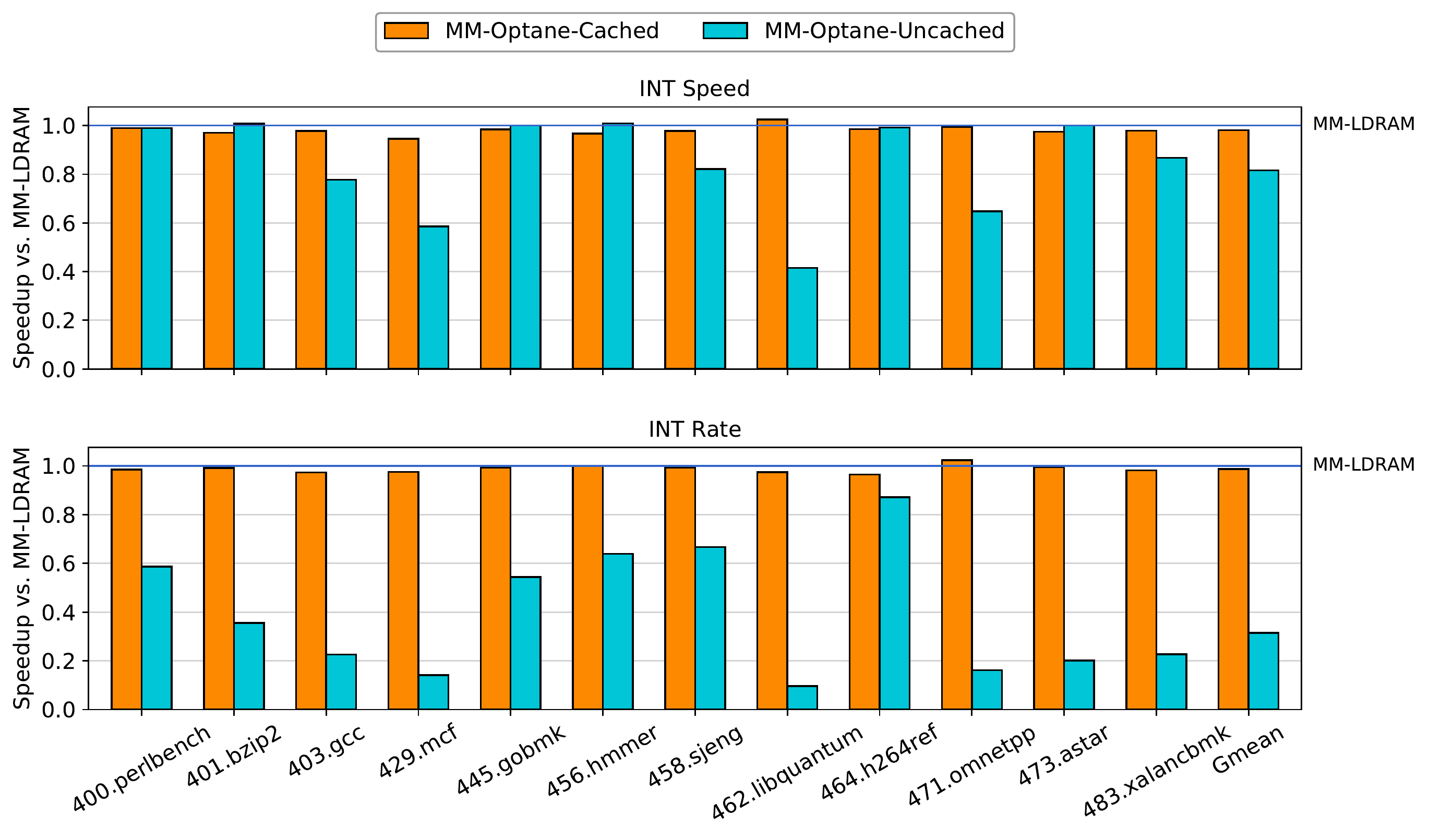,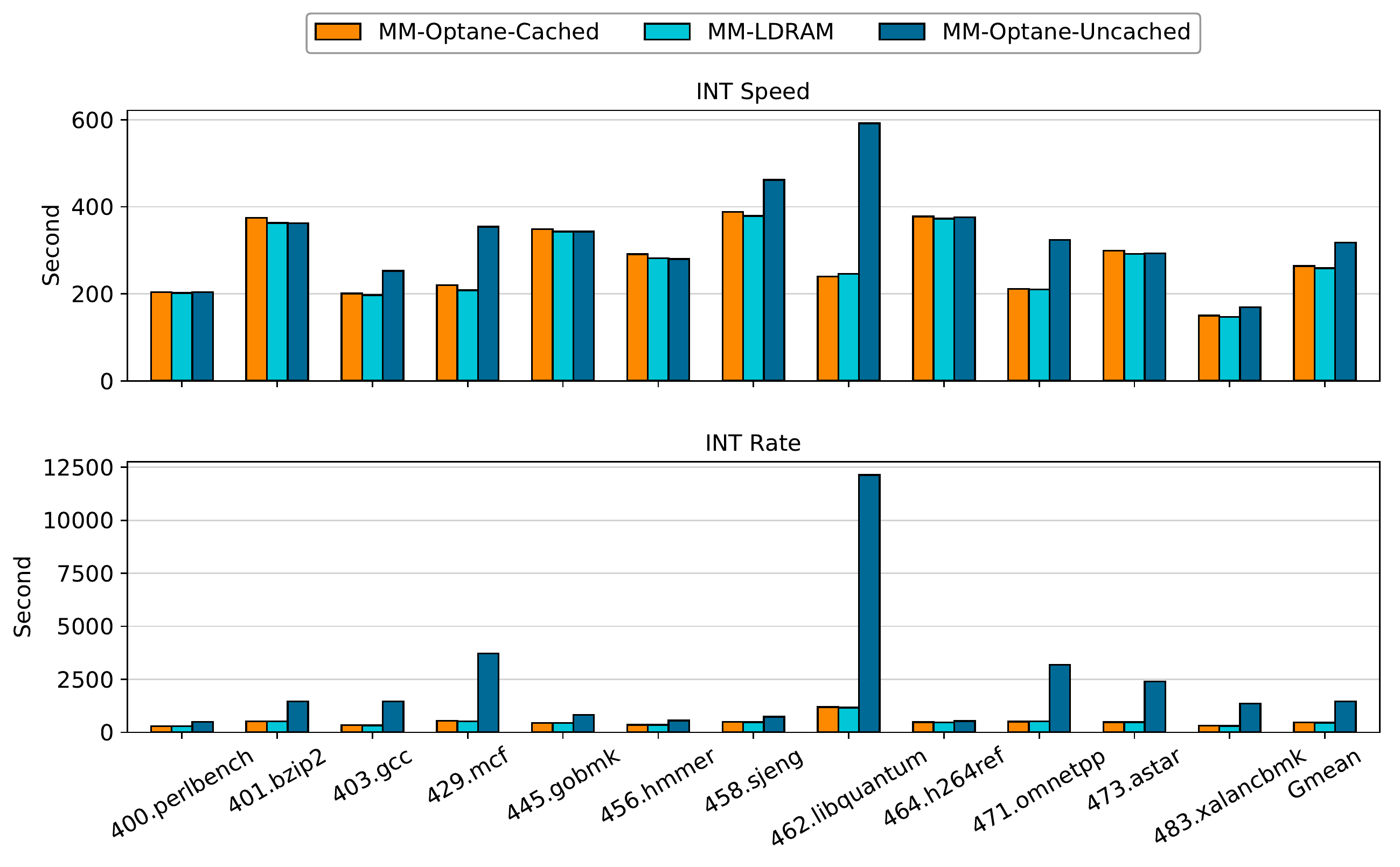,{\figtitle{SPEC 2006 integer suite} This graph 
shows (a) the speedup relative to \MMLDRAM{} for the
SPEC 2006 integer suite and (b) the execution time. Speed tests run the benchmark
single threaded, whereas the rate tests run the same benchmark on each 
core in a separate process (24 cores). Note that the rate test maxes out the bandwidth of uncached \XP{}
memory, but the DRAM cache effectively hides this issue
(see data
in \dataref{csvroot/spec/spec06_int_rate_ratio_normalized.csv}, \dataref{csvroot/spec/spec06_int_speed_ratio_normalized.csv}, \dataref{csvroot/spec/spec06_int_rate_time.csv}
and \dataref{csvroot/spec/spec06_int_speed_time.csv}).},fig:spec06int]

\doublerfigure[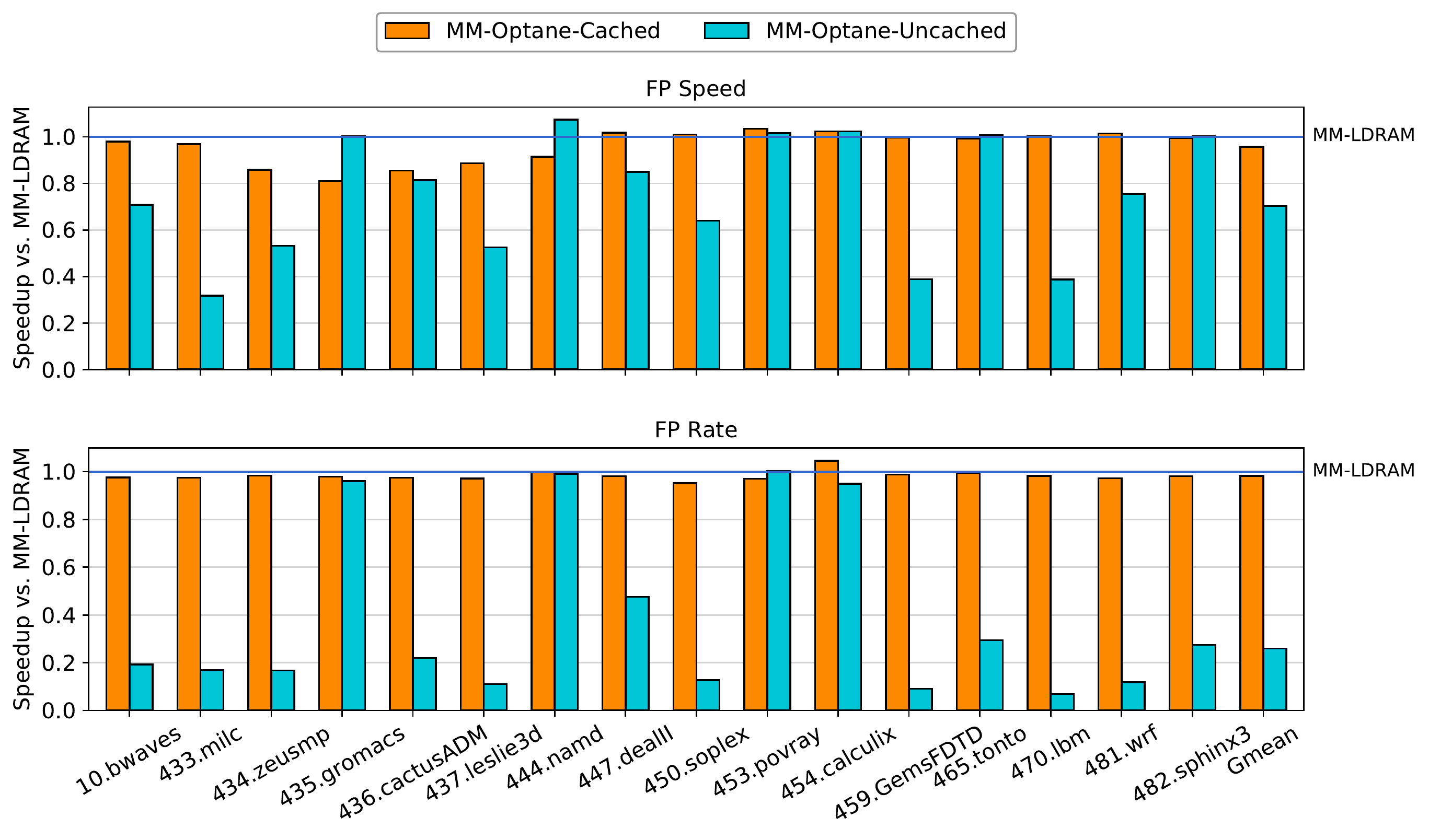,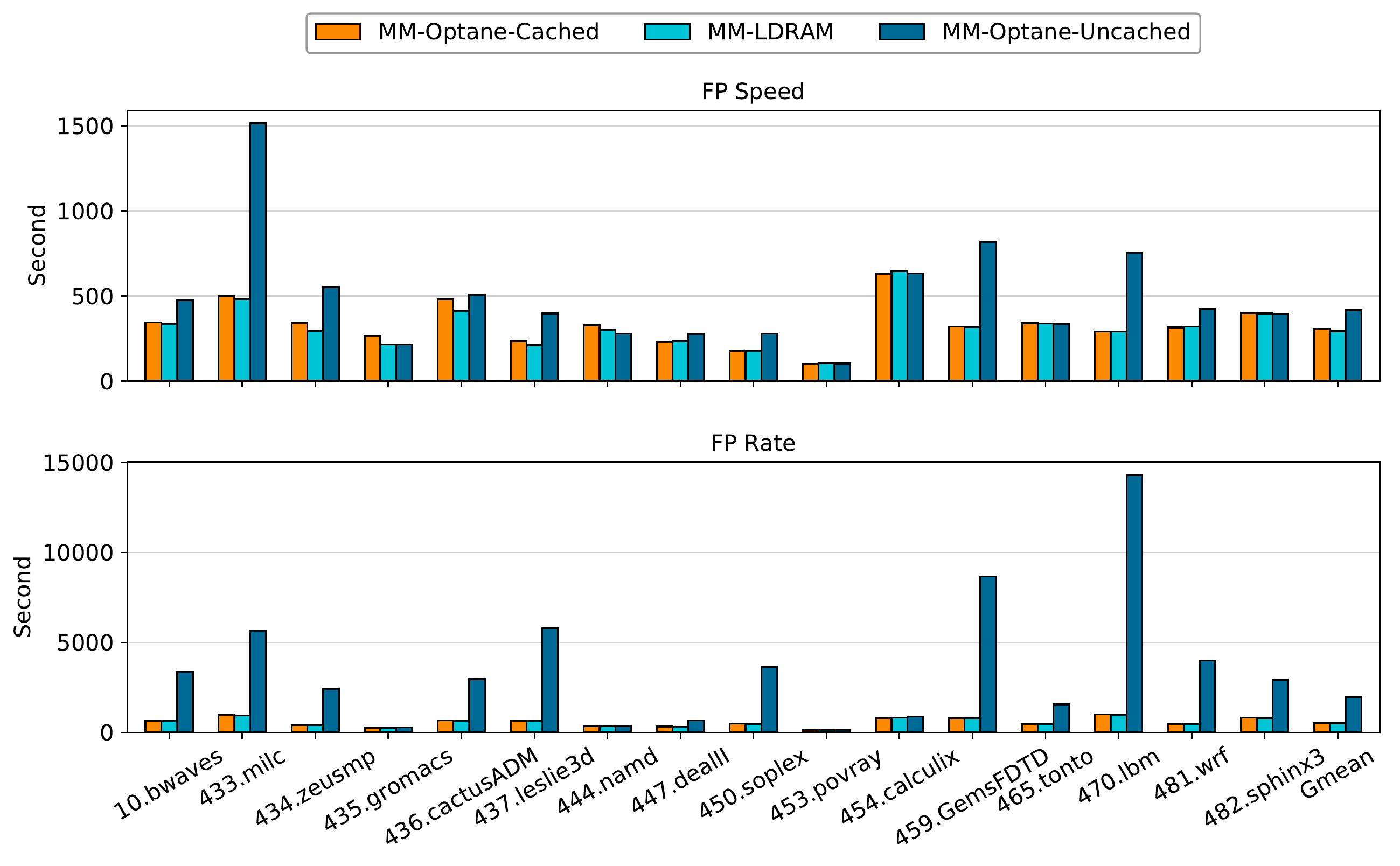,{\figtitle{SPEC 2006 floating point suite} This graph 
shows (a) the speedup relative to \MMLDRAM{} for the
SPEC 2006 floating point suite and (b) the execution time
(see data
in \dataref{csvroot/spec/spec06_fp_rate_ratio_normalized.csv}, \dataref{csvroot/spec/spec06_fp_speed_ratio_normalized.csv}, \dataref{csvroot/spec/spec06_fp_rate_time.csv} and \dataref{csvroot/spec/spec06_fp_speed_time.csv}).},fig:spec06fp]

\doublerfigure[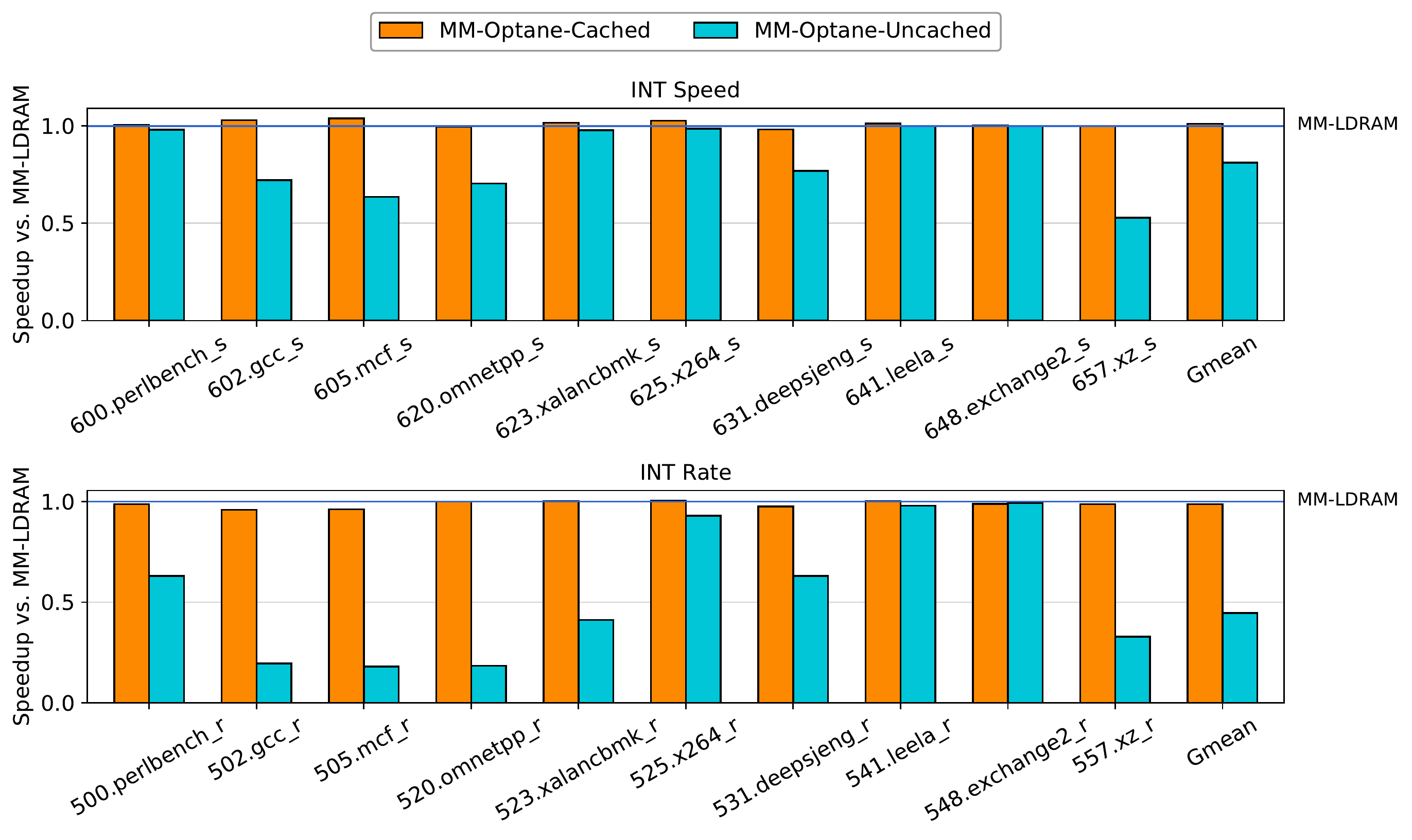,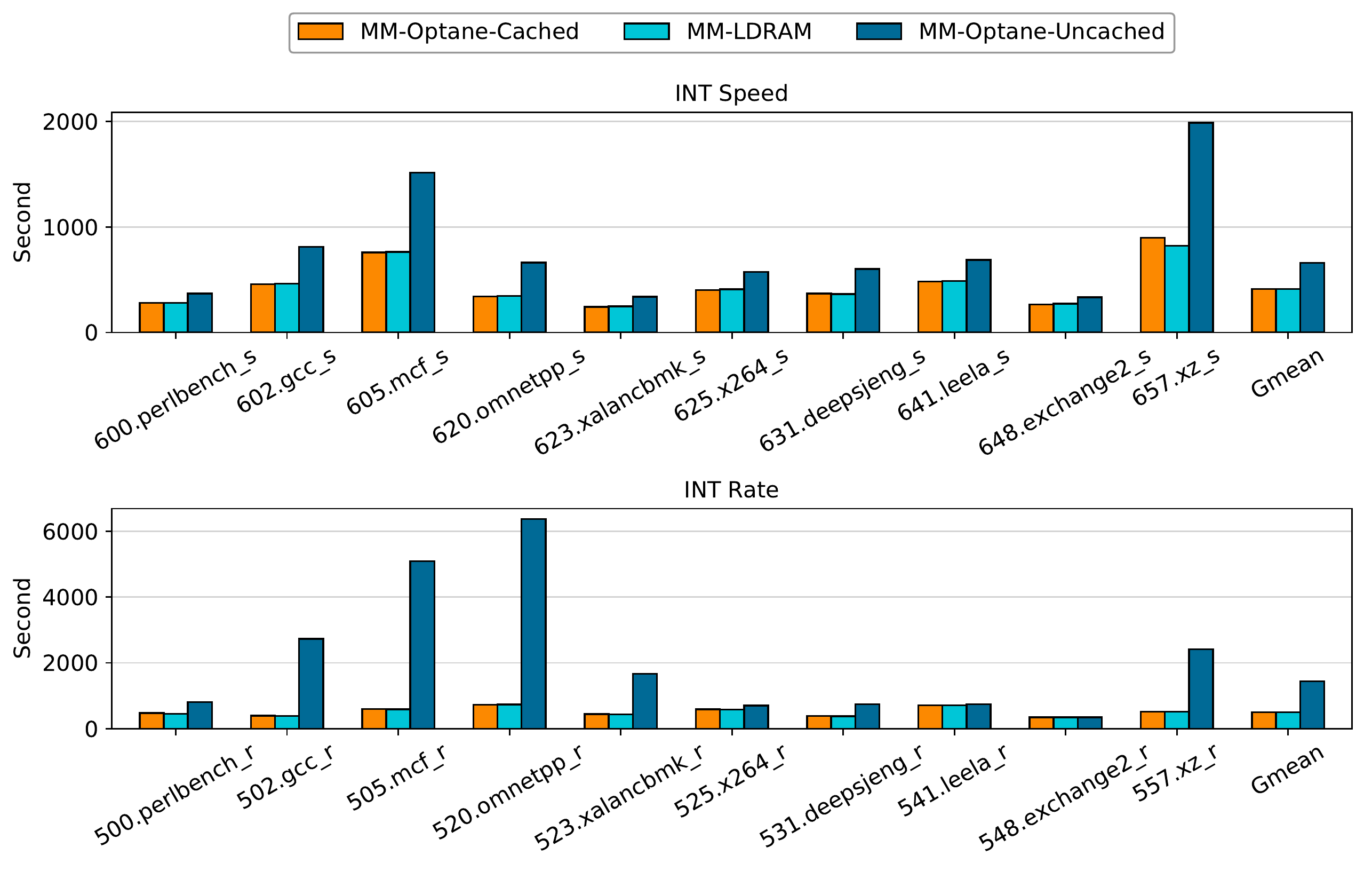,{\figtitle{SPEC 2017 integer suite} This graph 
shows (a) the speedup relative to \MMLDRAM{} for the
SPEC 2017 integer suite and (b) the execution time
(see data
in \dataref{csvroot/spec/spec17_int_rate_ratio_normalized.csv}, \dataref{csvroot/spec/spec17_int_speed_ratio_normalized.csv}, \dataref{csvroot/spec/spec17_int_rate_time.csv}
and \dataref{csvroot/spec/spec17_int_speed_time.csv}).},fig:spec17int]

\doublerfigure[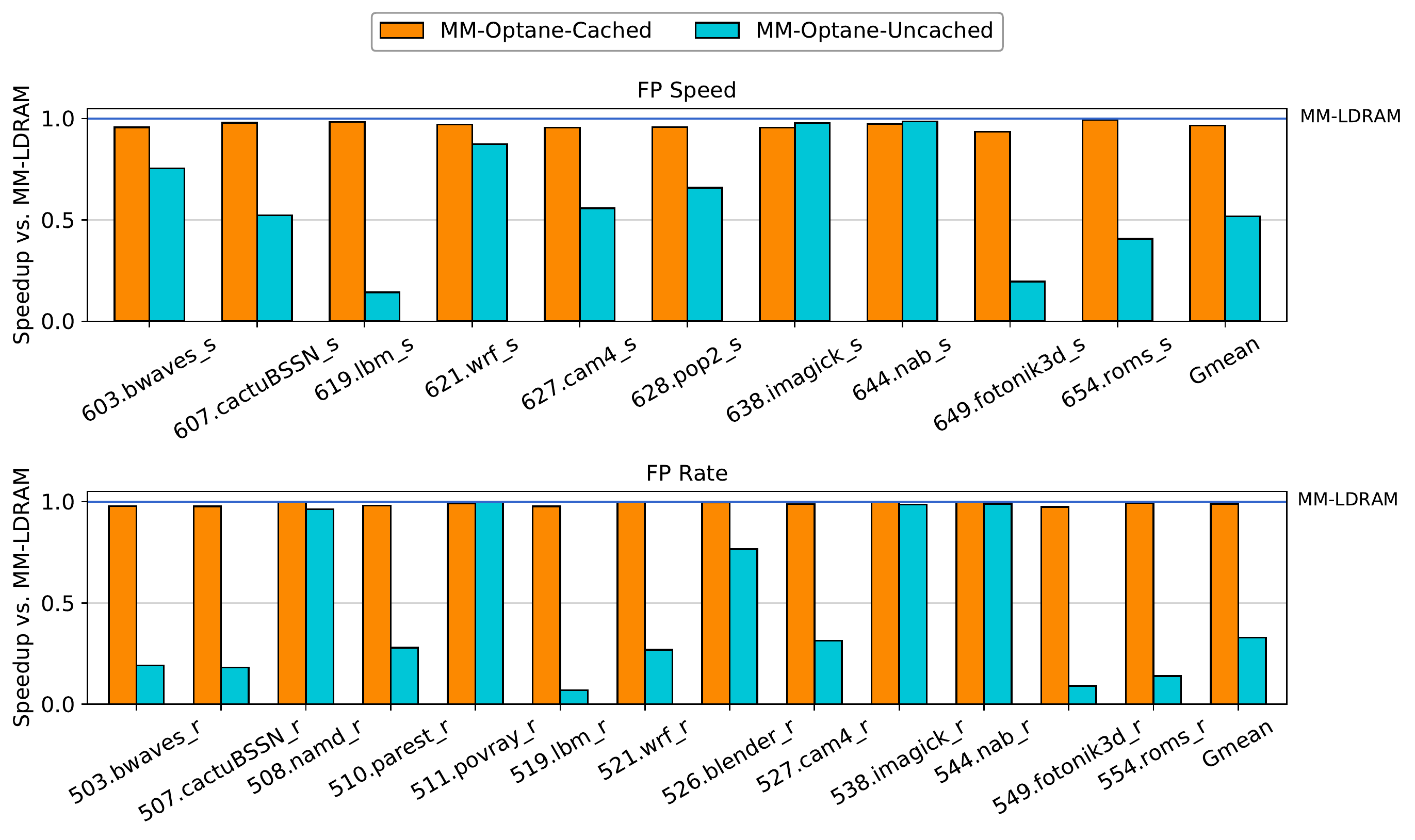,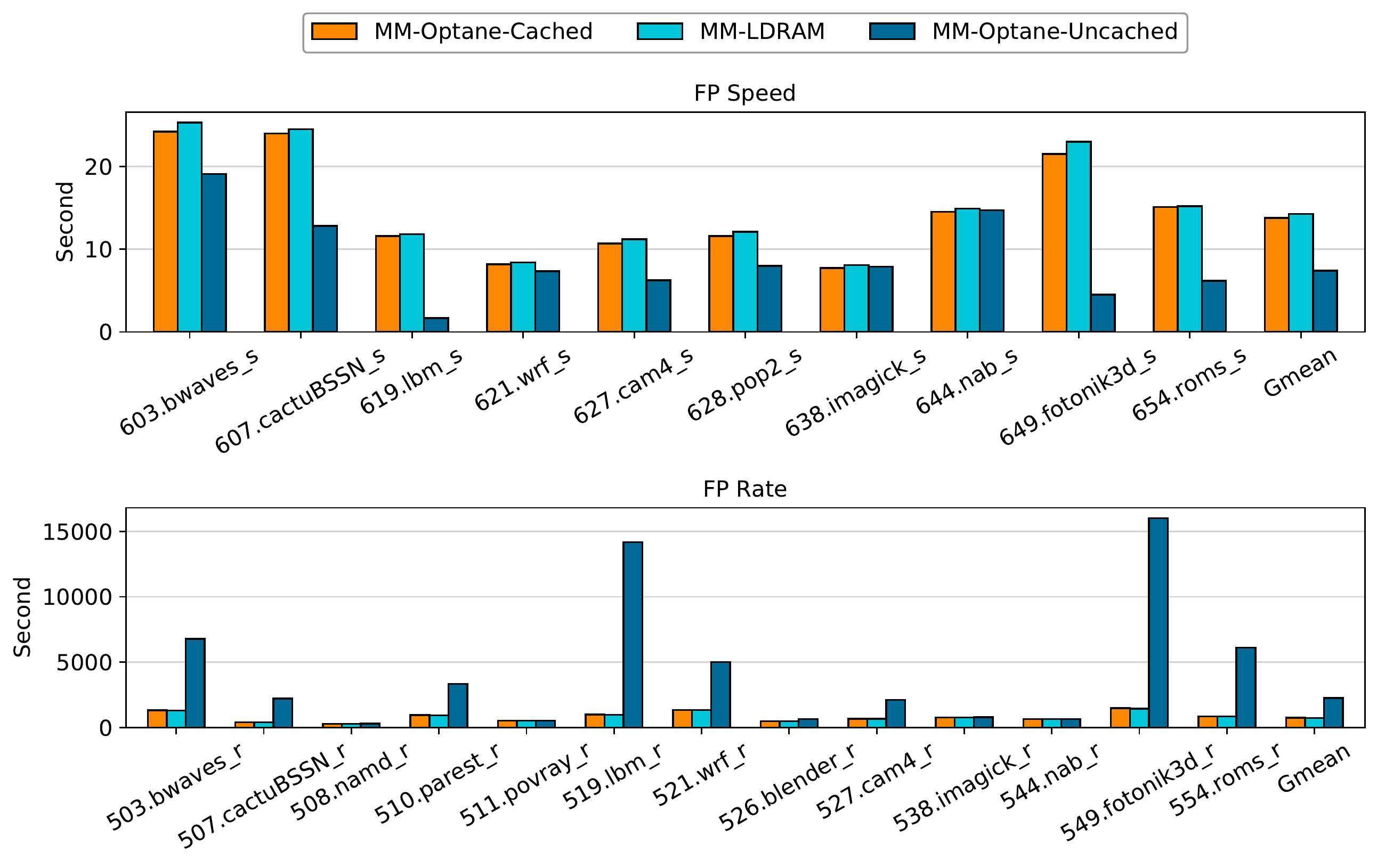,{\figtitle{SPEC 2017 floating point suite} This graph 
shows (a) the speedup relative to \MMLDRAM{} for the
SPEC 2017 floating point suite and (b) the execution time
(see data
in \dataref{csvroot/spec/spec06_int_rate_ratio_normalized.csv}, \dataref{csvroot/spec/spec17_fp_speed_ratio_normalized.csv}, \dataref{csvroot/spec/spec17_fp_rate_time.csv} and \dataref{csvroot/spec/spec17_fp_speed_time.csv}).},fig:spec17fp]

Our SPEC CPU tests demonstrate a number of points. First, in general, DRAM
outperforms cached \XP{} memory, and cached outperforms uncached \XP{}. Second,
uncached \XP{} memory is significantly slower in the rate test as multiple copies of the same
workloads in the rate test saturate the memory bandwidth. Third, cached \XP{} memory is
almost as fast as DRAM, which indicates the effectiveness of DRAM cache
for relatively small memory footprints.
Finally, certain workloads show better performance on
uncached \XP{} memory than DRAM in the speed test; we are still investigating
this result.

\takeaway{The DRAM cache is effective for workloads with small memory footprints.}{
With the \XPDIMM{} cached mode, workloads that fit in DRAM are unaffected by \XP{} memory's
higher latency and lower throughput.
}

\subsection{PARSEC}
\label{sec:parsec}

PARSEC~\cite{Zhan:2017:PMB:3053277.3053279} is a benchmark suite for testing
multi-processors. Similar to SPEC, PARSEC tests the CPU and memory
hierarchy of the system, however, unlike SPEC,
PARSEC is a multi-threaded benchmark suite. Users can specify the
number of threads, and we scale the thread count from one to sixteen for each benchmark throughout
our tests. We use the largest default input to create the largest possible memory
footprint. We tuned the iteration parameter to run the benchmark for an
adequate length of time, but we keep the rest of the configuration set to its default
values.

\wfigure[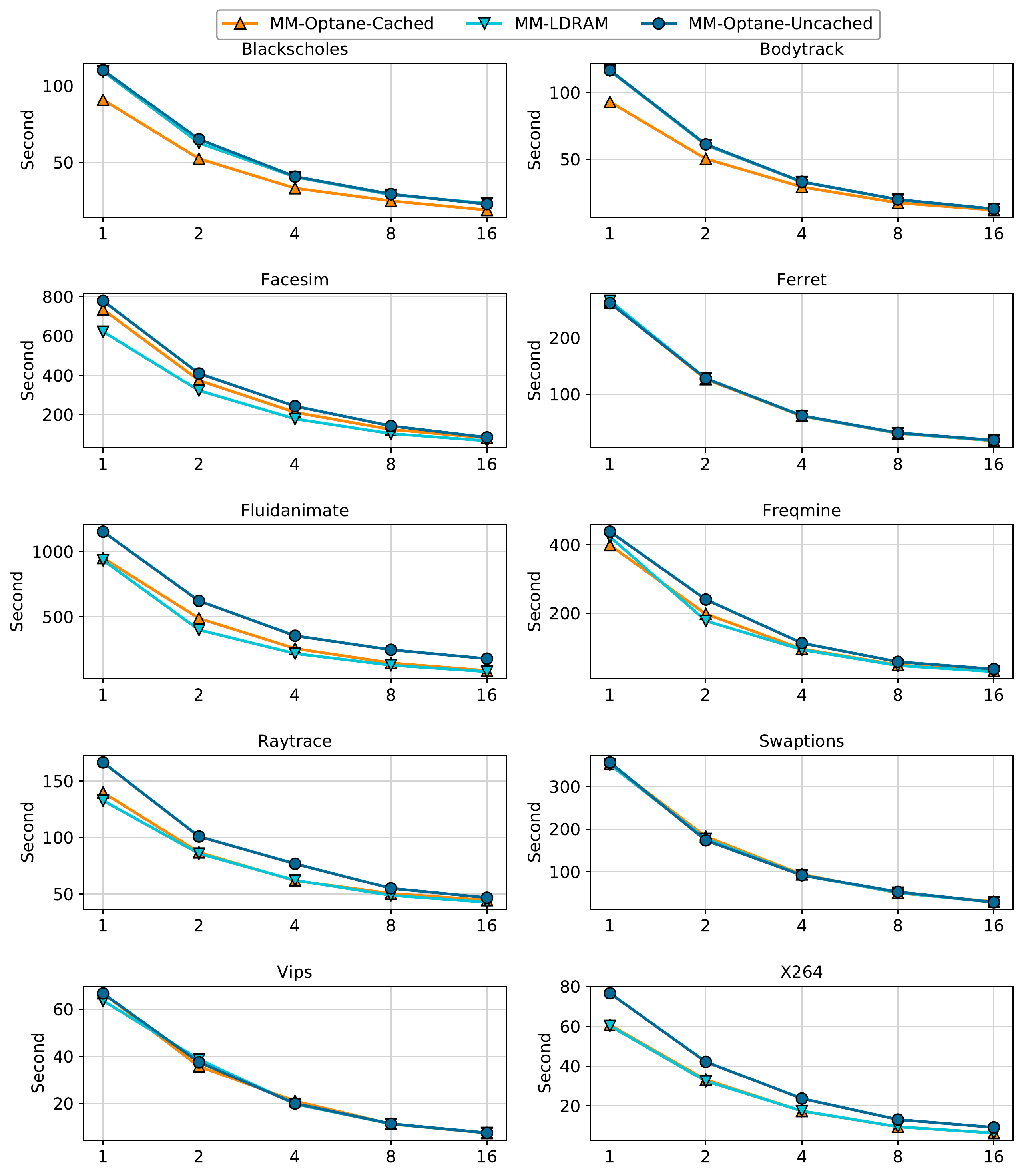,{\figtitle{PARSEC} These graphs
show runtime for the PARSEC benchmark suite run on varying numbers of threads.
Many benchmarks have a memory footprint that fits in the LLC, so they display
no difference between the memory types.
(see data
in~\dataref{csvroot/parsec/blackscholes.csv},
\dataref{csvroot/parsec/bodytrack.csv},
\dataref{csvroot/parsec/facesim.csv},
\dataref{csvroot/parsec/ferret.csv},
\dataref{csvroot/parsec/fluidanimate.csv},
\dataref{csvroot/parsec/freqmine.csv},
\dataref{csvroot/parsec/raytrace.csv},
\dataref{csvroot/parsec/swaptions.csv},
\dataref{csvroot/parsec/vips.csv},
and \dataref{csvroot/parsec/x264.csv}.)},fig:parsec]

As Figure~\ref{fig:parsec} shows, we only observe performance differences
on facesim, fluidanimate, raytrace, freqmine and x264, because the memory footprints
for the rest of the benchmarks can fit into the last level cache
on chip. Among those workloads that display differences, we observe a
performance gap between uncached \XP{} memory and DRAM that increases as the number of
threads increases and the \XP{} memory bandwidth gets saturated. 
In general, DRAM outperforms the other two memory
settings, while cached \XP{} has a close performance to DRAM,
indicating the DRAM cache's utility for small memory footprints. 

\takeaway{\XP{} memory's lower bandwidth can impact real-world applications.}{
\XP{} memory's bandwidth can be saturated with real-world multi-threaded applications,
resulting in a performance overhead when using uncached \XP{} as main memory.
}

\subsection{Memcached}
\label{sec:memcached}
Memcached~\cite{memcached} is a popular in-memory key-value
store used to accelerate web applications.  It uses slab allocation to allocate data and
maintains a single hash-table for keys and values.  We investigated memcached performance for
both different types of workloads (read or write dominant) and different total data sizes.

In our first experiment, to investigate how read/write performance
is impacted by memory type, we run two workloads: a GET-dominant (10\%
SET) workload, and a SET-dominant (50\%  
SET) workload. The key size is set at 128~Byte
and the value size is set as 1~KB, and the total memcached object storage memory size is set to 32~GB.
For each run, we start the memcached server with an empty cache.
We use the memaslap~\cite{memaslap} tool to generate the workload,
and set the thread count to 12 and the concurrency to 144 (each thread can have up to
144 requests pending at once). Both server and client threads are bound to dedicated cores on the same physical CPU.
\reffig{fig:memcached_rw} shows the throughput among different main memory 
configurations. This result demonstrates the real-world impact of \XP{}
memory's asymmetry between reads and writes, since the
DRAM cache is effective at hiding read latency, but has more trouble hiding write
latency.

\cfigure[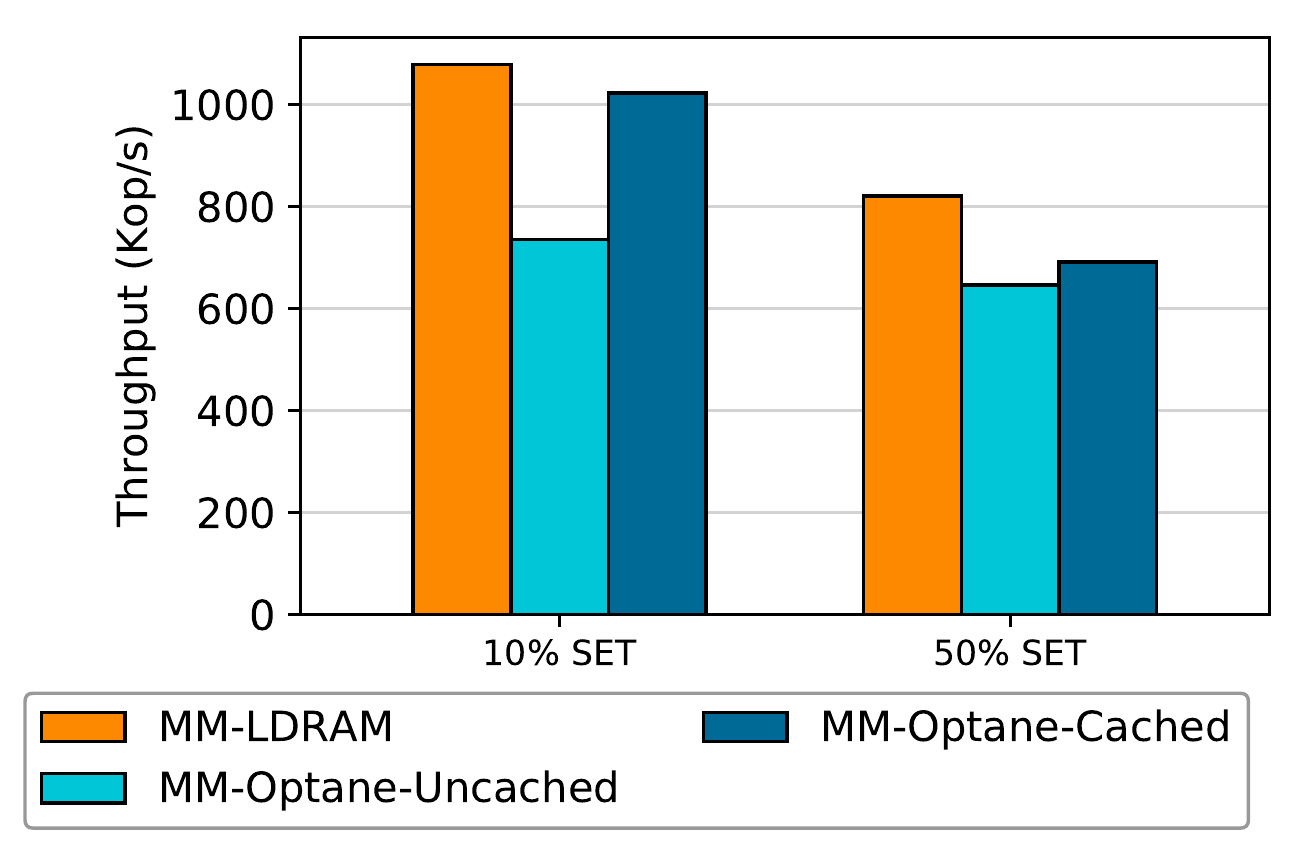,{\figtitle{Memcached on read/write workloads} This graph
shows memcached throughput for different mixes of operations.  Note that
the DRAM cache is effective in hiding read latency, but has more trouble hiding write
latency~(see data in~\dataref{csvroot/memory/memcached.csv}).},fig:memcached_rw]

In our second experiment, we vary the total size of the memcached
object store.  We run memcached with the 50\% 
SET workload as above and adjust the 
total size of the memcached store (between 32~GB and 768~GB). 
For each run, we add a warm-up phase before test execution.
The warm-up time and execution time are increased proportionally to the memcached store size. 

\reffig{fig:memcached_xpcached} shows two types of graphs.  The top shows the 
throughput of the different main memory configurations. 
The lower graph, in order to view the effectiveness of the DRAM cache, shows
the size ratio between client-requested accesses (that is, key and value size
as reported by memaslap)
and the total size of accesses that actually reached \XP{} memory in \MMPMEMC{} and 
\MMPMEMUC{} mode using device-level counters on the \XPDIMM{}.  
Note the machine has 192~GB DRAM on the local socket 
and another 192~GB on the remote socket, so at some point the
memcached store no longer only fits in the DRAM cache.
Due to write amplification, both within the application and within the
\XPDIMM{} itself,
\XP{} memory may experience more bytes written than total bytes requested from the client.

\ntwfigure[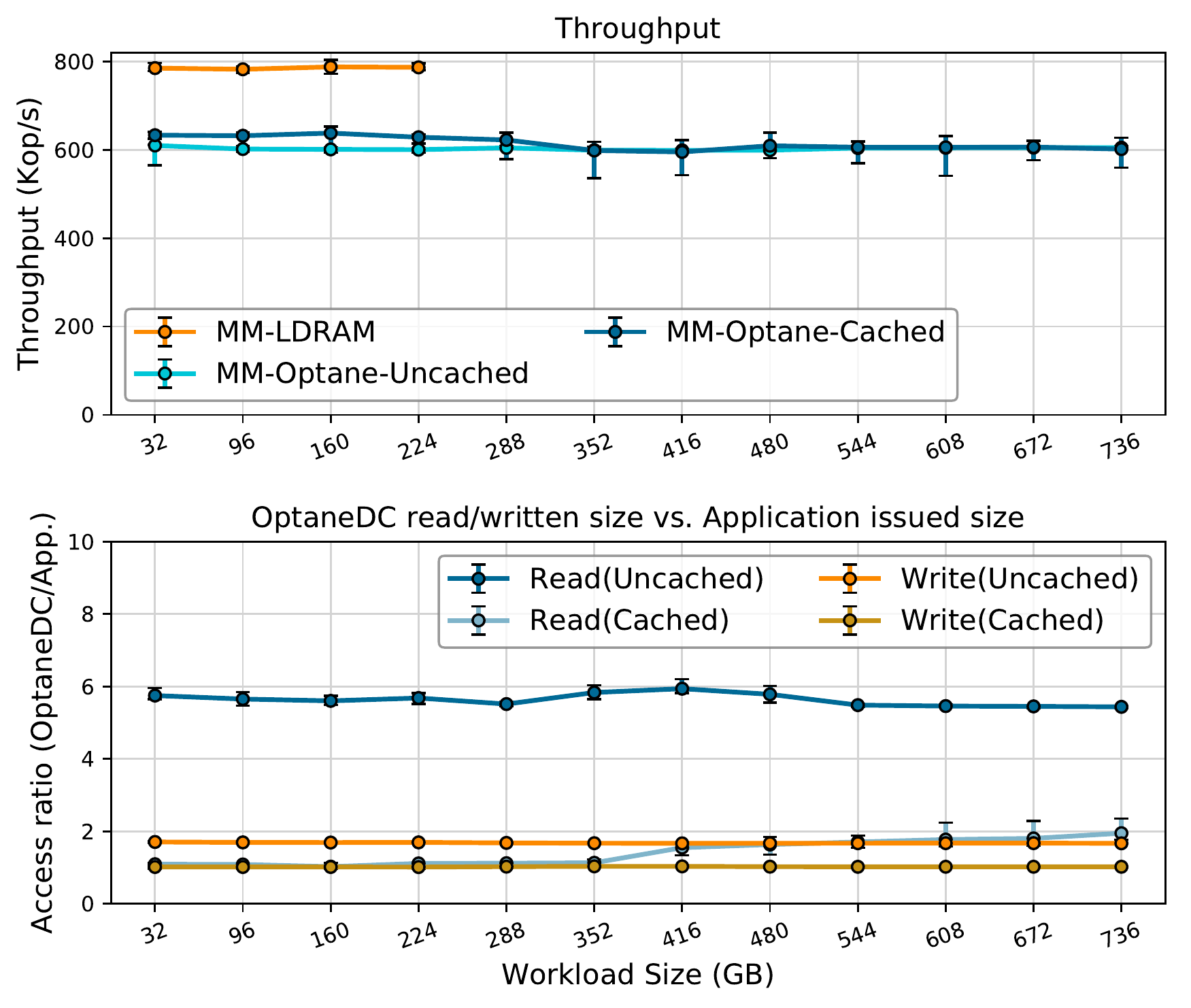,{\figtitle{Memcached 50\% SET throughput and memory access ratio}
The upper chart shows memcached throughput as the total size of the store grows. 
We ran the experiments 5 times and report the average with error bars covering the minimal and maximal values of each run.
Note that at 288~GB, the store no longer fits only in DRAM.  Also note that the DRAM cache
is ineffective at hiding the latency of \XP{} even when the store lies within
DRAM capacity.  The lower graph shows the proportion of application memory accesses that 
actually touch \XP{} memory in both \MMPMEMC{} and \MMPMEMUC{} mode.  
(see data in~\dataref{csvroot/memory/memcached_cache.csv} 
and~\dataref{csvroot/memory/memcached_ratio.csv}).},fig:memcached_xpcached]

\subsection{Redis}
\label{sec:redis-mm}
Redis~\cite{redis} is an in-memory key-value store widely used in website development 
as a caching layer and for message queue applications.
While Redis usually logs transactions to files, we turned off this capability
in order to tests its raw memory performance.
Our Redis experiment uses workloads issuing pure SETs followed by pure GETs.
Each key is an 8-byte integer and each value is 512~bytes,
and we run both the server and 12 concurrent clients on the same machine.  
As with memcached, for the \XP{} memory modes, we recorded the proportion
of memory accesses that were seen by \XP{} memory versus the combined value size
of the requests issued by the client.  \reffig{fig:redis_xpcached} shows the
result (thoughput in the top two graphs, and access ratio below).

In this experiment, \MMPMEMC{} is effective when the workload fits into DRAM.
The benefit of the cache on SET requests decreases as the 
workload size increases. As with memcached, the DRAM cache can effectively
reduce the data accesses to the actual \XP{} media.

\ntwfigure[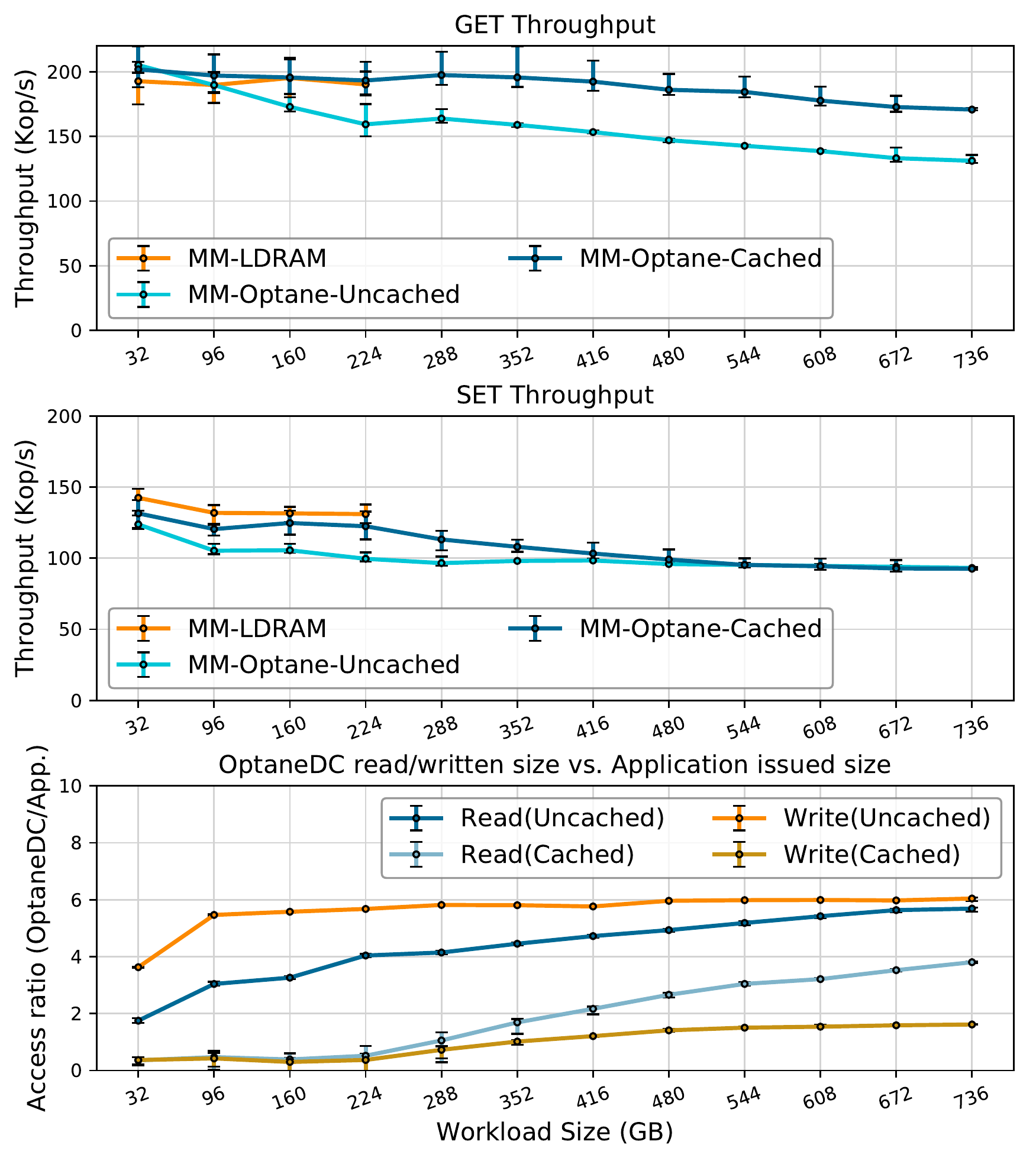,{
\figtitle{Redis throughput and memory access ratio}
The upper two charts show Redis throughput as the total size of the store grows for
workloads that are both read-dominant and write-dominant.  Note
that at 288~GB, the store no longer fits only in DRAM.  
The lower graph shows the proportion of application memory accesses that 
actually touch \XP{} memory in both \MMPMEMC{} and \MMPMEMUC{} mode. Due to access amplification
both within the application and \XPDIMM{},
\XP{} experiences significantly more bytes accessed than total value size.
(see data in~\dataref{csvroot/memory/redis_get.csv},
~\dataref{csvroot/memory/redis_set.csv} and
~\dataref{csvroot/memory/redis_ratio.csv}).},fig:redis_xpcached]

\section{\XP{} as Persistent Storage}
\label{sec:storage}

\XPCommercial{}s have the potential to profoundly affect the performance of storage
systems.  This section explores the performance of \XP{} as a storage technology
underlying various file systems.  For this section,
we use DRAM as the system memory, and use either
memory-based (e.g.\ \XP{} or DRAM)
or disk-based (e.g.\ Optane or Flash SSD) storage
underneath the file system.  These options correspond to
the system configurations \textbf{\PMLDRAM{}}, \textbf{\PMRDRAM{}},
\textbf{\PMLPMEM{}}, \textbf{\SSDOptane}, and \textbf{\SSDSATA}.  

We are interested in answering the following questions
about \XP{} memory as storage:
\begin{tightenum}
\item How well do existing file systems exploit \XP{} memory's performance?
\item Do custom file systems for NVMM give better performance than adaptations of block-based file systems?
\item Can using a load/store interface (DAX) interface to persistent memory improve performance?
\item How well do existing methods of emulating NVMM (namely, running the experiment on DRAM) actually work?
\end{tightenum}

We explore the performance of \XP{} memory
as a storage device using a number of different benchmarks. 
We first investigate basic performance by running raw file operations
in Section~\ref{sec:fops}, synthetic I/O in Section~\ref{sec:fio},
and emulated
application performance in Section~\ref{sec:filebench}.  Next,
in Sections~\ref{sec:rocksdb_storage} 
through~\ref{sec:mongodb_storage}, we explore
application performance with the workloads
listed in Table~\ref{tab:app-config}.

We evaluate seven file systems and file system configurations
with these benchmarks. Each benchmark runs on
all file systems, mounted on the three memory configurations
and the two SSD configurations (when compatible).

\boldparagraph{\extfs{}}  \extfs{} is a widely deployed Linux file system. This
configuration runs \extfs{} in normal (i.e., non-DAX) mode with
its default mount options and page cache.
\extfs{} only journals its metadata for crash consistency, but not data, which
means a power failure that occurs in the middle of writing a file page can
result in a torn page.

\boldparagraph{\extfs{-DJ}} This mode of \extfs{}
provides stronger consistency guarantees than its default setting by journaling both
file system metadata and file data updates. It ensures every
\texttt{write()} operation
is transactional and cannot be torn by a power
failure.

\boldparagraph{\extfsDAX{}} Mounting \extfs{} with the DAX option
bypasses the page cache.  Therefore, \extfsDAX{} accesses data directly in memory
(that is, on the \XP{} device).
It is not compatible with the data journaling feature, so \extfsDAX{} can not
provide consistency guarantees for file data writes.

\boldparagraph{\xfs{}} \xfs{} is another popular Linux file system. This
configuration uses the file system in its default (i.e., non-DAX) mode.
Similar to \extfs{}, \xfs{} also uses the page cache and does not
provide failure-atomic data writes to files.

\boldparagraph{\xfsDAX{}} This is the DAX mode for \xfs{}. Similar to
\extfsDAX{}, this mode does not use the page cache and also does not provide data
consistency guarantees.

\boldparagraph{\nova{}} \nova{}~\cite{novapaper, nova-fortis} is a purpose-built
NVMM file system. It implements a log-structured metadata and a copy-on-write
mechanism for file data updates to provide crash-consistency guarantees for
all metadata and file data operations. \nova{} only operates with \XP{} devices in
DAX mode, bypassing the page cache, and consequentially is incompatible
with block-based devices.

\boldparagraph{\nova{-Relaxed}} In this mode, \nova{} relaxes
its consistency guarantees for file data updates, by allowing in-place file
page writes, to improve write performance for applications that do not require
file data consistency for every write. This mode still guarantees metadata
consistency for all file system operations.

\subsection{File Operation Latency}
\label{sec:fops}

We begin by taking basic performance measurements
on our file systems.
We measure single-threaded system call latencies for the following file system operations:
\begin{itemize}[noitemsep]
\item Create: Create a file with \texttt{open()}, and \texttt{close()} it,
      without writing any file data or calling \texttt{fsync()}.
\item Append (4K): \texttt{open()} an existing file, \texttt{write()} 4~KB of data
      to its end, call \texttt{fsync()}, and \texttt{close()} the file.
\item Write (4K): \texttt{open()} an existing file, \texttt{write()} 4~KB of data
      to the middle, call \texttt{fsync()}, and \texttt{close()} the file. 
\item Write (512B): \texttt{open()} an existing file, \texttt{write()} 512~bytes of
      data to the middle, call \texttt{fsync()}, and \texttt{close()} the file. 
\item Read (4K): \texttt{open()} an existing file, \texttt{read()} 4~KB of data, and
      \texttt{close()} the file.
\end{itemize}

\reffig{fig:fileops} shows the measured operation latencies. Non-DAX file
systems (hatched bars) experience the longest latencies on the SATA SSD.
Their write performance, however, is the best on the Optane SSD,
implying their \texttt{write()} and \texttt{fsync()} paths are not optimized for
memory-type storage.

On DRAM and \XP{} devices, \extfsDAX{} and \xfsDAX{} show
better write latency numbers than their non-DAX counterparts.
Purpose-built \nova{} and \nova{-Relaxed} file systems
outperform conventional file systems for create and append operations. \nova{}'s
write latency is longer than \extfsDAX{}, especially for 512~byte writes, because
\nova{} performs page-level copy-on-write for file data consistency. \nova{-Relaxed}
regains performance by allowing in-place file data writes.

For memory-type storage devices, \XP{}'s longer latency than DRAM
affects all file systems, increasing their operation latencies by between 3.3\%
(\xfsDAX{}) and 156\% 
(\nova{}).

DAX-enabled file systems (\extfsDAX{}, \xfsDAX{}, \nova{}, and \nova{-Relaxed})
all have similar read latency numbers, and they all increase latency when moving from
DRAM to \XP{}. In comparison, read latencies of non-DAX file systems
are less affected because they still leverage the DRAM-based page cache to hide
the \XP{} memory's latency. 

\wfigure[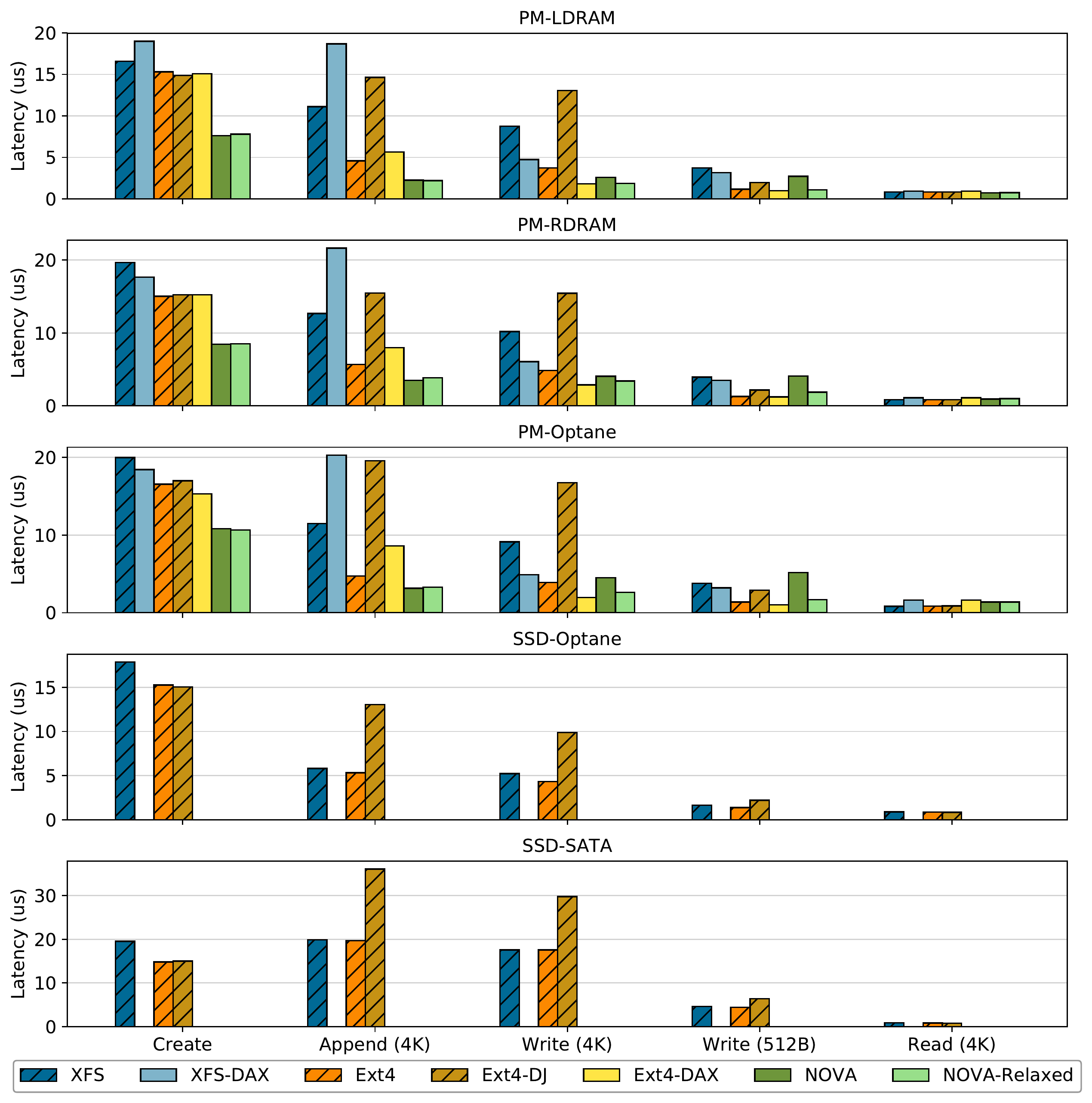,{
  \figtitle{File operation latency} This chart shows average file operation
	latencies across all storage types and file systems. Only non-DAX file
	systems (hatched bars) work with SSD storage devices.
	Among memory-backed configurations, \PMLPMEM{} has the longest latency numbers,
  especially for the write paths, indicating that using DRAM to emulate
	\XP{} memory overestimates performance. With \PMLPMEM{}, non-DAX file systems have shorter read
  latencies than DAX file systems because they leverage DRAM through the page cache.
  (see data in
   \dataref{csvroot/storage/fileops/fileops.ldram.csv},
   \dataref{csvroot/storage/fileops/fileops.rdram.csv},
   \dataref{csvroot/storage/fileops/fileops.pm-optane.csv},
   \dataref{csvroot/storage/fileops/fileops.ssd-optane.csv},
   and
   \dataref{csvroot/storage/fileops/fileops.ssd-sata.csv}).
},fig:fileops]

\takeaway{Non-DAX file systems can outperform DAX file systems on \XP{} because
non-DAX file systems benefit from the fast DRAM page cache.}{For non-DAX systems, the page cache
can serve to hide the read latency of \XP{} memory.}

\takeaway{The relatively long latency of \XP{} can amplify small 
inefficiencies in file system designs.}{\nova{}'s copy-on-write
mechanism for ensuring strong consistency of data writes incurs extra latency overhead.}

\subsection{FIO Bandwidth}
\label{sec:fio}

FIO is a versatile storage benchmark tool that can generate synthetic I/O
traffic to emulate practical workloads. We run FIO to measure the bandwidth of
basic read/write file operations.

We run FIO v3.11 using the ``sync'' ioengine with four types of read/write
workloads: sequential read, random read, sequential write, and random write.
All workloads use a 512~MB file size per thread and 4~KB read or write size
(``blocksize''). For write workloads, we issue an \texttt{fsync()} after
writing every 4~KB file data.  Each workload runs for 30 seconds, and the number
of threads vary from one to sixteen, with each thread accessing a different file.
FIO by default invalidates a file's page cache before performing each IO
operation, and therefore, for non-DAX file systems (\extfs{}, \extfs{-DJ}, and
\xfs{}) their read paths copy data twice: first from the storage media to the page
cache, and then from the page cache to a user buffer.
\reffig{fig:fio.combo} illustrates the FIO bandwidths with different file system
and storage configurations.

\wfigure[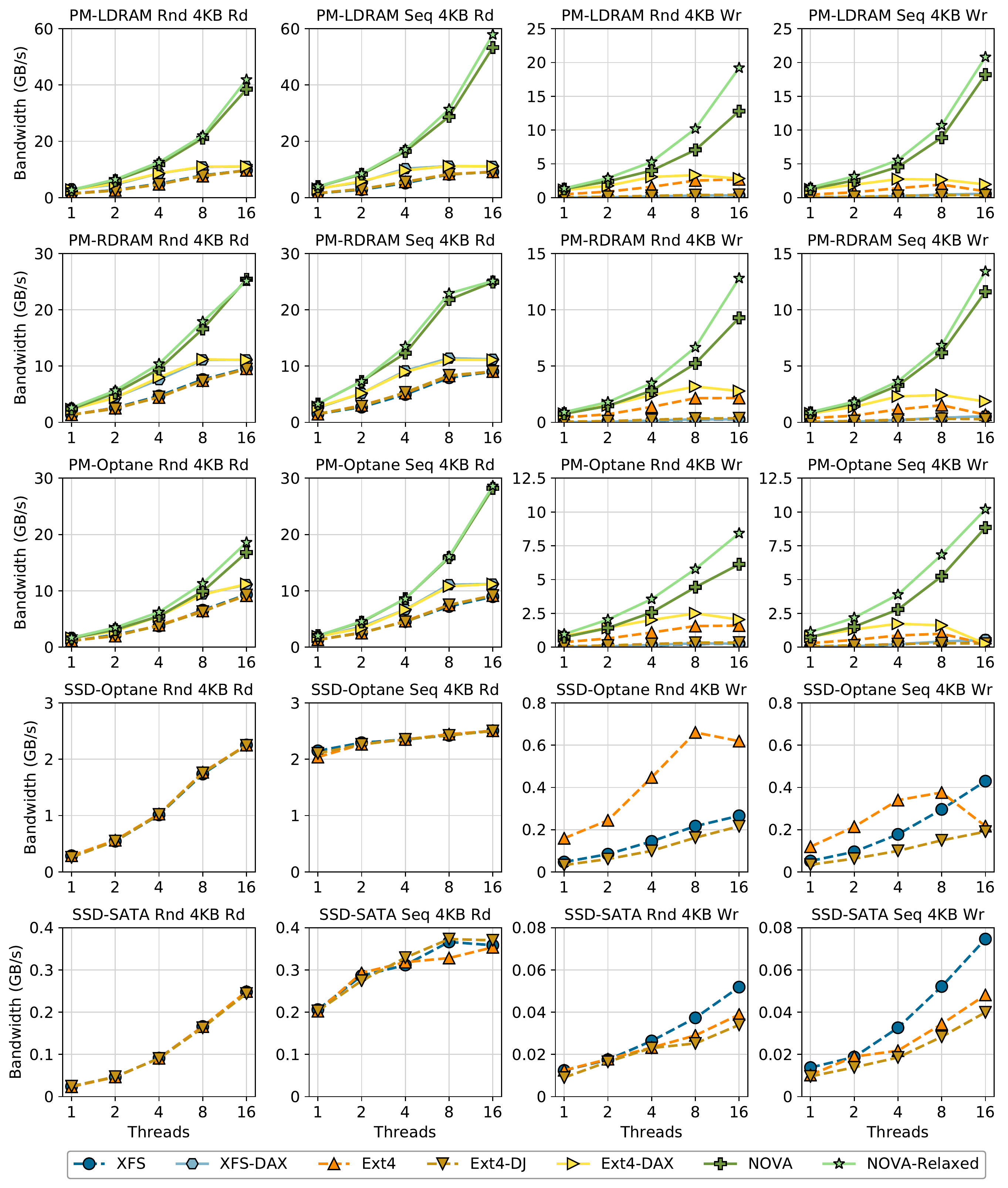,{
  \figtitle{FIO 4~KB read and write bandwidth}
  This graph shows the 4~KB read and write throughput of various file systems as a
  function of thread count on different kinds of storage. In memory-backed configurations,
  DAX-enabled file systems demonstrate better performance than non-DAX
  ones due to bypassing the page cache. \nova{} and \nova{-Relaxed} generally
  outperform other file systems and
  scale much better --- only these file systems demonstrate that \XP{} memory has much better
  performance than SSDs.
  (see data in \dataref{csvroot/storage/fio/fio_data.csv}).
},fig:fio.combo]

\subsection{Filebench}
\label{sec:filebench}

Filebench~\cite{filebench} is a popular storage benchmark suite 
that mimics the behavior of
common storage applications. We ran four of the predefined workloads,
and their properties are
summarized in \reftab{tab:filebench-config}. 

\begin{table}[h]
\centering
  \resizebox{.55\textwidth}{!} {
    \begin{tabular}{|l|l|l|l|l|}\hline
                     & Fileserver & Varmail & Webproxy & Webserver \\\hline
       nfiles        & 500~K      & 1~M     & 1~M      & 500~K \\\hline
       meandirwidth  & 20         & 1~M     & 1~M      & 20 \\\hline
       meanfilesize  & 128~K      & 32~K    & 32~K     & 64~K \\\hline
       iosize        & 16~K       & 1~M     & 1~M      & 1~M \\\hline
       nthreads      & 50         & 50      & 50       & 50 \\\hline
       R/W Ratio     & 1:2        & 1:1     & 5:1      & 10:1 \\\hline
    \end{tabular}
  }
  \caption{\textbf{Filebench configurations} These configurations are used for the experiments
	in \reffig{fig:filebench}.}
\label{tab:filebench-config}
\end{table}

\begin{tightenum}
\item \emph{fileserver} emulates the I/O activities of a file server with
write-intensive workloads. It performs mixed operations of creates, deletes,
appends, reads, and writes.
\item \emph{varmail} emulates a mail server that saves each email in a separate
file, producing a mix of multi-threaded create-append-sync, read-append-sync,
read, and delete operations.
\item \emph{webproxy} emulates the I/O activities of a a simple web proxy
server. The workload consists of create-write-close, open-read-close, delete,
and proxy log appending operations.
\item \emph{webserver} emulates a web server with read-intensive workloads,
consisting of open-read-close activities on multiple files and log appends.
\end{tightenum}

\wfigure[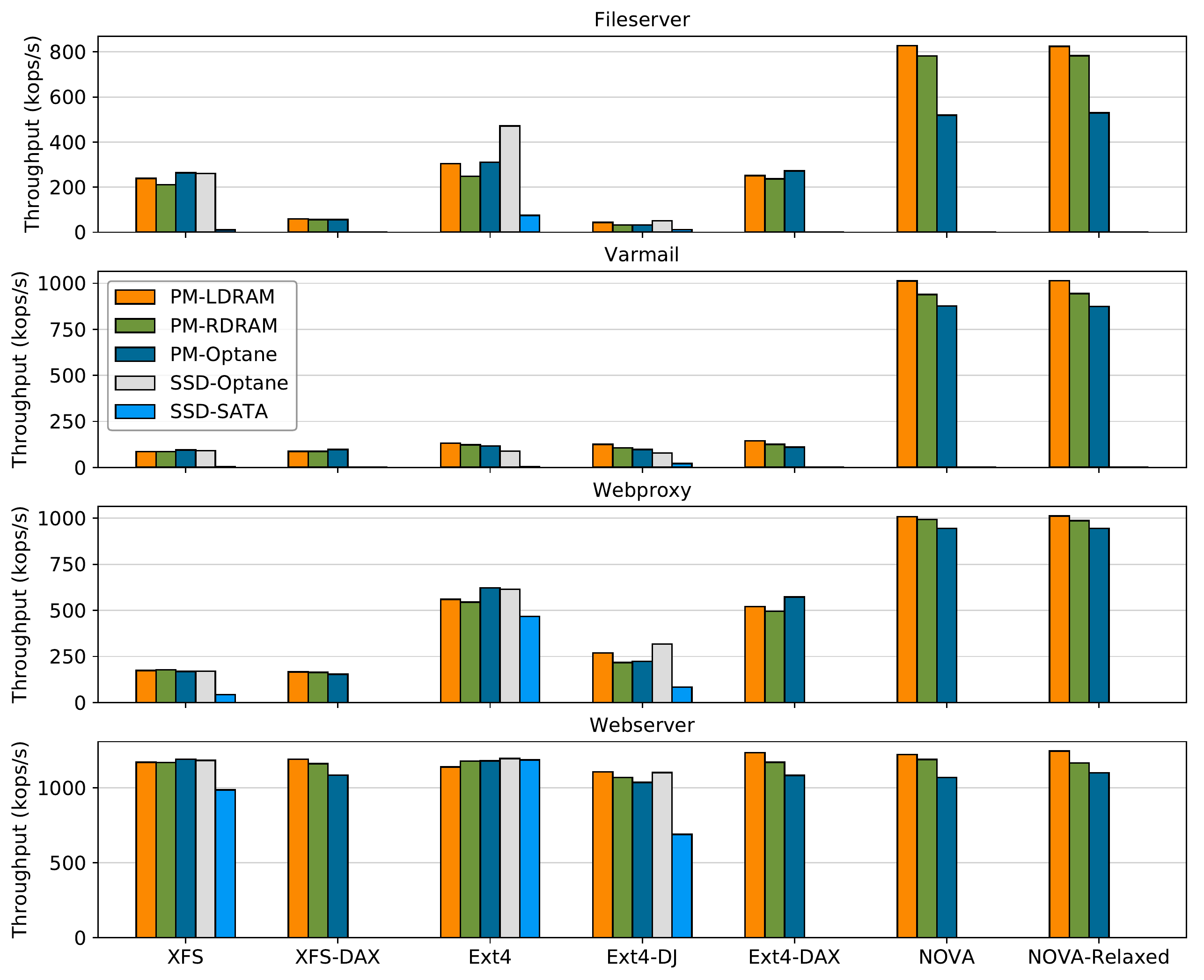,{
   \figtitle{Filebench throughput} This graph shows file system throughput
	 on a series of simulated workloads from the Filebench suite.  
	 In general, file systems perform similarly on
   read-intensive workloads (webserver), but \nova{} and \nova{-Relaxed} outperform
   other file systems when more write traffic is involved.
  (see data in
   \dataref{csvroot/storage/filebench/fileserver.csv},
   \dataref{csvroot/storage/filebench/varmail.csv},
   \dataref{csvroot/storage/filebench/webproxy.csv},
   and
   \dataref{csvroot/storage/filebench/webserver.csv})}
,fig:filebench]

\reffig{fig:filebench} presents the measured throughput using Filebench
workloads. The tested file systems perform similarly on read-intensive workloads
(e.g.\ webserver), however, \nova{} and \nova{-Relaxed} outperform others when more
write traffic is involved (e.g.\ fileserver and varmail). On average, \nova{} is
faster than other evaluated file systems by between 1.43\x{} and 3.13\x{}, and
\nova{-Relaxed} is marginally faster than \nova{}.  Interestingly, ext4 performs
better on block devices than even DRAM.  Investigation of this anomaly suggested
inefficiencies in ext4's byte-granularity code path are responsible.

\takeaway{Small random writes can result in drastic performance differences between
DRAM emulation and real \XP{} memory.}{ \PMLPMEM{} impacts \nova{}
and \nova{-Relaxed} most with the fileserver workload because it generates lots
of small random writes that consequently cause write amplification on
\XPDIMM{}s.}

\subsection{RocksDB}
\label{sec:rocksdb_storage}

Having taken simple measurements using microbenchmarks and emulated workloads
for basic system performance,
we transition to larger scale application workloads; detailed workload descriptions
and their runtime arguments can be found in Table~\ref{tab:app-config}.

\begin{table}
    \centering
    \begin{tabular}{lllll}
       Application   & Version & Type           & Benchmark       & Workload                    \\\hline
       RocksDB       & 5.4     & Embedded       & db\_bench       & K/V=16B/100B, 10M random SET, 1 thread     \\\hline
       Redis         & 3.2     & Client/server  & redis-benchmark & K/V=4B/4B, 1M random MSET, 1 thread  \\\hline
       Kyoto Cabinet & 1.2.76  & Embedded       & kchashtest      & K/V=8B/1KB, 1M random SET, 1 thread     \\\hline
       MySQL         & 5.7.21  & Client/server  & TPC-C           & W10, 1 client for 5 minutes           \\\hline
       SQLite        & 3.19.0  & Embedded       & Mobibench       & 1M random INSERT, 1 thread \\\hline
       LMDB          & 0.9.70  & Embedded       & db\_bench       & K/V=16B/96B, 10M sequential SET, 1 thread     \\\hline
       MongoDB       & 3.5.13  & Client/server  & YCSB            & 100k ops of Workload A,B, 1 thread \\\hline
    \end{tabular}
	\vspace*{1mm}
	\caption{\figtitle{Application configurations} These workload configurations are used for experiments
	in Sections~\ref{sec:rocksdb_storage} through~\ref{sec:mongodb_storage}}
	\label{tab:app-config}
	\vspace{-2mm}
\end{table}

RocksDB~\cite{rocksdb} is a high-performance embedded key-value store, designed by Facebook and inspired
by Google's LevelDB~\cite{leveldb}. RocksDB's design is centered around
the log-structured merge tree (LSM-tree),
which is designed for block-based storage devices, absorbing random writes and converting them to
sequential writes to maximize hard disk bandwidth.

RocksDB is composed of two parts: a memory component and a disk component. The memory component is a
sorted data structure, called the memtable, that resides in DRAM. The memtable absorbs new inserts and 
provides fast insertion and searches.  
When applications write data to an LSM-tree, it is first inserted to the memtable.
The memtable is organized as a skip-list, providing O(log n) inserts and searches. To ensure persistency,
RocksDB also appends the data to a write-ahead logging (WAL) file.
The disk component is structured into multiple layers with increasing sizes. Each level contains multiple
sorted files, called the sorted sequence table (SSTable). 
When the memtable is full, it is flushed to disk and becomes an SSTable
in the first layer. When the number of SSTables in a layer exceeds a threshold, RocksDB merges the 
SSTables with the next layer's SSTables that have overlapping key ranges. This compaction
process reduces the number of disk accesses for read operations.

\wfigure[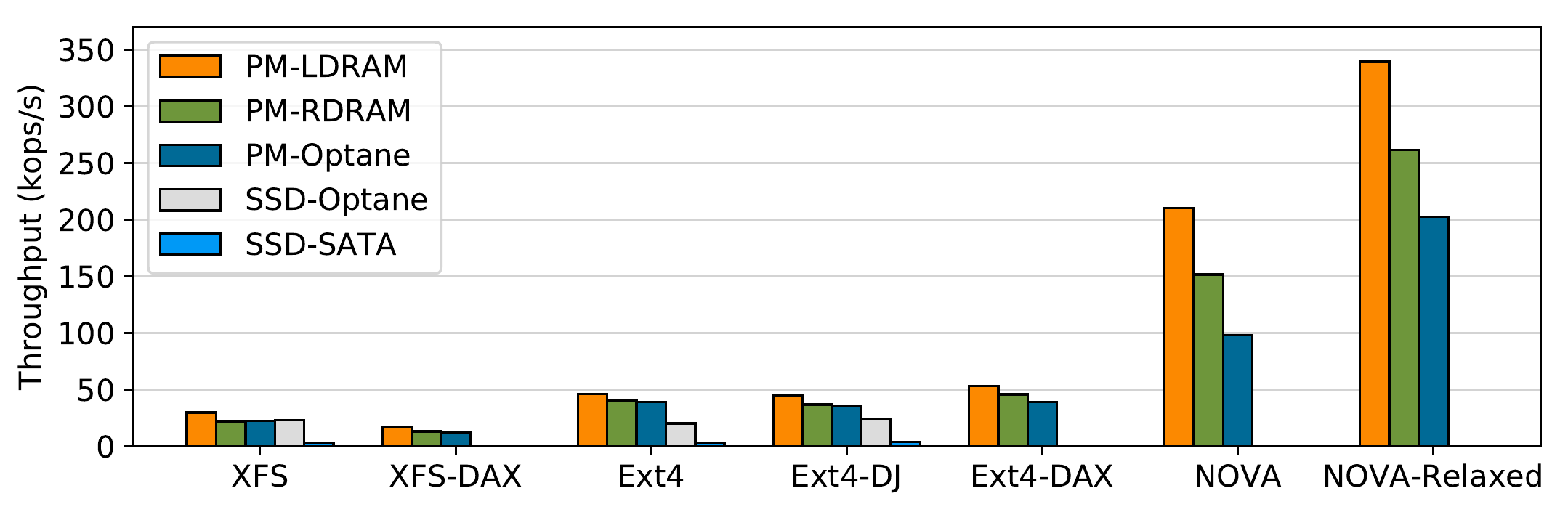,{\figtitle{RocksDB throughput} This graph shows throughput
on a write-dominant workload for the RocksDB key/value store.  The frequent use of syncs
in the application means that non-NVMM file systems incur significant flushing costs
and cannot batch updates,
whereas the NOVA-type file systems' fast sync mechanism drastically improves 
performance~\seedatain{csvroot/storage/app/rocksdb.csv}.},fig:rocksdb]

RocksDB makes all I/O requests sequential to make the best use of hard disks' sequential access strength.
It supports concurrent writes when the old memtable is flushed to disk, and only performs large writes to the
disk (except for WAL appends). However, WAL appending and sync operations can still impact performance
significantly on NVMM file systems. 
Using the \texttt{db\_bench} benchmark, we investigate the SET throughput with 
20-byte key size and 100-byte value size, syncing the database after each SET operation.
We illustrate the result in Figure~\ref{fig:rocksdb}.  Note that the frequent
use of sync operations in the application significantly hurts the performance
of most file systems, though NOVA-type file systems maintain their performance
through the use of an NVM-optimized sync operation.

\subsection{Redis}
\label{sec:redis_storage}

\wfigure[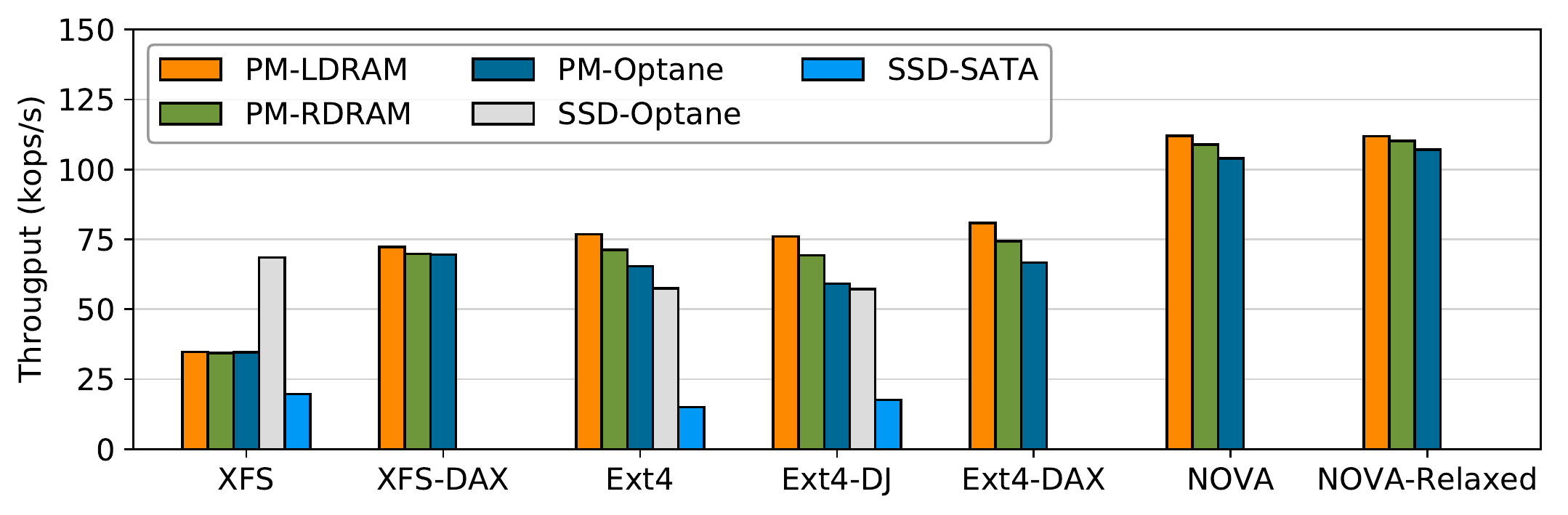,{\figtitle{Redis throughput} This graph
shows throughput on the Redis key-value store using a write-dominant
workload.  Like RocksDB, Redis issues frequent sync operations, and consequently
the NOVA-type file systems perform the 
best~\seedatainas{Graphs/redis-persistent/redis_storage.csv}{csvroot/storage/app/redis_storage.csv}.}
,fig:redis_storage]

Redis~\cite{redis} is an in-memory key-value store
widely used in website development as a caching layer and for message queue applications.
Redis uses an ``append-only file'' (AOF) to log all the write operations to the
storage device.  At recovery, it replays the log.  The frequency at which Redis
flushes the AOF to persistent storage allows the administrator to trade-off
between performance and consistency.

\reffig{fig:redis_storage} measures Redis's MSET (multiple sets) benchmark performance
where each MSET operation updates ten key/value pairs (190 bytes).
One MSET operation generates a 335~byte log record and appends it to the AOF.
Redis supports three \texttt{fsync} modes - ``always'', ``everysec'', ``no'' - for flushing
the AOF to the persistent storage.
For our experiment, we chose the ``always'' fsync 
policy where \texttt{fsync} is called after every log append. 
This version of Redis is ``persistent'' since it ensures that no data is lost after recovery. 
We measure this mode to see how the safest version of Redis performs on different NVMM file systems.
We put Redis server and client processes in the same machine for this experiment, though
the processes communicate via TCP.  As with RocksDB, the strong consistency requirement
of Redis and its frequent use of syncs results in a performance win for NOVA-type file systems.
Interestingly, XFS performs better on block-based devices than even DRAM.
Investigation into this anomaly suggested
inefficiencies in xfs's byte-granularity code path are responsible and manifest in
a few other benchmarks.

\subsection{Kyoto Cabinet}
\label{sec:kc_storage}

Kyoto Cabinet~\cite{kyotocabinet} (KC) is a high-performance database library.
It stores the database in a single file with database metadata at the head.
Kyoto Cabinet memory maps the metadata region, uses load/store instructions to
access and update it, and calls \texttt{msync} to persist the changes.  Kyoto
Cabinet uses write-ahead logging to provide failure atomicity for SET
operations.

\wfigure[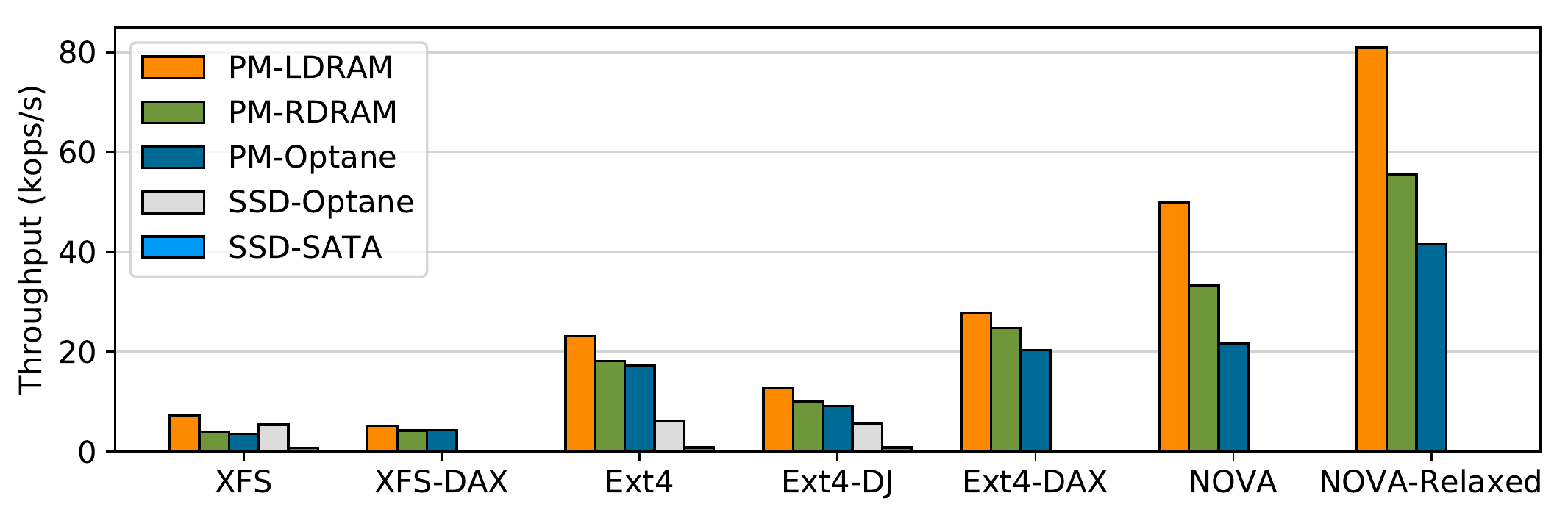,{\figtitle{Kyoto Cabinet throughput} This graph shows
the throughput of Kyoto Cabinet's HashDB on a write-dominant workload. 
As with the RocksDB and Redis experiments, \nova{-Relaxed} performs the best
due to a fast sync mechanism~\seedatain{csvroot/storage/app/kc.csv}.},fig:kyoto]

We measure the throughput for SET operations on Kyoto Cabinet's HashDB
data structure (\reffig{fig:kyoto}). 
HashDB is a hash table implementation where each bucket is the root of the binary search tree.  
A transaction on HashDB first appends an undo log record to the WAL
and then updates the target record in place.
During commit, it flushes the updated data using \texttt{msync} and 
truncates the WAL file to invalidate log records.
We use KC's own benchmark, \texttt{kchashtest order}, to measure 
the throughput of HashDB's with one million random SET transactions,
where,
for each transaction, the key size is 8 bytes, and the value size is 1024 bytes.
By default, each transaction is not persisted (i.e., not \texttt{fsync}'d) during commit, 
so we modified the benchmark such that every transaction persists at its end.
In these experiments, we uncovered a performance issue when \texttt{msync} is
used on transparent huge pages with DAX-file systems; performance dropped over
90\%.  
Switching off huge pages
fixed the issue --- DAX file systems for this benchmark are reported with huge pages turned off
and we are continuing to investigate the bug.

\subsection{MySQL}
\label{sec:mysql_storage}

\wfigure[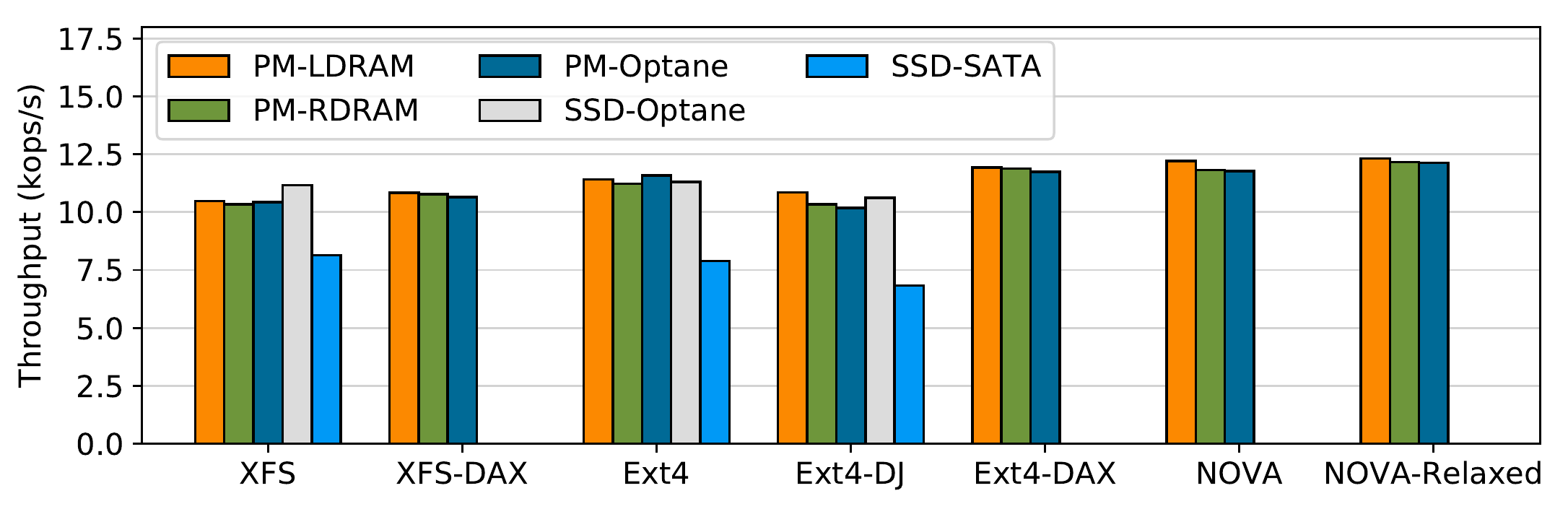,{\figtitle{MySQL running TPC-C} This experiment
demonstrates the popular MySQL's performance on the TPC-C benchmark.  Note that
performance across memory types remains surprisingly stable due to MySQL's
aggressive use of a buffer pool and checkpointing mechanism that avoid putting
the file system on the critical path as much as 
possible~\seedatain{csvroot/storage/app/mysql.csv}.},fig:mysql]

We further evaluate the throughput of databases on \XP{} with MySQL~\cite{mysql}, 
a widely-used relational database.
We measure the throughput of MySQL with
TPC-C~\cite{tpcc}, a workload representative of online transaction processing (OLTP). 
We use ten warehouses, and each run takes five minutes.  
Figure~\ref{fig:mysql} shows the MySQL throughput.  As MySQL's default settings
include aggressive use of the buffer pool and also a checkpointing mechanism
to avoid writing to persistence
regularly and to hide access latency, performance remains
surprisingly stable across file systems and storage device.

\subsection{SQLite}
\label{sec:sqlite_storage}

SQLite~\cite{sqlite} is a lightweight embedded relational database that is
popular in mobile systems.  SQLite stores data in a B+tree contained in a
single file.  To ensure consistency, SQLite can use several mechanisms to log
updates.  We configure it to use write-ahead, redo logging (WAL) since our
measurements show it provides the best performance.

\wfigure[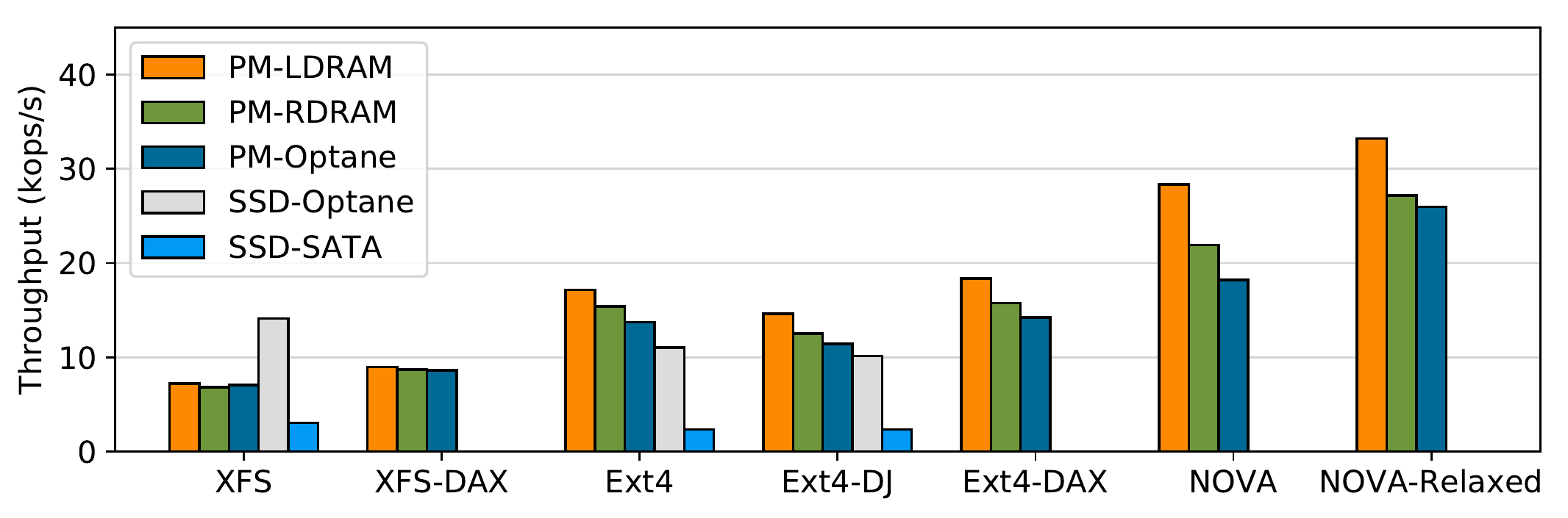,{\figtitle{SQLite throughput} This graph shows SQLite
throughput on a write-dominant workload.  \nova{-Relaxed}'s optimization
to allow in-place data updates to a file give it a significant performance
boost on this benchmark, since all accesses modify a single B+tree contained 
in a single file~\seedatain{csvroot/storage/app/sqlite.csv}.},fig:sqlite]
  
We use Mobibench~\cite{mobibench} to test the SET performance of SQLite in WAL
mode.  The workload inserts 100~byte values into a single table with one thread.
\reffig{fig:sqlite} shows the result.  \nova{-Relaxed} performs the best on this benchmark
and significantly improves over regular \nova{}.  This difference, which can be attributed solely
to the in-place update optimization, is significant in SQLite due to its randomly distributed
writes to a B+tree contained in a single file.

\subsection{LMDB}
\label{sec:lmdb}

Lightning Memory-Mapped Database Manager (LMDB)~\cite{lmdb} is a Btree-based,
lightweight database management library.  LMDB memory-maps the entire database
so that all data accesses directly load and store the mapped memory region.
LMDB performs copy-on-write on data pages to provide atomicity, a technique
that requires frequent \texttt{msync} calls.

We measure the throughput of sequential SET operations using LevelDB's \texttt{db\_bench} benchmark.
Each SET operation is synchronous and consists of 16-byte key and 96-byte value.
\reffig{fig:lmdb} shows the result.

\wfigure[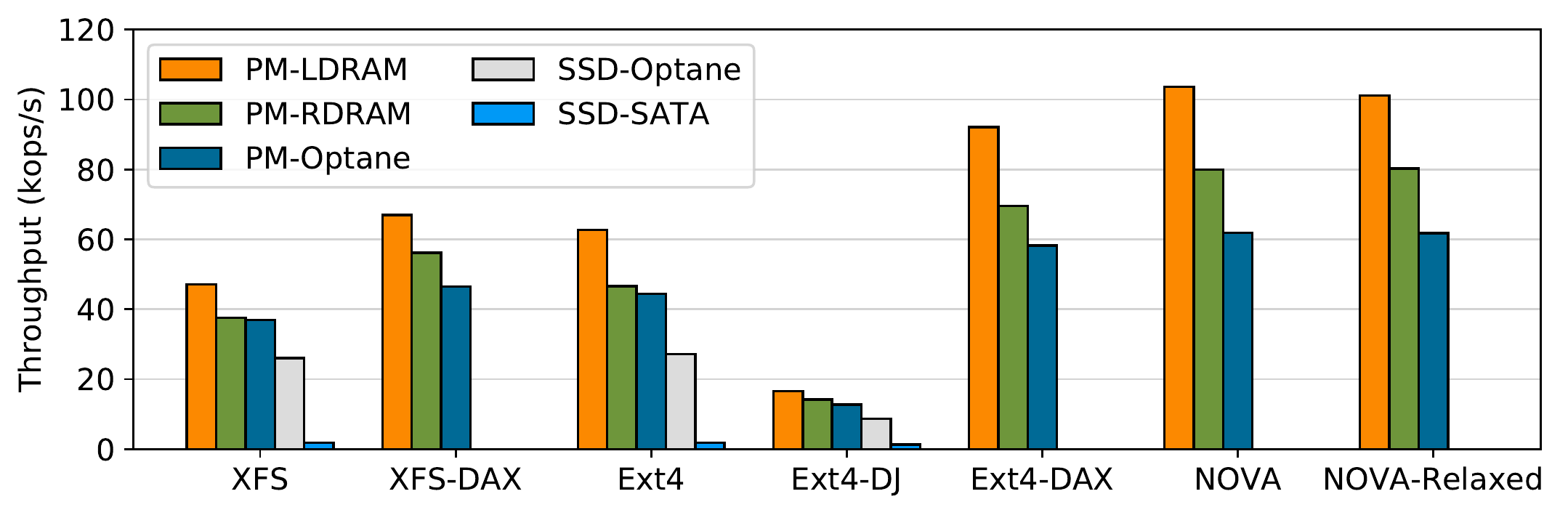,{\figtitle{LMDB throughput} This graph shows
the throughput of the LMDB key-value store on a write-dominant workload
that accesses keys sequentially~\seedatain{csvroot/storage/app/lmdb.csv}.},fig:lmdb]

\subsection{MongoDB}
\label{sec:mongodb_storage}

MongoDB is an open-source, NoSQL, document-oriented database program~\cite{mongodb}.
It supports pluggable storage engines, which are components of the database that manage storage
and retrieval of data for both memory and storage. In this section, we use MongoDB~3.5.13
with its default storage engine, \textit{WiredTiger} (WT).
The WT engine maintains data in memory, journals updates to the database to ensure immediate
persistence of committed transactions, and creates periodic checkpoints of the in-memory
data~\cite{mongodb-wiredtiger}.

We use the Yahoo Cloud Serving Benchmark (YCSB~\cite{ycsb}) to evaluate the performance
of MongoDB using its default engine. YCSB allows running
a write-dominant (YCSB-A with 50\% 
reads and 50\% 
updates) and a read-dominant (YCSB-B
with 95\% 
reads and 5\% 
updates) workload against MongoDB through a user-level client 
that interacts with the MongoDB server via TCP/IP. We have configured YCSB to 
populate the database with 100~K entries (26~byte keys and 1024~byte values) prior to 
executing 100~K operations (based on the workload characteristics) against the database. 

We run both server (MongoDB) and client (YCSB) processes on the same socket and report 
the single threaded throughput for YCSB-A and YCSB-B workloads.
\reffig{fig:mongodb_storage} shows the result.

\wfigure[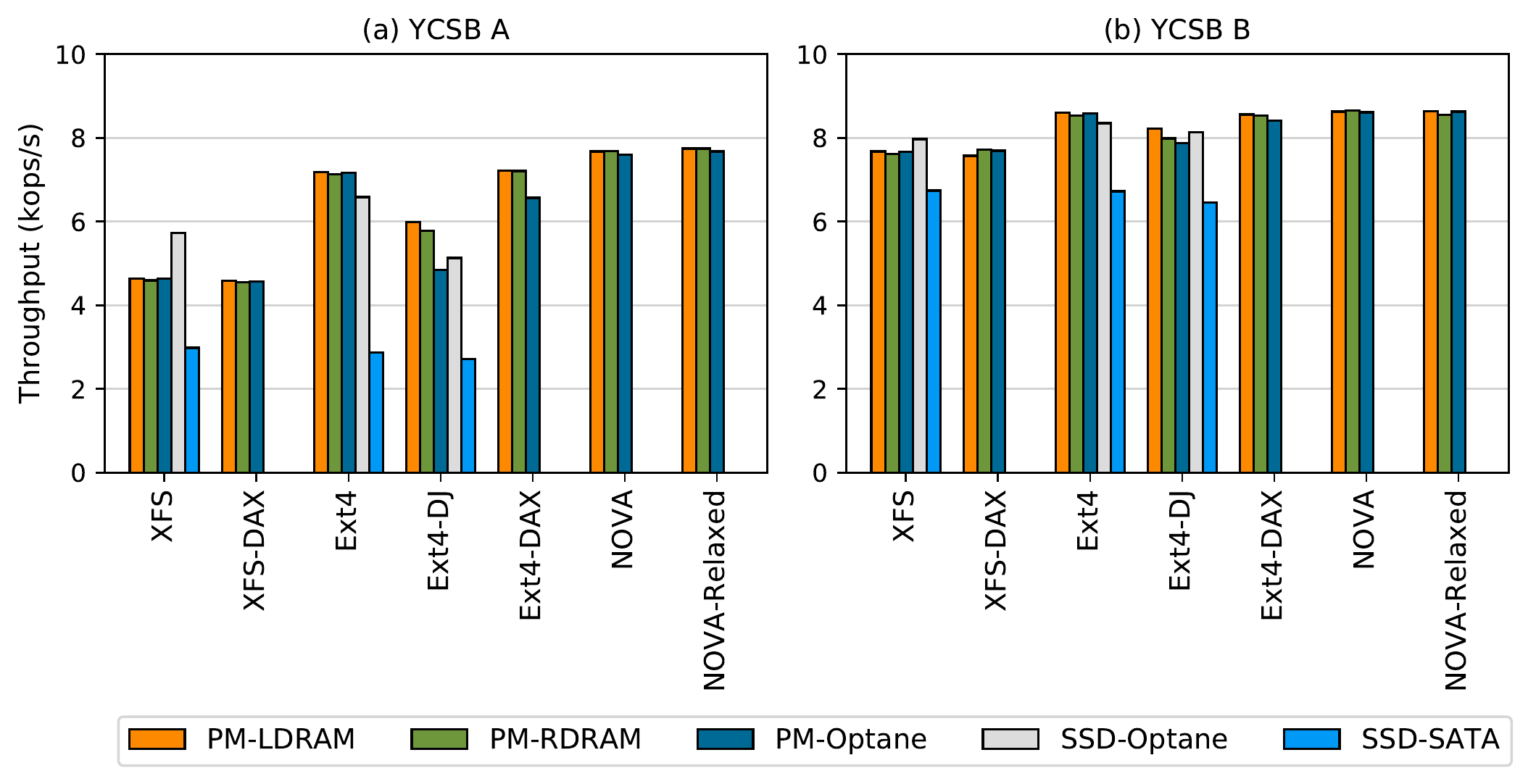,{\figtitle{MongoDB throughput with YCSB workloads} This graph shows
the single threaded throughput of MongoDB on a write-dominant workload (YCSB-A) in (a) and 
on a read-dominant workload (YCSB-B) 
in (b) (see data in~\dataref{csvroot/storage/app/mongodb_a.csv} and~\dataref{csvroot/storage/app/mongodb_b.csv}).},fig:mongodb_storage]

\takeaway{Applications generally perform slower on real \XP{} than on emulated persistent memory, and the gap grows when the file system is fast.}{
This result is expected given the latency differences observed in the previous sections.
}

\takeaway{Block-oriented file systems are not necessarily slower than their DAX counterparts
in real-world application benchmarks,
especially on read-oriented workloads.}{
This result seems to indicate the importance of using the DRAM page cache for boosting 
application performance.
}

\takeaway{Native NVMM file systems (\nova{}, \nova{-Relaxed}) generally provide better performance than adapted file systems throughout all applications we studied, especially those that use frequent sync operations.}
{Although this trend might not be the case for other types of applications or workloads, our result highlights the value of native NVMM file systems and efficient sync mechanisms.}

\section{\XP{} as Persistent Memory}
\label{sec:pmem}

While the \XPCommercial{} can be used as either a memory or storage device,
perhaps the most interesting, and novel, use case is when it is both;
that is, when it is a persistent memory device.  In this role,
\XP{} memory provides user space applications with direct access to persistent
storage using a load/store interface.  User space applications that
desire persistent memory access can mmap a file into their virtual address
space.  The application can then use simple loads and stores to access persistent data,
and use cache line flushes to ensure that writes leave the caches and become persistent
on \XP{} memory.
In this section, we investigate the performance of software designed to access
persistent memory from user space, without the need for an intervening file 
system.  Like the previous section on storage, we again expose the memory
as a pmem device, and use the relevant configurations 
(\textbf{\PMLDRAM{}}, \textbf{\PMRDRAM{}},
and \textbf{\PMLPMEM{}}).

\subsection{Redis-PMEM}
\label{sec:redis_pmem}

Our first persistent memory application is a modified version
of Redis~\cite{redis} (seen previously in 
Sections~\ref{sec:redis-mm} and~\ref{sec:redis_storage}). 
We used a forked repository of Redis 3.2~\cite{pmem-redis} that 
uses PMDK’s libpmemobj~\cite{intel-pmdk} for ensuring that its
state is persistent (and no longer uses a logging file, as was done
previously in 
Section~\ref{sec:redis_storage}). 
As with Section \ref{sec:redis_storage}, we use the
redis\-benchmark executable to measure the throughput. 
In order to compare the results side-by-side, 
we used the same configuration as the Section~\ref{sec:redis_storage}: 
4B for both key and value, 12 clients generated by a single thread, 
and a million random MSET operations.

\cfigure[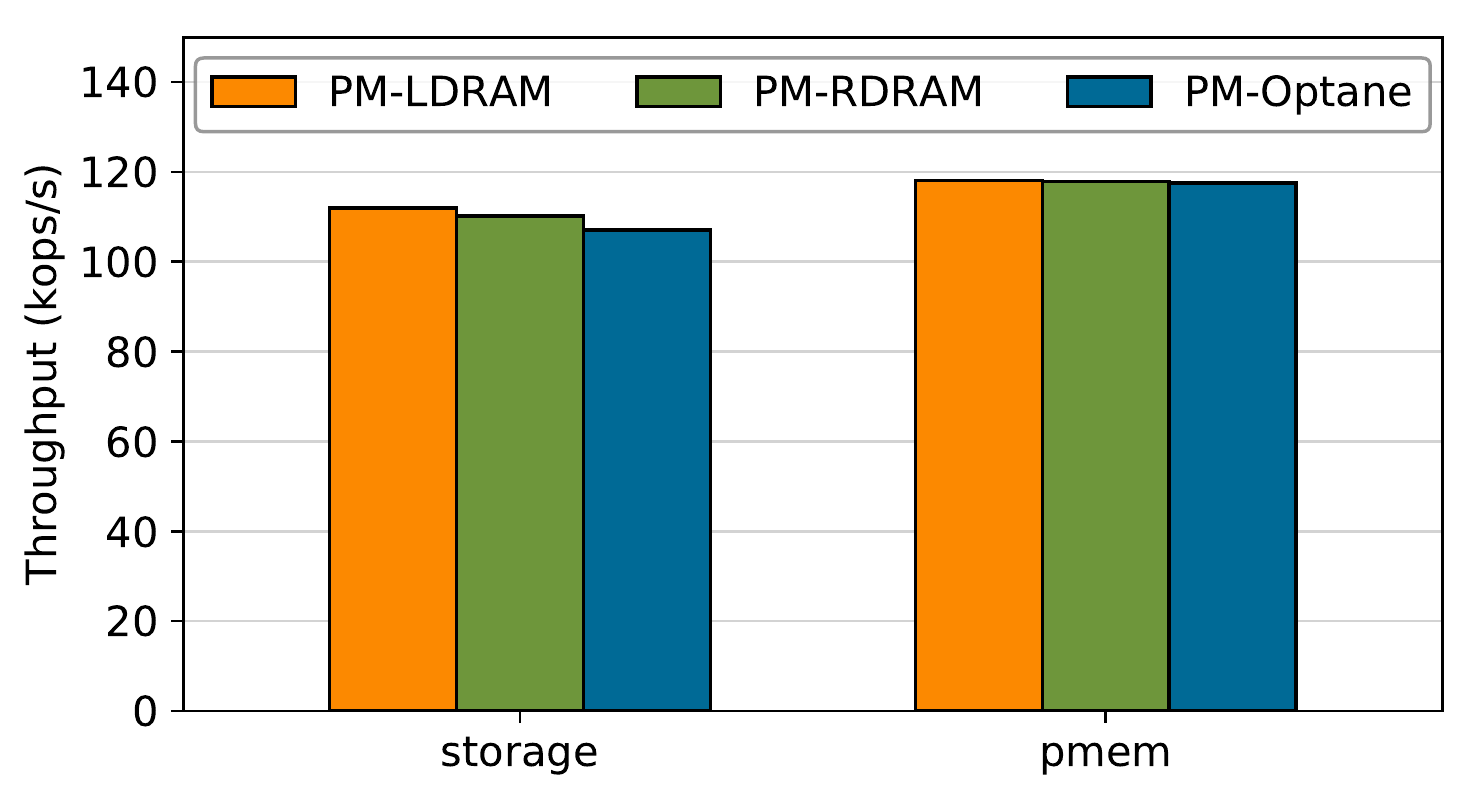,{\figtitle{Redis throughput on file systems or 
user-level persistence} This result compares Redis logging on an NVMM-aware file
system (NOVA-Relaxed) on the left to a persistent memory-aware version of Redis using Intel's PMDK
library to ensure that its state is persistent in user-space.  Notably,
the PMDK version on the right has better performance, indicating
the utility of user-space persistence that bypasses the file 
system~\seedatainas{Graphs/redis-persistent/redis_pmem.csv}{csvroot/pmem/redis_pmem.csv}.},fig:redis_pmem]

Figure~\ref{fig:redis_pmem} shows Redis's throughput
with two potential usages of the \XPDIMM{}. 
The left set of bars are a direct copy from section \ref{sec:redis_storage},
where Redis used a backing file on the NOVA-Relaxed file system to ensure data persistence. 
The right set is the PMDK version of 
Redis when using \XP{} as user-space persistent memory. 
Interestingly, the PMDK version of Redis outperforms
the file-backed Redis, even when the file-backed Redis is run on an NVMM-aware
file system.  This result indicates that custom user-space persistent
libraries are likely to be useful for performant applications, and in 
order for programmers to capture the promise of fast NVMM persistence,
application level changes may be required.  

\subsection{RocksDB-PMEM}
\label{sec:rocksdb_pmem}

Our next persistent memory application is a modified version
of RocksDB.
Since the volatile memtable data structure in RocksDB contains 
the same information as the write-ahead log (WAL) file, we can eliminate the latter 
by making the former a persistent data structure, thereby making
RocksDB an NVMM-aware user space application. 
Among the several data structures that RocksDB supports, 
we modified the default skip-list implementation and made it crash-consistent in NVMM 
using PMDK's libpmem library~\cite{intel-pmdk}.
In our RocksDB experiment, we used the same benchmark 
as in \refsec{sec:rocksdb_storage} and compare to the best
results that used the write-ahead log file (\nova{-Relaxed} for this benchmark).
Figure~\ref{fig:rocksdb_pmem} shows the throughput of both modes.

\cfigure[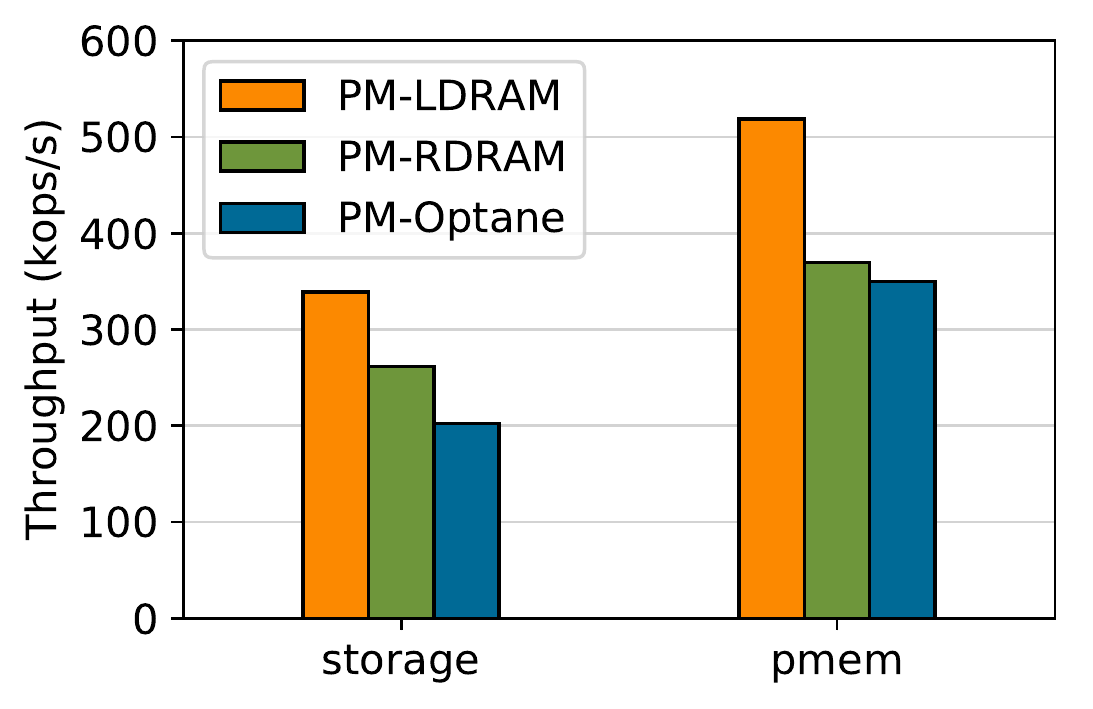,{\figtitle{RocksDB throughput with 
persistent skip-list} The performance of the persistent memory-aware RocksDB 
implementation with a persistent memtable
outperforms that of write-ahead-logging, volatile memtable architecture 
by a wide margin~\seedatain{csvroot/pmem/rocksdb.csv}.},fig:rocksdb_pmem]

The left set of bars (storage) is the result of volatile memtable backed by WAL using \nova{-Relaxed},
and the right set of bars (pmem) is the result of crash-consistent memtable made persistent in NVMM.
As with our Redis results in \refsec{sec:redis_pmem}, 
the persistent data structure provides better performance than
using both a volatile data structure and file-backed logging mechanism. 
Unlike Redis, which has network stack overheads, 
RocksDB is embedded software, and, consequently, its achieved gain is much larger (73\% 
on \PMLPMEM{}).
 
\subsection{MongoDB-PMEM}
\label{sec:mongodb_pmem}

Our third persistent memory application is MongoDB.
We extend the experiment setup in \refsec{sec:mongodb_storage} to measure the performance implications of 
replacing MongoDB's default storage engine (WT) with Intel's persistent memory storage engine for MongoDB 
(\textit{PMem}~\cite{intel-pmse}). The PMem engine uses Intel's PMDK~\cite{intel-pmdk} to 
transactionally manage MongoDB's data and to obviate the need to create snapshots and/or journal.

\reffig{fig:mongodb} shows the performance impact of using \PMLDRAM{}, \PMRDRAM{}, and 
\PMLPMEM{} to store MongoDB's data using either the default
WT storage engine (with a snapshots and journaling) or using the PMem storage engine. 
We run both server (MongoDB) and client (YCSB) processes on the same socket and report 
the single-threaded throughput for YCSB-A and YCSB-B workloads.

\wfigure[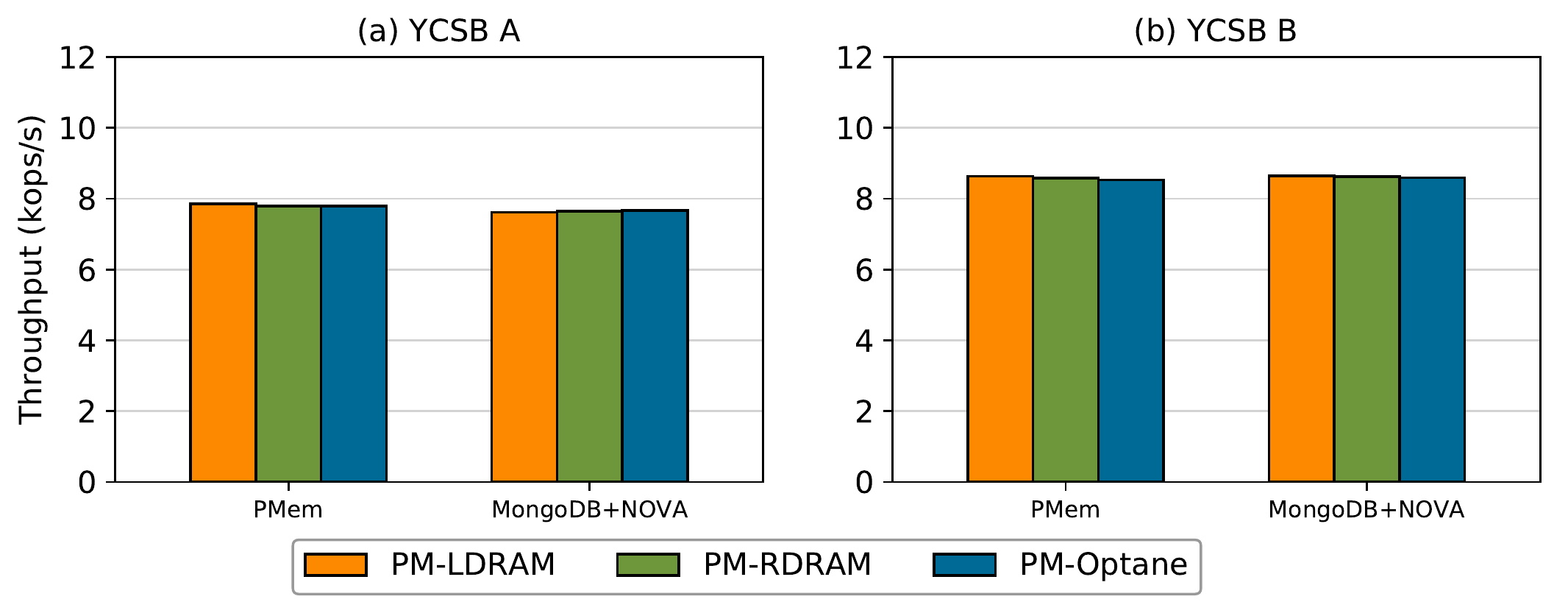,{\figtitle{Measuring the single-threaded throughput of MongoDB} 
using write-dominant (YCSB A) and read-dominant (YCSB B) workloads in presence of PMem and 
WiredTiger storage engines. The PMem engine outperforms MongoDB's WiredTiger for all 
configurations~\seedatain{csvroot/pmem/mongodb.csv}.},fig:mongodb]

\takeaway{Switching between \PMLDRAM{}, \PMRDRAM{}, and \PMLPMEM{} does not have a considerable 
impact on the performance (i.e., throughput) of running YCSB-A and YCSB-B workloads against 
MongoDB storage engines.}
{We believe this observation correlates to the high cost of the client-server communications 
between the YCSB client and MongoDB server as well as the software overhead of MongoDB's query 
processing engine.}

\takeaway{PMem storage engine provides similar performance to MongoDB's default storage engine 
(WiredTiger) for both write-dominant (YCSB A) and read-dominant (YCSB B) workloads.}

\subsection{PMemKV}
Intel's Persistent Memory Key-Value Store (PMemKV~\cite{intel-pmemkv}) is an NVMM-optimized 
key-value data-store. It implements various tree data structures (called ``storage engines'') 
to index programs data and uses the Persistent Memory Development Kit (PMDK~\cite{intel-pmdk}) 
to manage its persistent data.

We run our evaluation using PMemKV's benchmark tool 
to test the two available storage engines: \textit{kvtree2} and \textit{btree}. 
The kvtree2 engine adopts PMDK to implement a B+Tree similar to NV-Tree~\cite{188438}, where only 
the leaf nodes are persistent and the internal nodes are reconstructed after a restart. The btree 
engine employs copy-on-write to maintain a fully-persistent B+Tree.

\reffig{fig:pmemkv} reports average latency for five single-threaded runs for each 
configuration, with each run performing 2~million operations with 20~byte keys and 
128~byte values against a 16~GB memory-mapped file backed by \nova{}. 
Each configuration varies the operation performed: either 
random insert (fillrandom), sequential insert (fillseq), overwrite, random read (readrandom) 
and sequential read (readseq) operations.

\wfigure[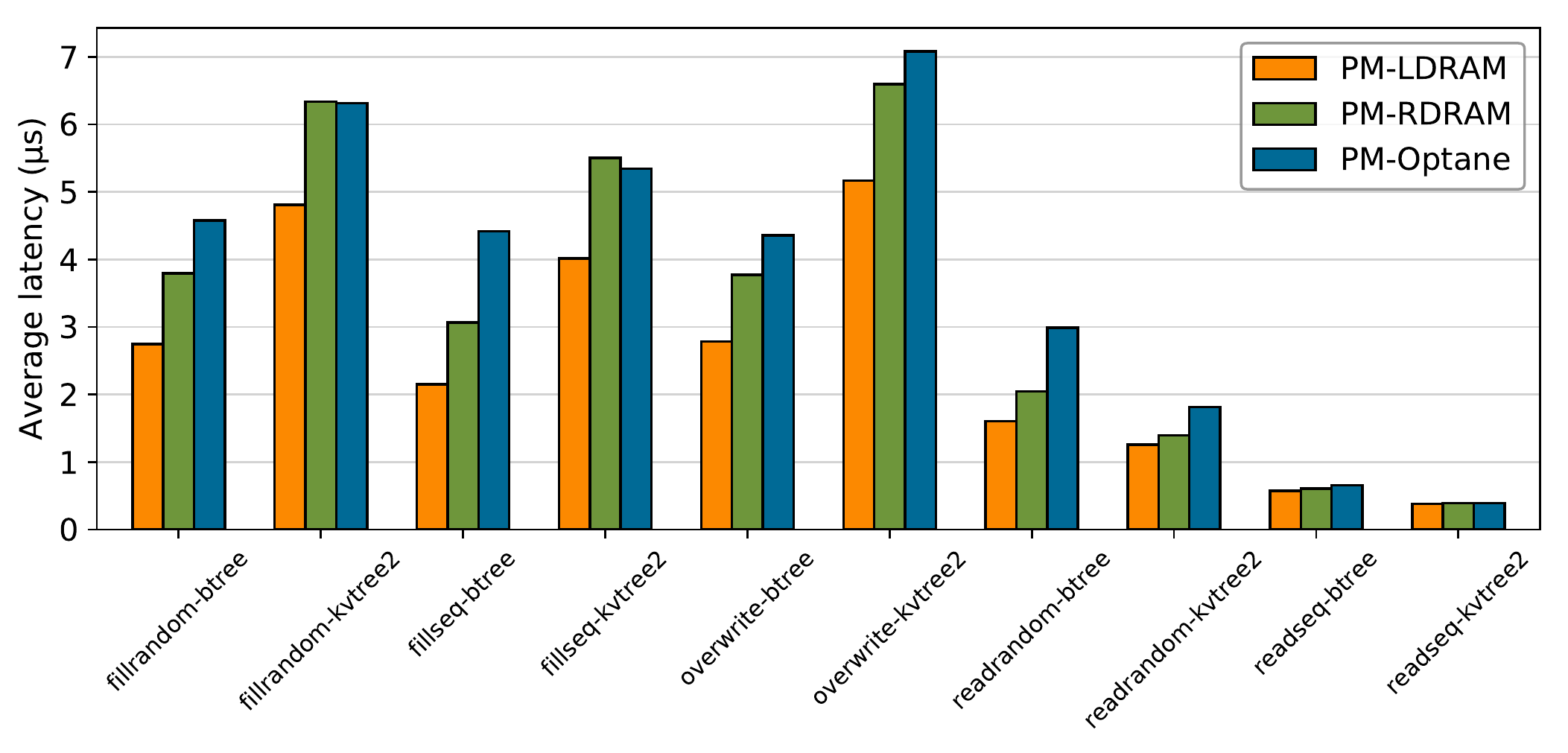,{\figtitle{Implications of \XP{} on Intel's PMemKV performance}: 
We report average latency of performing random insert, sequential insert, random read, sequential 
read and overwrite operations against PMemKV's storage engines (btree and kvtree2). Compared 
to \PMLDRAM{}, running PMemKV on \PMLPMEM{} shows similar latency for sequential read 
but up to 2.05$\times$ higher latency for write operations~\seedatain{csvroot/pmem/pmemkv.csv}.},fig:pmemkv]

\takeaway{For sequential reads in applications, \XP{} memory provides comparable latency to DRAM.}
{In comparison to \PMLDRAM{}, running PMemKV on \PMLPMEM{} increases the latency by 2\% to 15\% 
for sequential read and between 45\% and 87\% for random read operations.}

\takeaway{\XP{} incurs up to 2.05$\times$ higher latency 
for PMemKV write operations in comparison to \PMLDRAM{}.}
{This result agrees with the write performance gap between 
\XP{} memory and DRAM measured in \refsec{sec:basic}.}

\subsection{WHISPER}
The Wisconsin-HPL Suite for Persistence (WHISPER~\cite{whisper}) is a benchmark suite for non-volatile 
main memories. It provides an interface to run a set of micro and macro benchmarks 
 against a particular NVMM setup (e.g., \PMLDRAM{}, \PMRDRAM{}, and \PMLPMEM{}) 
and reports the total execution time of each benchmark. WHISPER also provides a knob to configure the 
size of the workloads to be \textit{small}, \textit{medium}, or \textit{large} --- we use
the large configuration in our test.  \reffig{fig:whisper} reports the execution 
time of running each benchmark normalized to its \PMLDRAM{} execution time as well as the average 
for all benchmarks.

\wfigure[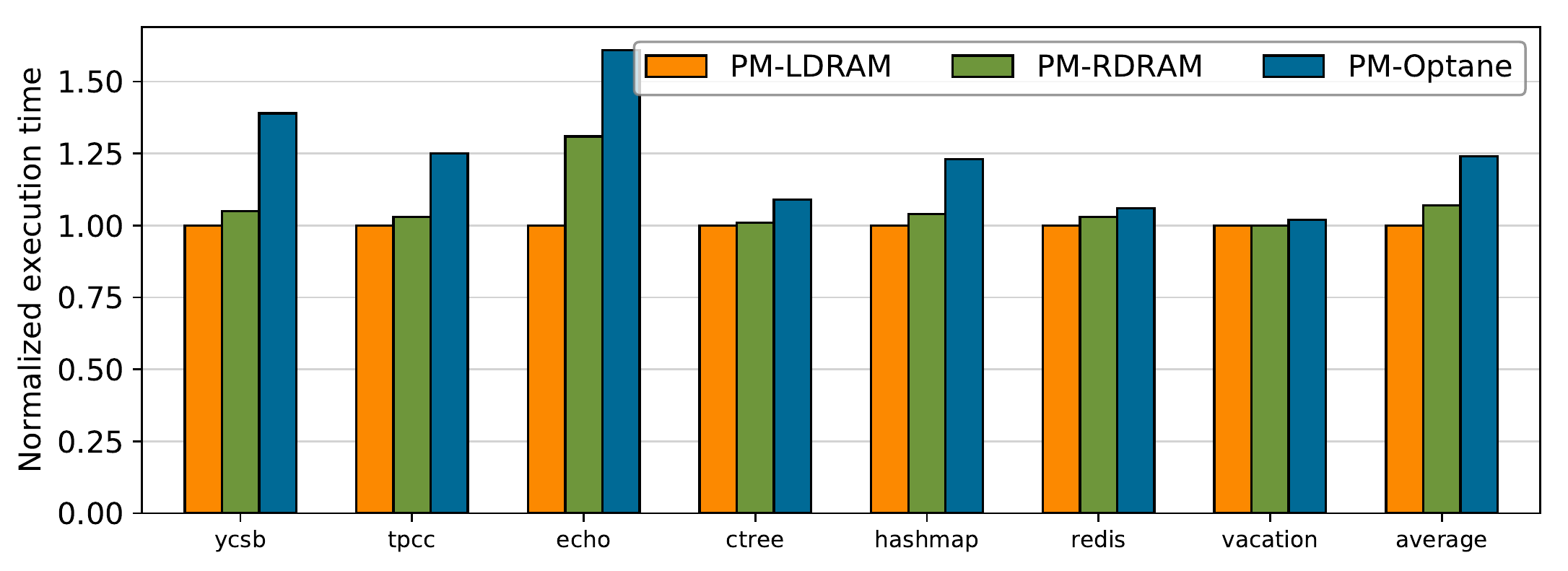,{\figtitle{Using WHISPER to measure the impact of \XP{} on the 
performance of applications.} In comparison to \PMLDRAM{}, \PMLPMEM{} and \PMRDRAM{} increase the 
execution time of WHISPER benchmarks by an average of 24\% and 7\%, 
respectively~\seedatain{csvroot/pmem/whisper.csv}.},fig:whisper]

\takeaway{In comparison to \PMLDRAM{}, \PMLPMEM{} increases execution time of WHISPER benchmarks by an 
average of 24\%.}
{This is an expected outcome due to the performance gap between \XP{} memory and DRAM.}

\takeaway{The performance difference between \PMLPMEM{} and \PMLDRAM{} is greatest for persistent data 
structures and lowest for client-server applications.}
{We observe that the portion of persistent memory accesses of each benchmark correlates to the 
gap between its \PMLDRAM{} and \PMLPMEM{} execution times.}

\subsection{Summary}

In summary, we offer a global look at applications run across
all different devices, file systems, and with user-space persistence
(Figure~\ref{fig:app_summary}).  This graph demonstrates not only
the wide range of options for providing persistent storage, but
also the benefits of deeply integrating \XP{} memory into the
system stack.  As we accelerate the storage media and remove
software overheads on the critical path to persistence, real-world
applications get significantly faster.  This figure represents the 
storage outlook of the near future as we migrate from old devices
and interfaces onto a far flatter and faster storage stack.

\wfigure[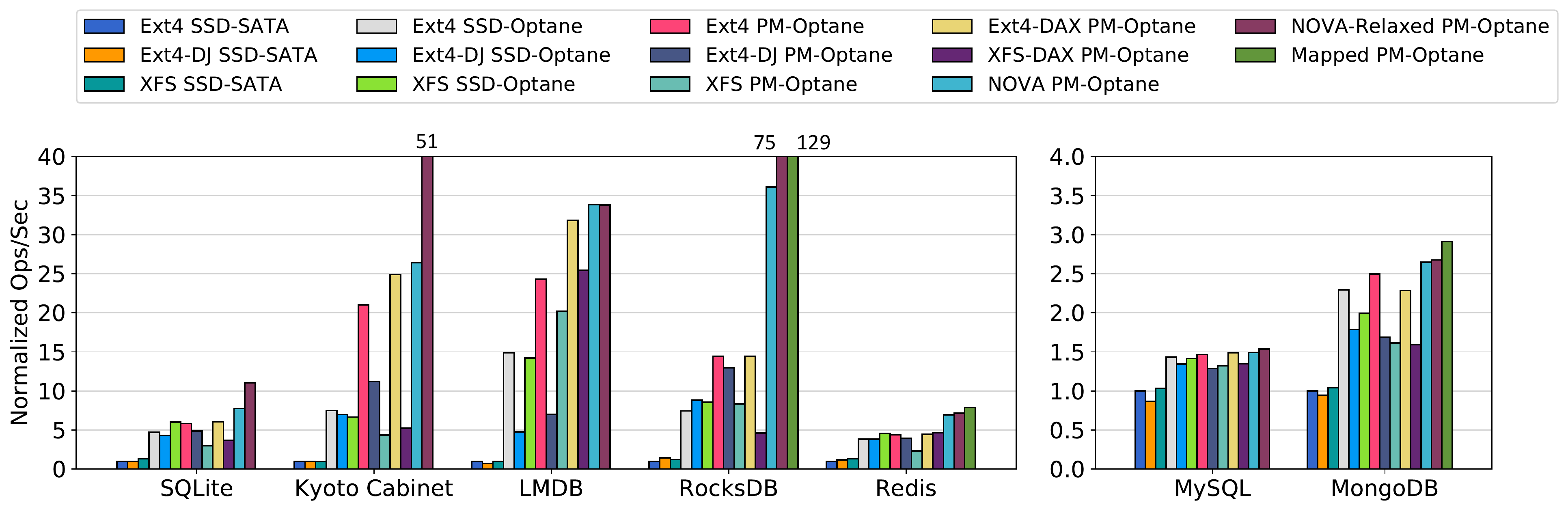,{\figtitle{Application throughput on \XP{} and
      SSDs} These data show the impact of more aggressively integrating \XP{}
    into the storage system.  Replacing flash memory with \XP{} in the SSD
    gives a significant boost, but for most applications deeper integration
    with hardware (e.g., putting the \XP{} on a DIMM rather than an SSD) and
    software (e.g., using an PMEM-optimized file system or rewriting the
    application to use memory-mapped \XP{}) yields the highest
    performance (see data in~\dataref{csvroot/storage/app/summary_1_full.csv} 
    and~\dataref{csvroot/storage/app/summary_2_full.csv}).},fig:app_summary]

\takeaway{Performance improves as \XP{} memory becomes more integrated into the
storage stack.}{The major performance difference between \XP{} memory and previous
storage media means that software modifications at the application level may reap 
significant performance benefits.}

\section{Conclusion}
\label{sec:conc}

This paper has provided a large sampling of performance
experiments on Intel's new \XPCommercial{}.  These experiments
confirm that the \XPDIMM{} creates a new tier of memory technology
that lies between DRAM and storage, and that its performance
properties are significantly different from any medium that
is currently deployed.

Our experiments, though early, were able to come to some
conclusions.  \XP{} memory, when used
in cached mode, provides comparable performance to DRAM for
many of the real world applications we explored and can greatly
increase the total amount of memory available on the system.  
Furthermore, \XP{} memory  provides significantly faster 
access times than hard drives
or SSDs and seem well positioned to provide a new layer
in the storage hierarchy when used in an uncached mode.  For many
real-world storage applications, using \XP{} memory and
an NVMM-aware file system will drastically accelerate performance.
Additionally, user-space applications that are NVMM-aware can achieve
even greater performance benefits, particularly when software overheads
are already low.  That said, it appears that previous research exploring
persistent memory software systems have been overly optimistic in assuming
that \XP{} memory will have comparable performance to DRAM
(both local and remote), and further work remains to be done in 
adapting these new designs to real \XP{} memory.

In closing, we hope that the data presented here will
be useful to other researchers exploring these new memory
devices.  Compared to what we now know about other memory
technologies, this report is only the beginning.  We believe
important questions remain both unasked
and unanswered, and that future work is necessary to complete
our understanding.

\bibliography{libpaper/common,paper}
\markforarxiv{libpaper/common.bib}
\markforarxiv{paper.bib}
\bibliographystyle{plain}

\appendix
\section{Observations}
\allrecommendations

\end{document}